%% file: main.tex
\documentclass[notitlepage,12pt]{article}

\usepackage{amssymb,amsmath,amsfonts,eurosym,geometry,ulem,graphicx,color,setspace,sectsty,comment,footmisc, achicago,pdflscape,array,hyperref,subfig,threeparttable, rotating, chngcntr, subfloat, pifont, mathtools, longtable, threeparttablex,cite}

\usepackage[font=small]{caption}
\newcommand{\squeezeup}{\vspace{-2.5mm}}
\renewcommand{\cite}{\citeasnoun}

\newcolumntype{L}[1]{>{\raggedright\let\newline\\arraybackslash\hspace{0pt}}m{#1}}
\newcolumntype{C}[1]{>{\centering\let\newline\\arraybackslash\hspace{0pt}}m{#1}}
\newcolumntype{R}[1]{>{\raggedleft\let\newline\\arraybackslash\hspace{0pt}}m{#1}}

\begin{document}

\begin{titlepage}
\title{Investor Emotions and Earnings Announcements\thanks{I am extremely grateful to Stefania Albanesi, Lee Dokyun, Sera Linardi, and Dan Berkowitz for their continued guidance and support throughout this project. Special thanks to StockTwits for sharing their data. I would also like to thank Graham Beattie, Andy Koch, Mark Azic, Doug Hanley, and Mallory Avery for helpful comments and suggestions. This research was supported in part by the University of Pittsburgh Center for Research Computing through the resources provided. All errors are my own. Preliminary results, I welcome feedback.}}

\author{Domonkos F. Vamossy\thanks{Department of Economics, University of Pittsburgh, d.vamossy@pitt.edu.}}
\date{\today}
\maketitle

\noindent

\begin{abstract}
	\singlespacing
	Armed with a decade of social media data, I explore the impact of investor emotions on earnings announcements. In particular, I test whether the emotional content of firm-specific messages posted on social media just prior to a firm's earnings announcement predicts its earnings and announcement returns. I find that investors are typically excited about firms that end up exceeding expectations, yet their enthusiasm results in lower announcement returns. Specifically, a standard deviation increase in excitement is associated with an 7.8 basis points lower announcement return, which translates into an approximately -5.8\% annualized loss. My findings confirm that emotions and market dynamics are closely related and highlight the importance of considering investor emotions when assessing a firm's short-term value. 

\noindent \\
\noindent\textbf{Keywords:} deep learning; investor emotions; capital markets. \\
\vspace{0in}\\
\noindent\textbf{JEL Codes: G41; L82.} \\
 \bigskip
\end{abstract}

\setcounter{page}{0}
\thispagestyle{empty}
\end{titlepage}
\pagebreak \newpage

\setstretch{1.4}

\section{Introduction} \label{sec:introduction}

Conventional wisdom often posits that changes to asset prices are the result of investor emotions.
For instance, \citeN{galbraith1994short} describes stock market bubbles as ``speculative euphoria'', while headlines such as ```Gut Feelings' Are Driving the Markets'' or ``How Emotion Hurts Stock Returns'' are common (e.g., \citeN{NYT1} and \citeN{NYT2}). Alan Greenspan, as a chairman of the Federal Reserve, famously remarked that the U.S. stock market exhibited an ``irrational exuberance'' when it experienced a rapid run-up in 1996. This observation portrayed a belief on his part that the increase had an origin in traders' positive emotions. In contrast, fear is cited as a force leading to sell-offs, price declines, and price variability. Market volatility indices, such as the CBOE's VIX, are often referenced as ``fear'' indices.

In this paper I test whether firm-specific investor emotions predict a firm's earnings and announcement returns. Specifically, I explore the following research questions: (1) Do investor emotions foreshadow earnings surprises? and (2) Do investor emotions predict announcement returns? Only recently has academic literature begun exploring the role emotions play in capital markets. Due to data difficulties regarding  the measurement of investor emotions, studies mainly relied on indirect proxies, or have been restricted to experimental evidence. By pairing a large, novel dataset with recent advances in text processing, I am able to overcome the data challenge inherent in studying investor emotions. I find that investors are typically excited about firms that end up exceeding analysts expectations, yet their enthusiasm results in lower announcement returns.\footnote{Throughout the paper I use the word excited, enthusiastic and happy interchangeably.}

To get to my answer, I use data from StockTwits, a social networking platform for investors to share stock opinions. A critical feature of this data is that it contains firm-specific messages, so I am able to compute firm-specific emotions. I employ a broad sample of over 4 million messages that span the decade starting in 2010. My analysis focuses on earnings announcements because they are recurring, paramount corporate events that are followed closely by capital market participants.

The primary challenge to studying my research questions is finding a way to quantify investor emotions. I overcome this by using deep learning and a large, novel dataset of investor messages.\footnote{For other applications of deep learning in economics see \citeN{albanesi2019predicting} and \citeN{vitaliy}. For reviews of machine learning applications in economics, see \citeN{mullainathan2017machine} and \citeN{athey2019machine}.} In particular, I construct emotion variables corresponding to seven emotional states: neutral, happy, sad, anger, disgust, surprise, fear. Emotion variables are generated by first quantifying the content of each message using textual analysis, and then averaging the textual analysis results across all messages by firm-quarter. My emotion variables, developed using StockTwits posts, are probabilistic measures, and hence the seven emotions sum up to 1. In addition to measuring the emotional content, I also distinguish between different types of messages by employing two classification schemes.\footnote{The emotions in this paper correspond to the seven emotional states specified in \citeN{breaban2018emotional}. I provide a detailed description of my classification schemes in the Appendix \ref{app:NLP}.} The first one isolates messages conveying information related to earnings, firm fundamentals or stock trading from general chat. The second separates messages conveying original information from those disseminating existing information. 

Once the emotion variables are constructed, I then use a fixed effects model, exploiting within firm variation in investor emotions, to test whether emotions predict a firm's earnings and announcement returns. I use firm fixed effects to isolate within firm variation. For instance, if a firm tends to have positive earnings surprises, this might make investors always more excited before announcements, and by including fixed effects, I can rule out that my results are driven by this. I also control for year, month, and day-of-the-week fixed effects, to rule out that my results are only driven by factors which effect emotions and returns across all firms simultaneously. I take a number of steps to mitigate additional concerns regarding the estimation. To ensure that I am not picking up reactive emotions, I look at the impact of pre-announcement emotions on earnings announcements, so there is a clear temporal separation between my independent and dependent variables. I tackle misattribution - the concern that my emotion measures are not capturing emotions correctly - by training an additional emotion model and use emotion variables obtained by this model for robustness checks and by investigating the impacts of contemporaneous emotions and asset prices, and find that my algorithm classifies messages as happier when they are talking about assets that have gone up in value. 

I document two main findings. First, that inter-firm investor emotions can predict the company's quarterly earnings. In particular, variation in how happy investors are is linked with marginally higher earnings surprises. Second, I find a negative relationship between the immediate stock price reaction to the quarterly earnings announcement and both within- and inter-firm variation in investor excitement. I show that this result is driven both by messages conveying original information and by those disseminating existing ones. When considering messages that convey information directly related to earnings, firm fundamentals, and/or stock trading relative to those messages which consist of other information, I find that the former has a slightly larger impact on announcement returns.  


I also confirm a behavioral finance theory, investigate heterogeneous impacts across firm and user types, and provide robustness checks. First, I provide support to theory from \citeN{shu2010investor}, positing a negative relationship between investor mood and expected returns. Second, I show that the predicted stock price reaction to earnings announcements by investor emotions is stronger for more volatile firms. This finding is in line with theory on investor sentiment, indicating that emotions and sentiment are closely related. Third, I find that it is investor emotions extracted from posts by retail investors and not institutions that best predict announcement returns. Last, to corroborate the negative relation between earnings announcement returns and investor enthusiasm, I use alternative emotion variables based on an emotion metadata compiled by other researchers. This finding remains significant with a comparable point estimate, confirming that it is indeed investor enthusiasm that drives my results.

This analysis contributes to the literature on behavioral finance in a variety of ways. First, I contribute to literature studying the connection between market behavior and emotional state. Existing research has shown that traders' moods can lead to price movements at the market level. Unlike my paper, a number of studies have leveraged indirect proxies to infer emotions. For instance, \citeN{kamstra2003winter} observe that returns are relatively low in the darker seasons of fall and winter, an outcome they presume is the result of the effect that weather has on mood.\footnote{Similarly, \citeN{hirshleifer2003good} find that good weather is correlated with higher stock returns, and appeal to a similar intuition to explain their results.} Studies relying on indirect proxies have severe limitations. For instance, \citeN{jacobsen2008weather} suggest that the findings reported in \citeN{kamstra2003winter} may be explained by a number of other factors related to the season (they illustrate with ice cream consumption and airline travel), thus questioning their conclusion that changes in investors' moods associated with the Seasonal Affective Disorder (SAD) directly influence stock market returns. On the other hand, research using direct emotion proxies has been limited to studying the relationship between investor emotions and daily stock returns, and aside from \citeN{li2016can}, has not used firm-specific emotions. \citeN{bollen2011twitter} find that Twitter mood predicts subsequent stock market movements, while \citeN{gilbert2010widespread} find that the level of anxiety of posts on the blog site Live Journal predicts price declines. I add to this literature by showing that firm-specific investor emotions predict both firms' announcement returns and earnings surprises.

The relationship between market behavior and emotional state has also been studied in controlled laboratory experiments. These papers explored the role of emotions in generating bubbles in experimental asset markets. For instance, \citeN{breaban2018emotional} measures emotions using traders' facial expressions and find that positive emotion is linked to higher prices and larger bubbles, while traders' fear before the market opens is associated with lower prices. Similarly, \citeN{andrade2016bubbling} document larger bubbles when induced investor enthusiasm is higher. I contribute to this literature by showing that emotions captured by investor messages on a social media platform behave similarly to emotions in the lab. Specifically, I find that stocks that enjoy high levels of investor enthusiasm leading up to the earnings announcement will experience smaller announcement returns. 

Another contribution of this paper is to literature studying the role social media plays in capital markets.\footnote{The importance of social media is voiced in studies exploring how companies exploit this channel as a means for investor communication. For instance, \citeN{blankespoor2014role} show that firms can reduce information asymmetry among investors by broadly disseminating their news, including press releases and other disclosures, to market participants using Twitter. \citeN{jung2018firms} find that roughly half of S\&P 1500 firms have created a corporate presence on either Facebook or Twitter.}$^{,}$\footnote{Adding to the literature on StockTwits and Twitter is research that has examined investors' use of Internet search engines, financial websites, forums, and other social media platforms. This research has provided mixed evidence on whether this information helps predict future earnings and stock returns. Using Google search volume as a proxy for investors' demand for financial information, \citeN{da2011search} find that increased Google searches predict higher stock prices in the near-term followed by a price reversal within a year. \citeN{drake2012investor} show that the returns-earnings relation is smaller when Google search volume before earnings announcements is high. \citeN{antweiler2004all} and \citeN{das2007yahoo} both find that the volume of posts on message boards, such as Yahoo! or Raging Bull, is associated with stock return volatility, but not stock returns. \citeN{chen2014wisdom} demonstrate that information in user-generated research reports on the Seeking Alpha investing portal helps predict earnings and long-window stock returns following the report posting date.} These studies have investigated whether social media content can predict the overall movement of the stock market. For instance, \citeN{mao2012correlating} find that the daily number of tweets that mention S\&P 500 stocks is significantly associated with the changes in that same index. This literature has also analyzed how Twitter and/or StockTwits activity influences investor response to earnings. \citeN{curtis2014investor} find that high levels of activity correlate with greater sensitivity to earnings announcement returns and earnings surprises, while low levels of social media activity are associated with significant post-earnings-announcement drift. \citeN{cookson2020don} show that even though it is unlikely that investor trades from those on StockTwits move the market, disagreement measured by these messages robustly forecast abnormal trading volume. I add to these papers by showing that emotions extracted from social media can help predict a firm's earnings and announcement returns.

The closest connection to my paper is with \citeN{bartov2017can}, who find that aggregate opinion from Twitter can help predict a firm's forthcoming quarterly earnings and announcement returns. I examine a different feature of the environment, and ask what features of opinion help predicting the company's earnings and the market response to earnings? To do so, I construct my emotion variables to leverage aspects of textual content ignored by traditional sentiment models by incorporating emojis and emoticons and create a multi-dimensional object. The low correlation between sentiment and emotions illustrates that they contain different information.

My final contribution is to research on the value of diversity\footnote{This hypothesis originates from \citeN{hong2004groups}, who show that a diverse group of intelligent decision-makers reaches reliably better decisions than a less diverse group of individuals with superior skills and concludes that under certain conditions, ``diversity trumps ability''. Interestingly, traditional information intermediaries, such as financial analysts, tend to herd to the consensus viewpoint (\citeN{jegadeesh2010analysts}) and produce inefficient earnings forecasts (\citeN{abarbanell1991analysts}), perhaps because they belong to a rather small and homogeneous group (\citeN{welch2000herding}). This is relevant to the research questions of this paper, because StockTwits has a diverse set of investors with widely different investment philosophies.} and the wisdom of crowds\footnote{The wisdom of crowds refers to the phenomenon that aggregated information provided by many often results in better predictions than those made by any single group member, even when that member is an expert. \citeN{surowiecki2004wisdom} presents numerous case studies and anecdotes to illustrate the principle. One such example comes from the work of Sir Francis Galton: after observing a weight-judging competition at a county fair in 1906, Galton found that the crowd accurately predicted the weight of an ox when their guesses were averaged. The average guess was closer to the ox's true weight than most the individual predictions, including estimates coming from cattle experts, butchers, and farmers. A similar outcome was witnessed in \citeN{berg2008results}, which revealed the remarkable ability of the Iowa Electronic Markets to predict high-profile elections, outperforming polls conducted by experts. Recent research that builds on the wisdom of crowds concept shows that the content of tweets can be used to predict: (1) earnings announcement returns (\citeN{bartov2017can}), and (2) future returns around Federal Open Market Committee (FOMC) meetings (\citeN{azar2016wisdom}).} hypotheses. First, in line with the value of diversity hypothesis, I find that the predictive power of emotions are diminished when considering only groups of (more) homogeneous users by segmenting them by sharing similar investment horizons, trading experience, trading approach, popularity and account types. Second, I add to wisdom of crowds literature by showing that investor messages are better predictors when surrounded by higher user engagement.

The question whether firm-specific emotions from social media help predict firms' earnings and announcement returns has been left unexplored thus far in the literature. This is the very question that I subdivide into two main points of examination in my paper. My first research question investigates whether stock specific investor emotions, obtained from individual messages written prior to the earnings announcement, predict the company's earnings surprise. This is the case, for instance, if investors are excited about firms that end up exceeding analysts expectations. If the opposite holds, i.e. investors are systematically enthusiastic about firms disappointing expectations, it would add to the list of behaviors retail investors exhibit that adversely affects their financial well-being (e.g., \citeN{barber2013behavior}). My second research question examines the relation between stock specific investor emotions, obtained from individual messages written prior to the earnings announcement, and the market response to earnings. To address this question, I control for the earnings surprise. This allows me to explore the nature of the StockTwits information that predicts stock returns. If the information conveyed by emotions is above and beyond earnings realizations, then the coefficient on emotions will continue to be significant even after controlling for the content of the report.

The two research questions outlined delineate different aspects on the role investor emotions play. The first one examines if investor emotions are informative. If there is an earnings surprise, not all information has been aggregated in earnings expectations. If the earnings surprise can be predicted by emotions, then emotions carry this missing information. The second one investigates whether investor emotions influence price dynamics. If investor enthusiasm contains information beyond earnings realizations, then the relationship between announcement returns and enthusiasm should be nonzero. These two research questions are important since they address whether and how investor emotions influence the way new information is incorporated into asset prices. I find that the value relevance of emotions for stock returns stems not only from predicting the earnings surprise, but also from other information relevant to stock valuation not accounted for by unobservable time-invariant stock characteristics, time patterns, or by control variables used in prior research. Results of this paper suggest that investor emotions are important determinants of stock returns.

The remainder of the paper is organized as follows. Section \ref{sec:theory} provides a theoretical framework, Section \ref{sec:data} describes the data; Section \ref{sec:strategy} provides the empirical strategy; Section \ref{sec:findings} presents my primary results; Section \ref{sec:additional_findings} confirms the results of theoretical work by \citeN{shu2010investor}, explores heterogeneous effects and conducts a sensitivity analysis. Section \ref{sec:conclusion} concludes.

\section{Theoretical Framework}\label{sec:theory}

The theoretical framework for this paper comes from \citeN{shu2010investor}.\footnote{An alternative theory is provided by \citeN{duxbury2020emotions}, who present an emotion-based account of buy and sell preferences in asset markets. Specifically, they leverage psychological research (e.g., \citeN{loewenstein2001risk}) and propose that when the price of a single asset increases (decreases) above its purchase price, anticipatory hope increase (decreases).}  \citeN{shu2010investor} modifies the Lucas model (\citeN{lucas1978asset}), and shows how investor mood variations affect equilibrium asset prices and expected returns. Specifically, equity prices correlate positively with investor mood, with higher asset prices associated with better mood. In contrast, expected asset returns correlate negatively with investor mood. Given this, we expect to find positive contemporaneous relationships between investor enthusiasm and excess returns, while a negative relationship between pre-announcement investor enthusiasm and announcement returns. I provide a simple framework with a potential mechanism in Section \ref{app:model}.

\section{Data}\label{sec:data}

\subsection{Data Sources}

\subsubsection{StockTwits Data}

My investor emotion dataset comes from StockTwits, which was founded in 2008 as a social networking platform for investors to share stock opinions. StockTwits looks similar to Twitter, where users post messages of up to 140 characters (280 characters since late 2019), and use ``cashtags'' with the stock ticker symbol (e.g., \$AMZN) to link ideas to a particular company. Although the app does not directly integrate with other social media platforms, participants can share content to their personal Twitter, LinkedIn, and Facebook accounts.

My original dataset spans the decade of 2010; starting from 1 January, 2010 until 31 December, 2019. In total, there are 117,354,459 messages by 416,249 unique users mentioning 9,742 tickers. For each message, I observe sentiment indicators as tagged by the user (bullish, bearish, or unclassified), sentiment score as computed by StockTwits, ``cashtags'' that connect the message to particular stocks, like count, and a user identifier. The user identifier allows me to explore characteristics of the user, such as follower count. For most users, I also have information on self-reported investment philosophy that can vary along two dimensions: (1) Approach - technical, fundamental, momentum, value, growth, and global macro; or (2) Holding Period - day trader, swing trader, position trader, and long term investor.\footnote{I group technical with momentum and value with fundamental for my heterogeneity explorations. For investment horizon, I explore day traders and long term investors.} Users of the platform also provide their experience level as either novice, intermediate, or professional. Leveraging textual analysis, I also distinguish between institutional and retail investor accounts. This user-specific information about the style, experience, type and investment model employed is useful to explore heterogeneity in investor emotions.

I restrict my sample to cover stocks traded on NASDAQ/NYSE, and remove messages that appear automated.\footnote{I define automated messages as messages posted over 1,000 times by the same user over the period 2010-2019.} I focus on messages that can be directly linked to particular stocks, so I restrict attention to messages that only mention one ticker. Last, I require at least two users posting per stock for the duration over which I compute averages to discard noisy signals. I summarize my sample restrictions in Table \ref{tab:sample_restrictions}. 

\input{Tables/sample_restrictions}

\begin{figure}[htbp] 
\squeezeup \squeezeup
\begin{center}
	 \includegraphics[scale=0.45]{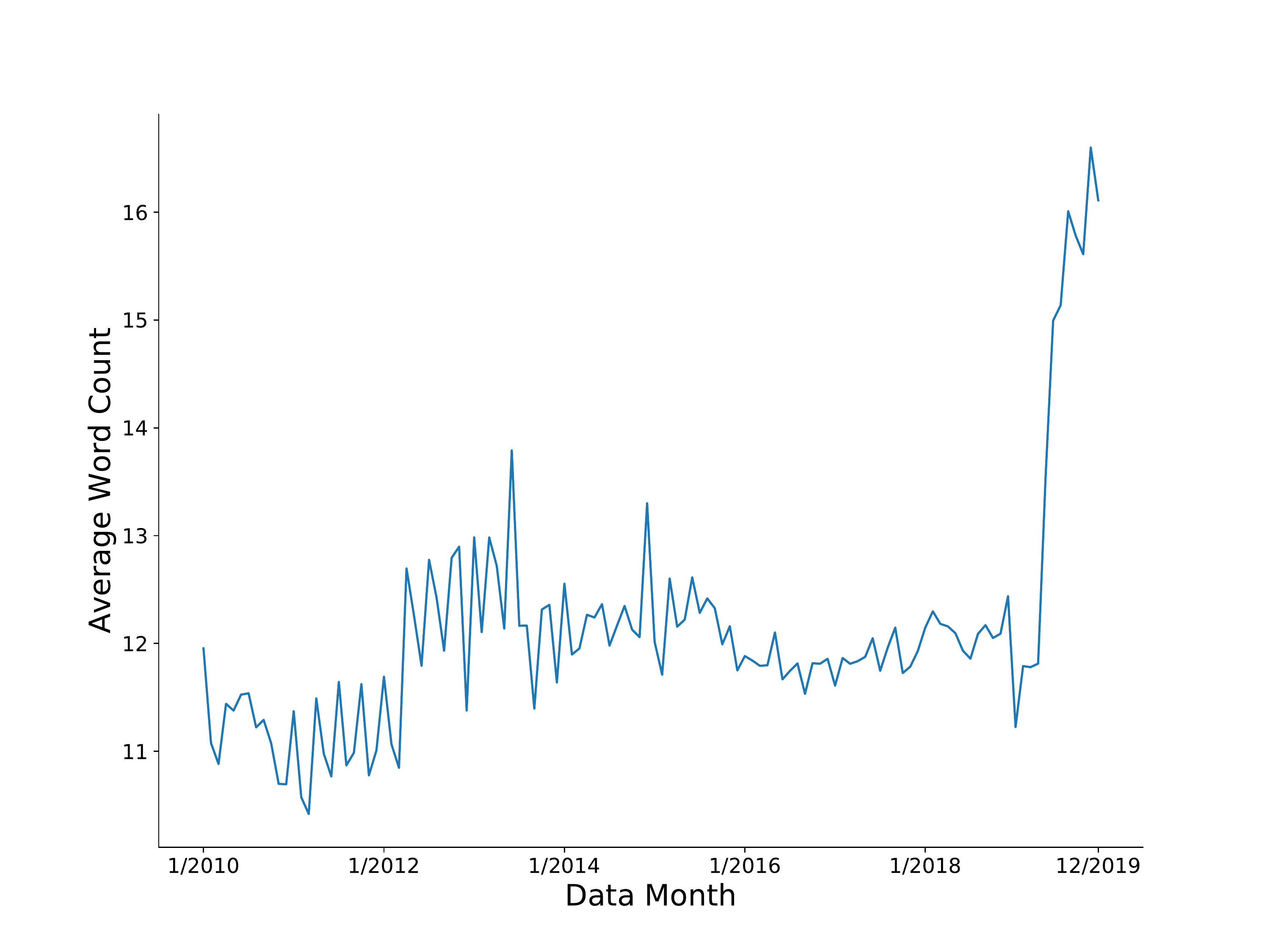}
	\caption{Time Series of Average Post Length}\label{fig:post_length}
\end{center}
\squeezeup \squeezeup
\begin{flushleft}
	\footnotesize{Notes: Similarly to Twitter, StockTwits introduced longer messages in late 2019 (280 characters).}
\squeezeup \squeezeup 
\end{flushleft}
   
\end{figure}

I plot the average word count per messages over time in Figure \ref{fig:post_length}, displaying a relatively stable trend with a spike in late 2019. This spike is due the character limit extension from 140 characters to 280 characters. Given that the average post length peaks at 16, and since I use the first 30 words to extract the emotion from messages, this likely does not effect the estimation.

\begin{figure}[htbp] 
\squeezeup \squeezeup 
\begin{center}
	 \includegraphics[scale=0.475]{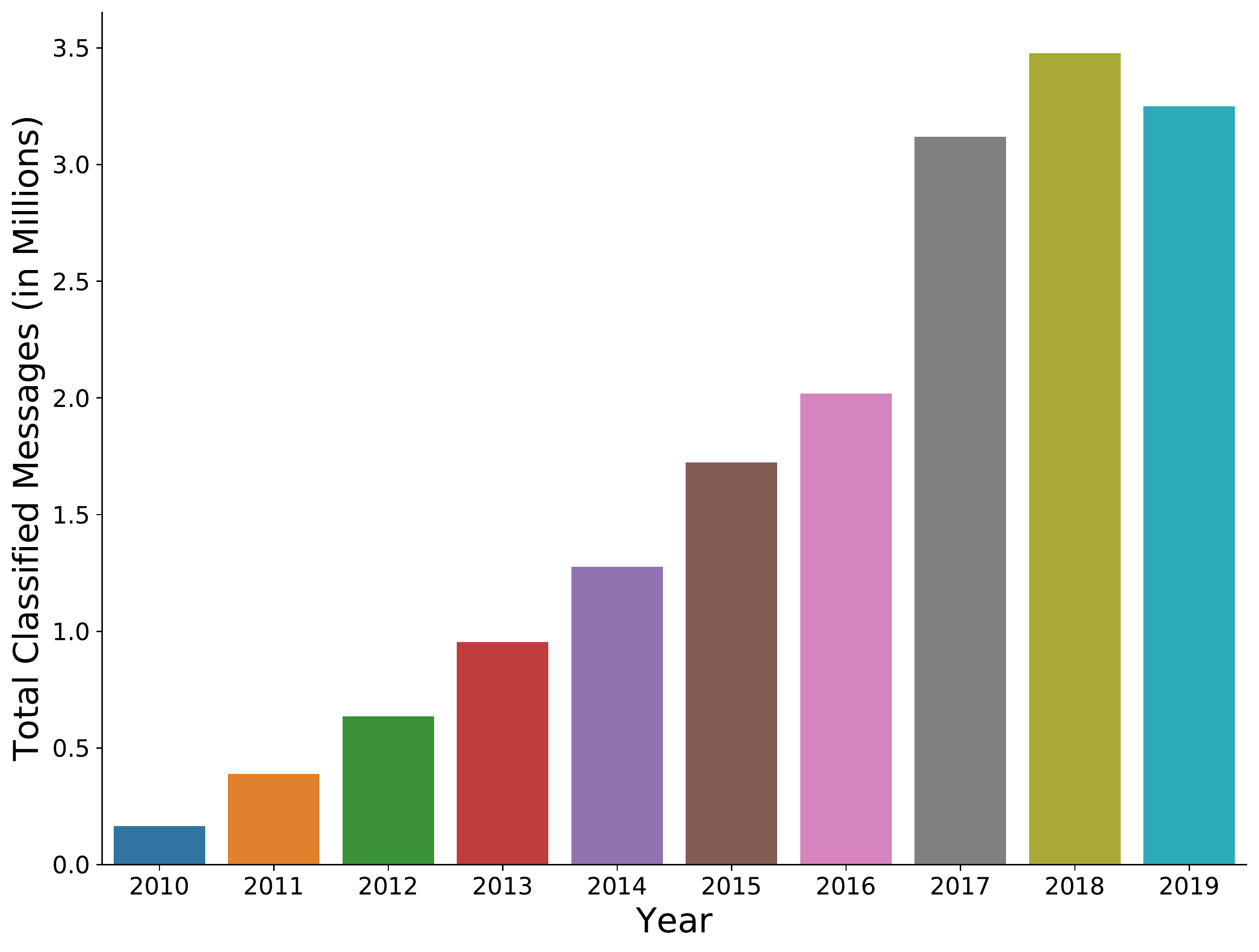}
	\caption{StockTwits Messages Over Time}\label{fig:annual_data}
\end{center}
 \squeezeup \squeezeup 
\end{figure}

Figure \ref{fig:annual_data} portrays the number of messages over time in my data, indicating substantial growth in the early years in the data, which plateaus around 2016. I control for the growing nature of my sample and the changing nature of my posts by including time fixed effects in my analysis.

\begin{figure}[htbp] 
\squeezeup \squeezeup
\begin{center}
	 \subfloat[]{\includegraphics[scale=0.315]{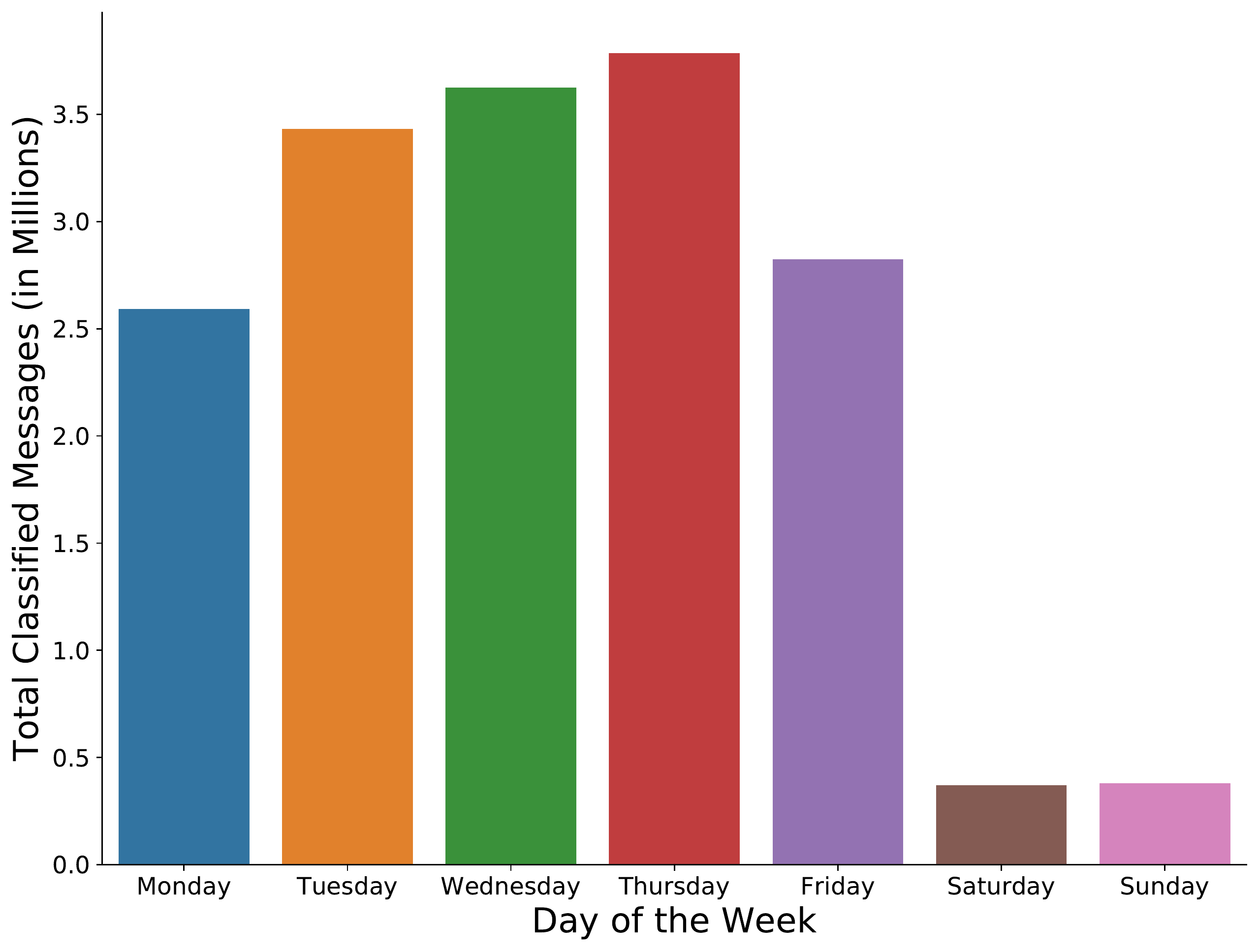}}
	\subfloat[]{\includegraphics[scale=0.315]{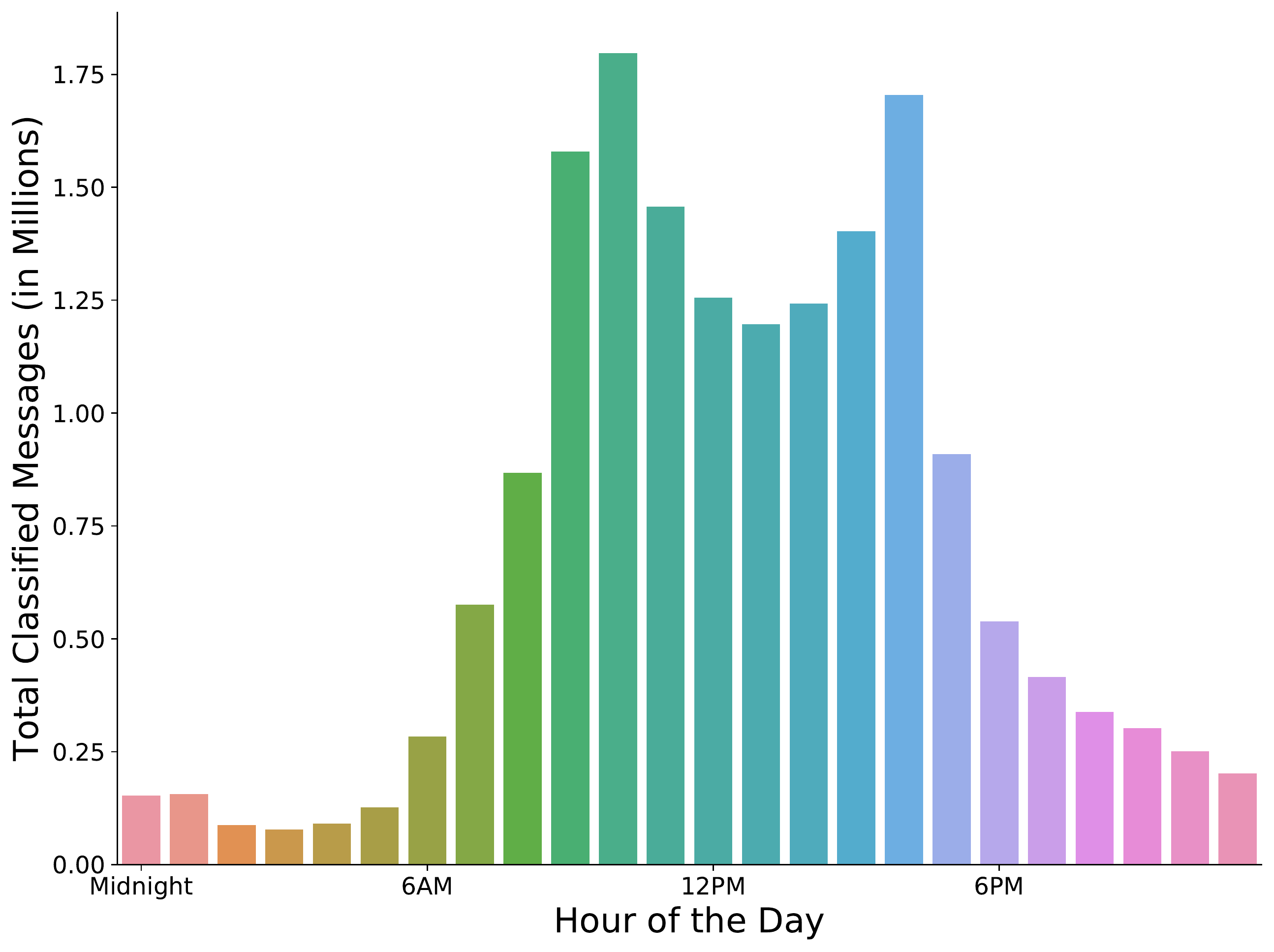}}

	\caption{Distribution of Messages}\label{fig:post_distributions}
\end{center}
\squeezeup \squeezeup  
\begin{flushleft}
	\footnotesize{Panel (a) portrays the day-of-the-week, while Panel (b) depicts the hour-of-the-day distribution of messages.}
\squeezeup \squeezeup  
\end{flushleft}
\squeezeup \squeezeup
\end{figure}

I also explore when investors post the messages. In particular, I examine whether they post messages concurrently with daily news so that it reflects hour by hour changes in beliefs, or in the evening after work, when they have more free time and then it is more of a reflective general analysis. In Panels (a) and (b) of Figure \ref{fig:post_distributions} I plot the distribution of messages by the day of the week and by the hour of the day respectively. We can clearly see that most posting activity on the platform happens when the markets are open (Monday-Friday and between 9am and 4pm). This behavior is consistent with investors updating their beliefs in real time as financial events unfold. 

\clearpage 

\begin{figure}[htbp] 
\squeezeup \squeezeup
\begin{center}
	 \includegraphics[scale=0.475]{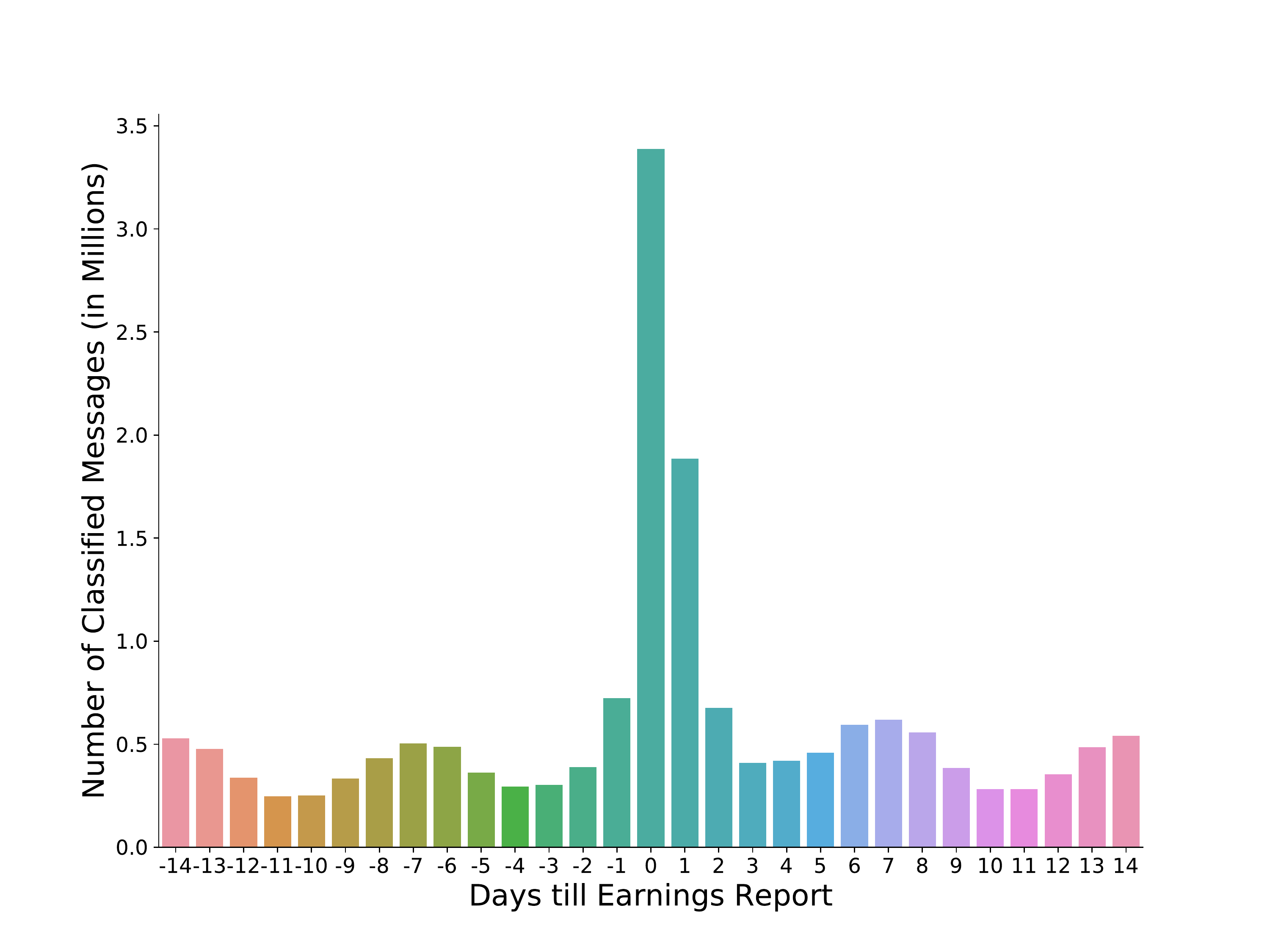}
	\caption{Posts around Earnings Announcements}\label{fig:announcement_distributions}
\end{center}
 \squeezeup \squeezeup
\end{figure}

Last, I plot the average message volume across firms surrounding earnings announcements in Figure \ref{fig:announcement_distributions}. It indicates that social media activity increases a week before the earnings announcement and peaks on the earnings announcement day. Specifically, social media activity increases by a factor of 3 on the announcement day compared to from the week prior. This dramatic increase is in line with studies documenting abnormal attention surrounding earnings announcements (e.g., \citeN{lawrence2016yahoo}). In total, I analyze 81,886 firm-earnings announcement observations spanning 4,467,461 messages.

\subsubsection{Pricing Data}

\begin{figure}[!h] 
\begin{center}
	 \subfloat[]{\includegraphics[scale=0.5]{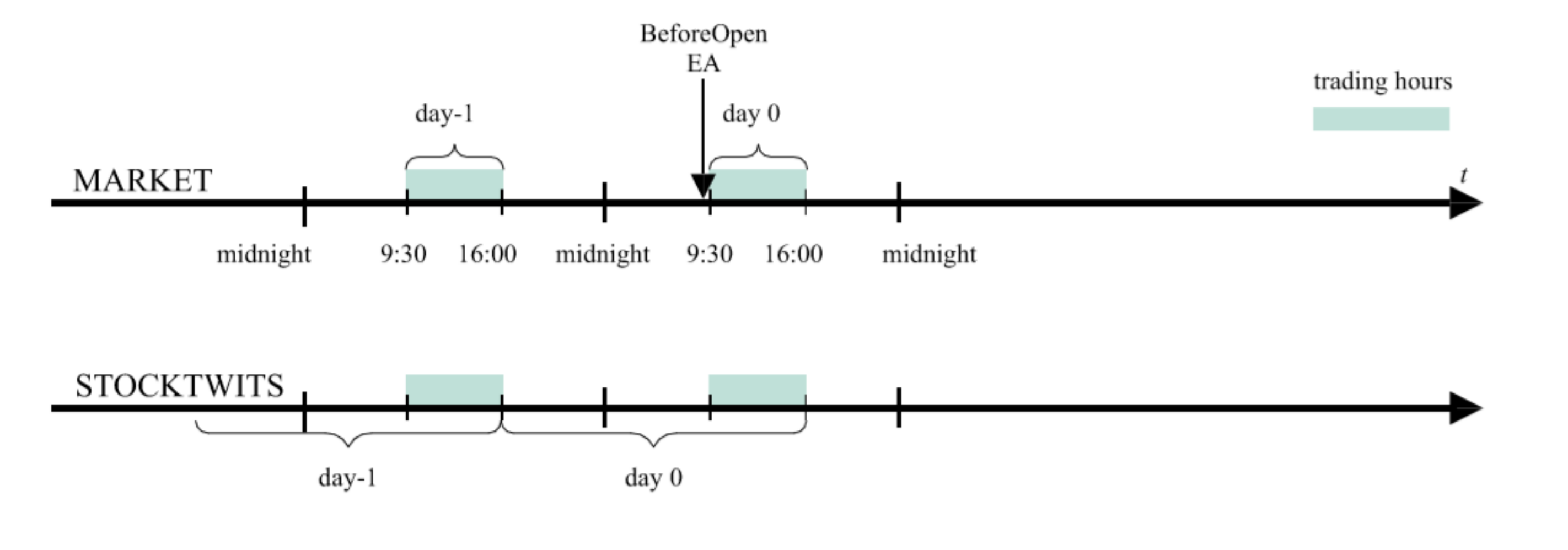}}
	 
	\subfloat[]{\includegraphics[scale=0.5]{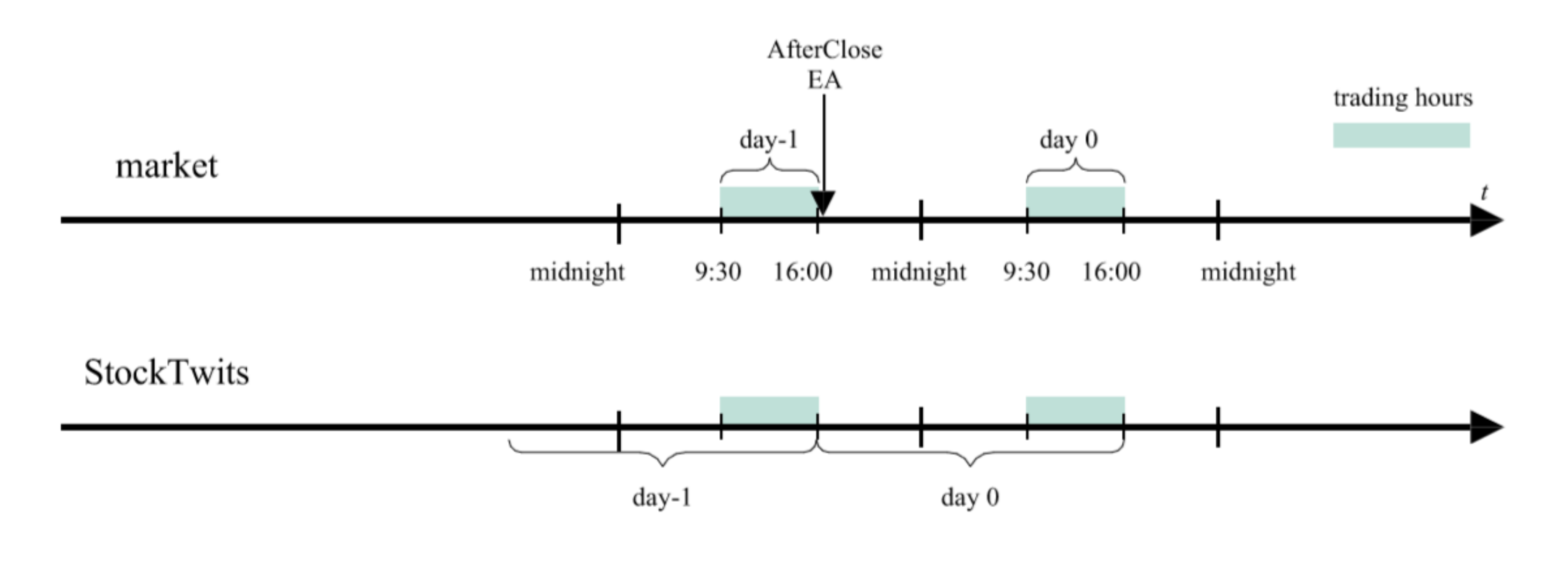}}

	\caption{Event Windows for Firms Announcing Before vs. After the Market Opens}\label{fig:event_calculation}
\end{center}
\squeezeup \squeezeup 
\begin{flushleft}
	\footnotesize{Notes: Panel (a) displays the estimation for firms reporting before the market opens, while Panel (b) portrays it for firms with announcements after the market closes.}
\squeezeup \squeezeup 
\end{flushleft}
 
\end{figure}

Price and volume-related variables are obtained from CRSP, accounting information is obtained from Standard and Poor's COMPUSTAT, analyst and earnings announcement related information is obtained from I/B/E/S, and institutional ownership data is from Thomson Reuters Institutional Holdings (13F). I match this data with StockTwits, and compute days till earnings announcements based on \citeN{gabrovvsek2017twitter}. I illustrate this in Figure \ref{fig:event_calculation}.

\subsection{Text Analysis}

I now briefly describe the text analysis methodologies used in this paper. For an in depth discussion, see Appendix \ref{app:text_processing}, \ref{app:NLP}, and \ref{app:comparison}.

\subsubsection{Messages as Indicators for Emotion}

In order for my emotion measure to be useful, it must reveal the true state of investors. Thus, before using the data, I must rule out that users are trying to manipulate the stock market by posting fake opinions. For instance, if a user believes the stock price will go down and thus wants to sell the stock, she could post positive messages that might increase the price temporarily and thus would allow her to sell at a higher price. This would invalidate my measure, as I would capture her emotion as happy, even though her current emotional state might not be. This does not, though, seem to be an important concern in my data for a number of reasons. First, there is anecdotal evidence that users post on platforms to attract followers, gain internet fame, or find employment. In all those cases, it is incentive compatible for them to provide their honest opinion about the stock. Second, I also investigate the pricing impacts concerning S\&P 1500 firms, which have large market caps that make it unlikely that individual investors could move prices.

\subsubsection{Measuring Emotion}

The primary challenge underlying my research design is the estimation of emotion. To overcome this, I use textual analysis to quantify the emotion expressed in investor messages. I leverage a large set of emojis and emoticons along with emotionally charged words to generate a dataset of investor messages with corresponding emotions. I then use a standard bi-directional GRU model with word embeddings (see \citeN{chung2014empirical}) to obtain a probabilistic assessment for each message in the data.

\subsubsection{Measuring Information Content}

To further explore the channels whereby emotions operate, I also compute the emotional state of messages separately as they relate to fundamental information (``fundamental'') or whether they look like general social media chat (``chat''). I provide examples of messages and their predicted emotion probabilities for a set of ``fundamental'' and ``chat'' posts in Table \ref{tab:twit_examples}. 

I also distinguish between messages by whether they provide original information (``original'') or they disseminate existing ones (``dissemination''). A message is considered original if (1) it is not a retweet of another user's message and (2) it does not include a hyperlink.

\subsubsection{Measuring Sentiment}

StockTwits uses an unclassified supervised learning model to generate a sentiment score for messages, and reports this score and statistics of this on its platform. I found that a large fraction of messages receive a score of 0, meaning that the message is either unclassified or has no forward looking sentiment. To be consistent with prior research, I use a Naive Bayes model  (\citeN{bartov2017can}).\footnote{I use the Naive Bayes classifier developed by \url{https://textblob.readthedocs.io/en/dev/_modules/nltk/classify/naivebayes.html}, and classify messages with a predicted probability just under 0.49 as negative, and just over 0.51 as positive, and hence, my neutral class contains messages with a sentiment between 0.49 and 0.51.}

\subsection{Differences between Emotion and Sentiment}

Since early investor sentiment studies, such as \citeN{de1990noise}, research has revealed that investor sentiment and emotion are closely related. Examples include optimism (pessimism) or hope about (fear of) the future. As \citeN{shiller2003efficient} suggested, excessive price volatility in asset markets may indicate that investors' decisions are influenced by such optimism or pessimism. \citeN{tetlock2007giving} provides empirical support to \citeN{de1990noise} by documenting that high media pessimism exerts downward pressure on prices through short-term spikes in trading volume. Still, there are three important distinctions between my emotion measure and sentiment. 

First, the main difference is definition. Unlike emotion, investor sentiment is defined as ``a belief about future cash flows and investment risks that is not justified by the facts at hand'' (\citeN{baker2007investor}). Now, whether a model not trained specifically on social media data can extract this component is not within the scope of this paper. Nonetheless, to alleviate such concerns I also train a deep learning based sentiment model trained on messages pre-tagged by the author of the post as ``bullish'' or ``bearish''. 

The second is dimensionality. While investor sentiment is a one dimensional object, my investor emotion is a multi-dimensional construct. This allows me to pinpoint what features of messages seem to matter more. For instance, both fear and anger are likely classified as negative, yet an angry message is different from a fearful message (see Table \ref{tab:twit_examples} for examples), and it is conceivable that firms with angry messages perform differently than firms with fearful messages. 

Third, unlike my emotion model which incorporates emojis and emoticons, the sentiment model built on the Naive Bayes classifier assigns a score of 0.5 (i.e., neutral) for each of the emoticons and emojis included in my dictionaries, both in its original format (i.e., ``:)'') and its changed format (i.e., ``happyface''). For instance, the message ``I am :)'' would be classified as happy with the emotion model, and neutral with the sentiment model. Therefore, the sentiment model measures the content of only words, ignoring potentially important information. This issue, however could be fixed by training a sentiment model that incorporates emojis and emoticons. If that would be the case, then the emotion model could be thought of as a higher dimensional sentiment measure.

\subsection{Descriptive Statistics}

I first define the variables used in my analyses in Table \ref{tab:variable_definitions}. 

\input{Tables/variable_definitions}

Table \ref{tab:summary_stats} presents the descriptive statistics for the analysis variables. I find a high fraction of neutral messages, with a mean of 72.7\% and a median of 74.9\%, mainly driven by firms with few posts (weighting firm-quarter observations by number of posts yields a mean of 51.86\% for neutral). Looking at my Naive Bayes based sentiment variable, I observe a positive skewness, with a mean of 1.867, and a median of 1.773. This might suggest a ``good-news'' bias in twits, following from investors being more likely to share their optimism on social media than pessimism. My earnings surprise variable: standardized unexpected earnings (SUE) have a mean and median of 1.052 and 0.707 respectively. This suggests that firms in my sample exceeded analysts expectations more than disappointed. My measure of abnormal returns around earnings announcements, has a slightly negative mean, $-$0.09\%, and a median of  $-$0.05\%.

\input{Tables/summary_stats} 
\input{Tables/correlations}
\clearpage 

Table \ref{tab:corr_matrix} presents pairwise correlation coefficients among my analysis variables. The variables include my StockTwits based emotion measure, earnings surprise (SUE), abnormal stock returns around earnings announcements (EXRET), and my control variables. My emotion measures in Table \ref{tab:corr_matrix} show low pairwise correlations with investor sentiment (below 0.1 for each, using Spearman and Pearson, respectively), suggesting that they may be capturing different aspects of investor opinion. The small pairwise correlation coefficients among my control variables indicate that there is little evidence of a multi-collinearity.

\squeezeup 
\section{Empirical Strategy}\label{sec:strategy}

\squeezeup
\subsection{Emotions and Earnings Surprises}

I start with addressing my first research question: do investor emotions predict the company's earnings? This would be the case when investor emotions contain information relevant to company's future earnings. In particular, it is conceivable that positive investor emotions indicate performance exceeding prior expectations. To test this question I estimate the following model:

\begin{equation}\label{eq:surprise}
Y_{ift} = \alpha_i + \sum_{j=1}^{j=6} \beta_j EMOTION_{iftj} + \gamma X_{ift} + \delta_t + \delta_f + \epsilon_{ift} 
\end{equation}

Here, the dependent variable is the earnings surprise, measured using standardized unexpected earnings (SUE) for firm f during announcement t. My test variables, EMOTION$_{iftj}$, is the average firm-specific emotion extracted from individual messages written 10 trading days before until 2 trading days before the announcement. Specifically, EMOTION$_{iftj}$ for $j \in \{\text{happy, sad, fear, disgust, angry, surprise}\}$ is a probabilistic measure of the average emotion from StockTwits, where the benchmark group is the neutral. In Equation (\ref{eq:surprise}), the hypothesis that the average emotion from individual messages is predictive of the upcoming earnings surprise implies $|\beta_j| > 0$ for some j. 

The control variables (X$_{ift}$) include: the lagged earnings surprise from the previous quarter to control for the positive autocorrelation in earnings surprises (SUE$_{i,t-1}$); \citeN{carhart1997persistence} four-factor buy-and-hold abnormal stock returns for the firm over the window $[-10,-2]$ to control for information outside of the realm of StockTwits that may have reached the capital market prior to the earnings release (EXRET$_{-10,-2}$); firm size (Size); market-to-book ratio (MB); number of analysts in the consensus I/B/E/S/ quarterly earnings forecast (ANL); institutional investor holding (INST); where applicable, indicator variable for the fourth fiscal quarter (Q4); an indicator variable for past quarterly loss (Loss). These last seven variables control for effects shown by prior research to explain the cross-sectional variation in earnings surprises. I include firm ($\delta_f$) and time ($\delta_t$) fixed effects (year, month, day of the week) to account for firm-specific and time patterns in earnings surprises that my controls might not account for. Along the lines of prior research (e.g., \citeN{petersen2009estimating}), I cluster standard errors by firm, because the errors may be correlated over time at the firm level. 


\subsection{Emotions and Announcement Returns}

I now address my second research question: Can the emotions extracted from StockTwits messages predict quarterly earnings announcement stock returns? Certainly, if emotions are irrelevant, then the answer is no. Given \citeN{shu2010investor}, I expect a negative association between pre-existing enthusiasm and announcement returns. To test this question empirically, I examine the relationship between abnormal stock returns (EXRET) in the three days around earnings announcements, $[-1,1]$, where day 0 is the earnings announcement date, and investor emotions in a nine-trading-day period leading to the earnings announcement, $[-10,-2]$. To this end, I estimate Equation (\ref{eq:surprise}) with the dependent variable being \citeN{carhart1997persistence}'s buy-and-hold abnormal stock returns for firm f during announcement t over the three-day window $[-1,1]$, $EXRET_{ift}$. 

The prediction that pre-announcement emotional states are informative of earnings announcement returns imply that $|\beta_j| > 0$ for some $j \in \{\text{happy, sad, fear, disgust, angry, surprise}\}$. This would be the case if, as discussed in \citeN{bartov2017can}, the market uses stock recommendations and analyst earnings forecasts in forming its earnings expectations and stock prices, but does not extract information as they are released from other, less prominent sources, such as StockTwits. Based on \citeN{shu2010investor}, I expect to find $\beta_{happy} < 0$. 


Here, the control variables (X$_{ift}$) leverage the findings of prior research: For instance, I include excess returns from ten days before the announcement until two days before the announcement to control for momentum in stock returns. This ensures that the effects I attribute to emotional states are not driven by momentum of pre-announcement returns. I include institutional ownership as a control variable, to acknowledge that the marginal investor who sets stock prices is a sophisticated investor whose equity valuations and earnings expectations may not only rely on analyst forecasts and recommendations. The other four variables are used to control for effects shown by prior research to explain the cross-sectional variation in stock returns around earnings announcements. I also include my realized earnings surprise variable of the current quarter (SUE) to explore the nature of the StockTwits information that predicts stock returns. If the information conveyed by emotions is above and beyond earnings realizations, then the coefficient on emotions will continue to be significant even after controlling for SUE. Once again, I include firm ($\delta_f$) and time ($\delta_t$) fixed effects to account for firm-specific and time patterns in earnings surprises that my controls might not account for. I cluster standard errors by industry-quarter, using Fama-French 48-industry groupings, because the errors may be correlated in the same calendar period across firms in the same industry.

\subsection{Estimation Concerns}

It is conceivable that firms that tend to have positive earnings surprises make investors always more excited before announcements. To rule out that my results are driven by this, I use firm fixed effects. I also control for year, month, and day-of-the-week fixed effects, to ensure that my results are not driven by factors which effect emotions and returns across all firms simultaneously. I take a number of steps to mitigate additional concerns regarding the estimation. First, to guarantee that I am not picking up reactive emotions, I look at the impact of pre-announcement emotions on earnings announcements, so there is a clear temporal separation between my independent and dependent variables. Second, I tackle misattribution - the concern that my emotion measures are not capturing emotions correctly - by training an additional emotion model and use emotion variables obtained by this model for robustness checks and by investigating the impacts of contemporaneous emotions and asset prices, and find that my algorithm classifies messages as happier when they are talking about assets that have gone up in value. One caveat of my analysis is that I do not control for traditional media coverage, and hence, I cannot exclude the possibility that it is the emotions invoked from traditional media coverage that drive my results.

\section{Primary Findings}\label{sec:findings}

\subsection{Emotions and Earnings Surprises}\label{sec:earnings_surprise}

I start with addressing my first research question: do investor emotions predict the company's earnings? Before I exploit within-firm variation, Columns (1-2) of Table \ref{tab:reg_cross_section} documents relationship between investor emotions and earnings surprises without firm fixed effects. As the results show, emotions alone can only explain some of the variation in earnings surprises (0.8\%). I find that a standard deviation increase in happiness results in a 1.8\% standard deviation increase in earnings surprises ($0.4359*0.162/3.8609$). Still, I find that investors on the platform are more enthusiastic about firms that end up beating expectations. Upon including control variables, only happy remains significant, and its effect size is reduced by approximately 30\%. 

\input{Tables/cross_sectional_reg}

I next account for unobservables by including firm, year, month, and day of the week fixed effects. Table \ref{tab:reg_pre_announcement_mood_surprise} presents the results. When I estimate the entire sample, I find that within-firm variation in anger can be useful in predicting earnings surprises (Column (1)). The significance disappears when I restrict the sample to S\&P 1500 firms (Column (2)), or when I only include messages that do not contain information about earnings or firm fundamentals (Column (3)). Looking at messages pertaining to stock fundamentals, I find a negative relationship between sad and earnings surprises, i.e., a within-firm standard deviation increase in sad is associated with a 0.9\% within-firm standard deviation decrease in earnings surprise (Column (4)). Next, the predictive power is only present for messages containing original information (Column (5)), and not for those disseminating existing ones (Column (6)). Taken together, my results provide support that investor emotions extracted from social media marginally help predicting earnings surprises.

\input{Tables/surprise_preannouncement_reg}

\subsection{Emotions and Announcement Returns}\label{sec:earnings_returns}

I now address my second research question: Can emotions extracted from StockTwits messages predict quarterly earnings announcement stock returns? I first present the results from estimating Equation (\ref{eq:surprise}) without firm fixed effects in Columns (3-4) of Table \ref{tab:reg_cross_section}. Column (3) suggests a negative relationship between emotions and abnormal returns around earnings announcements, as the coefficients on fear, anger, and happy are significantly negative. When controls from prior research explaining the cross-sectional variation in stock returns around earnings announcement are included (Column (4)), only the effect of happy remains statistically significant. Considering the results in Column (4), these impacts are not negligible; a standard deviation increase in excitement decreases announcement returns by 7.2 basis points per three trading days ($-0.4423*0.162$), an approximately -5.8\% annualized loss. 

\begin{figure}[h] 
\begin{center}
	\includegraphics[scale=0.925]{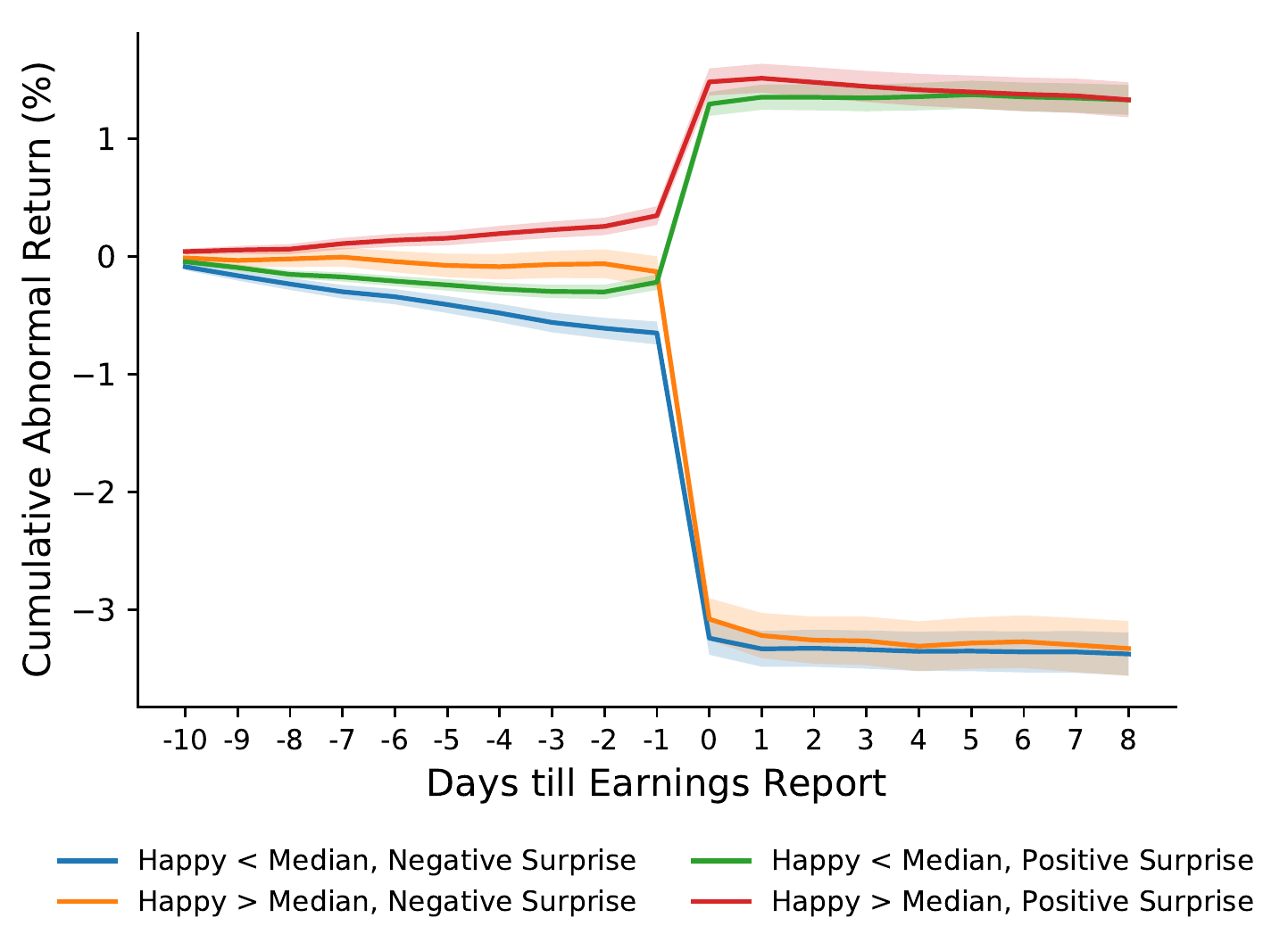}
	\caption{Emotion and Announcement Returns.}\label{fig:announcement_returns}
\end{center}
\squeezeup \squeezeup 
\begin{flushleft}
	\footnotesize{Notes: Relationship between pre-announcement happiness (i.e., $[-10,-2]$) and cumulative abnormal returns. Abnormal returns are winsorized at the 1\% and 99\% level.}
\squeezeup \squeezeup 
\end{flushleft}
\end{figure}

I illustrate this finding graphically in Figure \ref{fig:announcement_returns}. To do so, I split my sample into firms with below versus above median pre-announcement enthusiasm, and into firms with positive versus negative earnings surprise. Figure \ref{fig:announcement_returns} shows substantially different paths of cumulative abnormal returns for firms enjoying high levels of investor excitement. Specifically, when comparing firms that exceed expectations (red line versus green line), I find smaller price adjustments for firms having over the median share of happy posts. In contrast, when looking at firms that disappoint expectations (yellow line versus blue line), I document larger price adjustments. Interestingly, firms that exceed expectations but have below the median share of happy messages trend similarly to firms that disappoint expectations while having above the median share of happy messages until the announcement is released. 

Table \ref{tab:reg_pre_announcement_mood_returns} shows that this relationship holds even with firm fixed effects (Column (1-6)), for larger firms (Column (2)), and the impacts are larger when user engagement is higher (Column 3)). Columns (4-5) repeat the analysis using measures of emotions disaggregated between messages that convey earnings or trade-related information (fundamental) and messages that provide other information (chat). I find that emotions extracted from messages specifically mentioning firm fundamentals and earnings have larger point estimates. Next, contrasting Column (6) and Column (7), I find that both messages containing original information and those disseminating existing information drive my results. Since I control for realized earnings surprise, my findings suggest that the value relevance of emotions provided by StockTwits for stock returns stems not only from predicting the earnings surprise, but also from other information relevant to stock valuation not accounted for by unobservable time-invariant stock characteristics or by time patterns. 

\input{Tables/announcement_preannouncement_reg}

A comparison of the results in Tables \ref{tab:reg_cross_section}-\ref{tab:reg_pre_announcement_mood_returns} presents an interesting contrast. Investor enthusiasm extracted from messages matters both for predicting earnings surprises
and the market reaction to earnings news. In particular, it seems that investors are excited
about firms that exceed expectations (i.e., positive relationship between happy and SUE),
but their enthusiasm may lead to short-term overpricing, and hence, when I compare firms
announcing similar earnings surprises, ones that experienced higher investor enthusiasm
tend to experience lower announcement returns. While the emotion-earnings relationship
only holds without firm fixed effects, both within-firm and inter-firm variation in emotions
are indicative of the market reaction to earnings news.

\section{Additional Findings}\label{sec:additional_findings}

\subsection{Testing the Theoretical Framework}\label{sec:mechanism}

To help corroborate the theoretical work of \citeN{shu2010investor}, I further analyze the link between investor emotions and excess returns. I first estimate the relationship using contemporaneous emotions and excess returns during windows $[-10,-2]$ and $[-1,1]$. Then to confirm the negative relationship between investor excitement and expected returns, I examine the impacts of earnings announcement emotions ($[-1,1]$) on post-announcement returns ($[2,4]$). If the theory holds, I expect to see similar impacts between earnings announcement emotions and post-announcement returns as I saw for pre-announcement emotions and announcement returns. That is, when comparing companies with similar earnings surprises, ones that experienced higher enthusiasm should experience lower post-announcement returns. 

I now examine the effects of pre-announcement investor emotions on contemporaneous excess returns by estimating Equation (\ref{eq:surprise}) with $EXRET_{-10,-2}$ being my dependent variable, while excluding controls for earnings surprise. Table \ref{tab:reg_pre_announcement_mood_prereturns} presents the results. As expected, I find a large positive (negative) association between positive (negative) emotional states and excess returns. This relationship is smaller for larger firms (Column (2)), larger when user engagement is higher (Column (3)), and holds for messages of all types (Columns (4-7)). 

\input{Tables/pre_announcement_preannouncement_reg}
\input{Tables/announcement_announcement_reg}
\input{Tables/announcement_post_announcement_reg}

To provide further support that my emotion measures are capturing investor emotions accurately, I also estimate Equation (\ref{eq:surprise}) with contemporaneous (i.e., same window) emotion variables. Even after controlling for the content of the report, I find a large positive (negative) association between positive (negative) emotional states and announcement returns (Table \ref{tab:reg_announcement_day}). Given Table \ref{tab:reg_pre_announcement_mood_prereturns} and Table \ref{tab:reg_announcement_day}, my measure of happy must be picking up happiness, because when excess returns are high people should be happy, which is what my measure shows. I abstain from analyzing the contemporaneous effects further, since these could be reactive and not predictive (i.e., I do not know whether emotion leads or lags price movements). 

I next relate announcement emotions (window: $[-1,1]$) to post-announcement returns (window: $[2,4]$). The results in Table \ref{tab:reg_post_announcement} show a negative relationship between happy and post-announcement returns; effects that are smaller for larger firms (Column (2)), and larger when user engagement is higher (Column (3)); only messages related to earnings or firm fundamentals are driving the results (Column (4-5)). Taken together, these results confirm \citeN{shu2010investor}: negative relationship between investor enthusiasm and expected returns, while positive association between investor enthusiasm and contemporaneous returns.

\squeezeup
\subsection{Heterogeneous Effects}

\subsubsection{Emotions, Expectations and Volatility}

Theory on investor sentiment posits that younger, smaller, more volatile, unprofitable, non–dividend paying, distressed stocks are most sensitive to investor sentiment. Conversely, ``bond-like'' stocks are less driven by sentiment (see, \citeN{baker2007investor}). To examine whether emotions behave similarly to sentiment, I interact my happy variable with a dummy variable intended to capture high volatility stocks. In line with this, I find larger point estimates for more volatile firms (Column (2) of Table \ref{tab:reg_pre_announcement_mood_returns_heterogeneity}). 

I also explore the effect of emotions for firms exceeding versus disappointing expectations, and find larger impacts for firms disappointing expectations (Column (3) vs. Column (4) of Table \ref{tab:reg_pre_announcement_mood_returns_heterogeneity}).

\input{Tables/announcement_preannouncement_heterogeneity_reg}

\subsubsection{User Characteristics}

\citeN{hong2004groups} show that a diverse group of intelligent decision makers reach reliably better decisions than a less diverse group of individuals with superior skills. I investigate this by segmenting my messages coming from traders with similar investment horizons (long-term, short-term), trading approaches (value, technical), trading experiences (amateur, intermediate, professional), popularity levels (users with followers in the 95th percent versus the rest), and account type (institutional vs. human). 

I report heterogeneity across user types in Table \ref{tab:reg_pre_announcement_mood_returns_heterogeneity_users} and document a few interesting observations. First, in line with the value of diversity hypothesis, I find that the emotions of homogeneous groups are less informative in predicting announcement returns. Second, the relationship between happiness and returns are negative in most specifications, and is statistically significant in over half of them. Last, it is the variation in excitement expressed by traders, and not by institutions that predicts returns (Columns (10-11)). 

\input{Tables/announcement_preannouncement_heterogeneity_users}

\subsection{Sensitivity Analysis}\label{sec:robustness}

I report the results of the sensitivity analysis in Table \ref{tab:reg_robustness}.

\squeezeup 
\subsubsection*{Four-year sample}

I compare my point estimate on sentiment in Column (1) with \citeN{bartov2017can} using their empirical specification. One difference between my four year sample and their is that it starts a year later. Yet, I find similar coefficients (0.0638 versus 0.0599). Controlling for the sentiment variable only marginally affects the coefficients on emotions, and does not impact the statistical significance on happy.

\squeezeup 
\subsubsection*{Alternative Dependent Variable}

My main specification for excess returns is defined in Table \ref{tab:variable_definitions}, but as I show in Columns (2-3), my results are robust to alternative specifications. 

\squeezeup 
\subsubsection*{Extending the Window Length}

My primary analyses concerns the window just leading up to earnings announcements ($[-10,-2]$). I now expand the window to $[-20,-2]$. I find slightly larger coefficients Column (4), suggesting that investor emotions measured over longer-term horizons are also relevant.

\squeezeup 
\subsubsection*{Alternative Classification}

Arguably the most important part of my robustness checks, I now explore my Twitter based model in Column (5). The point estimate on happy is comparable to the one obtained by the StockTwits based model. This finding provides strong support that it is indeed investor enthusiasm that helps predicting the market response to earnings reports.

\input{Tables/announcement_preannouncement_robustness}

\squeezeup 
\subsubsection*{Alternative Weighting}

As a validity check, I consider two alternative weighting schemes. First, I investigate abandoning the weighting scheme entirely, and hence, messages are weighted equally (Column (6)), and second, I weight each message by the number of likes it received, 1+$\log(1+$\# of likes to be specific (Column (7)). The results, are largely unaltered under these alternative specifications. That is, I continue to find that average investor excitement from messages is associated with lower announcement returns.

\section{Conclusion} \label{sec:conclusion}

In this paper, I study the impact of firm-specific emotions on quarterly earnings announcements. I demonstrate that investor emotions can help predict the company's quarterly earnings. I find that both within- and inter-firm variation in investor enthusiasm is linked with lower announcement returns. In particular, I find that both messages that convey original information, and those disseminating existing information drive my results. When considering messages that carry information directly related to earnings, firm fundamentals, and/or stock trading and those covering other information, I find that the former has a larger impact on announcement returns. 


The link between emotions and market behavior has interesting policy implications. It demonstrates that there is a concrete foundation for the idea that central banks, governments, firms, and the media should consider the effects of announcements and data release on the emotional state of market participants and how this might, in turn, affect market prices. Such impacts would arise alongside the influence of new information on economic fundamentals that might affect asset prices accordingly. While the effects of information on fundamentals can be identified with well-established techniques in finance and economics, studying the emotional component requires new tools. In my view, the methods described herein constitute a step forward in this direction.

\singlespacing
\bibliography{ms}
\bibliographystyle{achicago}

\clearpage

\setstretch{1.4}
\clearpage

\appendix
\clearpage

\counterwithin{figure}{section}
\counterwithin{table}{section}
\addcontentsline{toc}{section}{Appendices}

\section*{Appendix}\label{app:app}

\section{A Simple Model of Investor Emotion}\label{app:model}
The theoretical framework of this paper is motivated by \citeN{epstein2008ambiguity}. I include this simple model to illustrate how emotion can affect asset prices. There are three dates, labeled 0, 1, and 2. I focus on news about one particular asset (asset A). There are $\frac{1}{n}$ shares of this asset outstanding, where each
share is a claim to a dividend: 
\begin{equation} \label{eq:dividend}
d= m + \epsilon^a + \epsilon^i
\end{equation}
where m denotes the mean dividend, $ \epsilon^a$ isan aggregate shock, and $\epsilon^i$ is an idiosyncratic shock that affects only asset A.\newline

\noindent \textbf{Assumption 1.} Shocks are mutually independent and normally distributed with mean zero.
\[
\epsilon^i \sim \mathcal{N}(0,\,\sigma_i^{2})\, 
\] 
\[
\epsilon^a\sim \mathcal{N}(0,\,\sigma_a^{2})\,
\] 

I summarize the payoff on all other assets by a dividend:
 \begin{equation}\label{eq:market}
\tilde{d} = \tilde{m} + \epsilon^a + \tilde{\epsilon^i}
\end{equation}

There are $\frac{n-1}{n}$ shares of other assets outstanding and each pays $\tilde{d}$. The market portfolio is then a claim to $\frac{1}{n} d + \frac{n-1}{n} \tilde{d}$.When $n=1$, asset A is the market. Aside from this special case, asset A can be interpreted as a stock in a single company (for n large). In this scenario, $\tilde{d}$ can be interpreted as the sum of stock payoffs for other companies. For what follows, I assume a symmetric case of n stocks that each promise a dividend of the form Equation (\ref{eq:dividend}), with the aggregate shock being identical, while the idiosyncratic shocks being independent across companies. I use this symmetric case for simplicity and tractability, however, the precise nature of $\tilde{d}$ is irrelevant for most of my results. \newline

Dividends are revealed at date 2. At date 1, the representative agent receives two noisy signals $(s_1, s_2)$, informing her about the aggregate and the idiosyncratic shock. This captures the idea that the investor is able to access news updates (sector and company specific). 
\begin{equation} 
s_1 = \epsilon^i + \epsilon_1
\end{equation}
\begin{equation}
s_2 = \epsilon^a + \epsilon_2
\end{equation}
\textbf{Assumption 2.} Signals are imprecise; $\epsilon_1$ and $\epsilon_2$ are mutually independent and normally distributed with mean zero. \[
\epsilon_1 \sim \mathcal{N}(0,\,\sigma_1^{2})\,
\] 
\[
\epsilon_2\sim \mathcal{N}(0,\,\sigma_2^{2})\,
\] 

The investor tries to infer $\epsilon^i + \epsilon^a$ from the two signals $(s_1,s_2)$. The set of one-step-ahead beliefs about $s_1$ and $s_2$ at date 0 consists of normals with mean zero and variance $\sigma_i^2+\sigma_1^2$ and $\sigma_a^2+\sigma_2^2$ respectively. The set of posteriors about $\epsilon^i + \epsilon^a$ is calculated using standard rules for updating normal random variables. For fixed $\sigma_{i_{(i=1,2)}}$, let $\gamma_i$ denote the regression coefficient\footnote{It is common to measure the information content of a signal relative to the volatility of the parameter (\citeN{epstein2008ambiguity}).}:

\begin{equation}
\gamma_1(\sigma_1) = \frac{cov(s_1, \epsilon^i)}{var(s_1)} = \frac{\sigma_i^2}{\sigma_i^2+\sigma_1^2}
\end{equation}
\begin{equation}
\gamma_2(\sigma_2) = \frac{cov(s_2, \epsilon^a)}{var(s_2)} = \frac{\sigma_a^2}{\sigma_a^2+\sigma_2^2}
\end{equation}

For fixed $\sigma_i$, the coefficient $\gamma_i(\sigma_i)$ determines the fraction of prior variance in $\epsilon^a$ and in $\epsilon^i$ that is resolved by the signal. Given $(s_1, s_2)$, the posterior density $\epsilon^a + \epsilon^i$ is also normal. In particular\[
\epsilon^i + \epsilon^a \sim \mathcal{N}(\gamma_1 s_1 + \gamma_2 s_2,\,(1-\gamma_1)\sigma_i^{2}+(1-\gamma_2)\sigma_a^{2})\,
\] 
\noindent \textbf{Assumption 3.} There is a representative agent who does not discount the future and cares only about consumption at date 2. Her utility function is represented by:
	\begin{equation}
	u(c) = -e^{-\rho c}
	\end{equation}

\subsection{Bayesian Benchmark}
The price of asset A equals the expected present value minus a risk premium that depends on risk aversion and covariance with the market. It is straightforward then to calculate the price of asset A at dates 0 and 1: 
	\begin{equation}
	q_0^{Bayesian} = m - \rho cov\Big(d, \frac{1}{n} d + \frac{n-1}{n} \tilde{d}\Big) = m - \rho \Big( \frac{1}{n} \sigma_i^2+ \sigma_a^2 \Big)
	\end{equation}
	\begin{equation}
	q_1^{Bayesian} = m + \gamma_1 s_1 + \gamma_2 s_2 - \rho \Big[\frac{1}{n}(1-\gamma_1)\sigma_i^2 +(1-\gamma_2)\sigma_a^2\Big]
	\end{equation}
	
At date 0, the expected present value is simply the prior mean dividend m. At date 1, it is the posterior mean dividend $m+\gamma_1 s_1 + \gamma_2 s_2$, as it now depends on the value of the signals (given that the signal is informative: $\gamma_i > 0$). The risk premium depends only on time (and is independent of $s_1$ and $s_2$): it is smaller at date 1, since the signal resolves some uncertainty. At either date, it is composed of two parts, one is driven by the variance of the aggregate shock ($\epsilon^a$), and the other one equals the variance of the idiosyncratic shock ($\epsilon^i$) multiplied by the market share of the asset: $\frac{1}{n}$. As n becomes large, idiosyncratic risk is diversified away and does not matter for prices.

\subsection{Investor Emotion}
In absence of signals, the investor is guided by her emotions at date 0. As it has been shown in the literature, the investor overprices the asset when in a good mood (\citeN{breaban2018emotional}).\newline 

I use $\eta$ to represent the investors emotional state at date 0, such that $\eta \in (-1,1)$. For simplicity, I assume that her emotional state only affects the valuation of asset A. In particular, the investor overweights the mean dividend when pricing asset A by $1+\eta$.\footnote{This can be easily extended. Say $\eta_a$ is the emotion parameter for asset A, while $\eta_m$ is the emotion parameter for all other assets in the market.}. In this environment, the price of asset A in period 0:
	\begin{equation}
	q_0^{EM} = m(1+\eta)- \rho cov\Big(d, \frac{1}{n} d + \frac{n-1}{n} \tilde{d}\Big) = m(1+\eta) - \rho \Big( \frac{1}{n} \sigma_i^2+ \sigma_a^2 \Big)
	\end{equation}

As there is information to process at date 1, the investor loses her emotional attachment and relies on the signals. 
The date 0 price, however exhibits a premium (discount) due to emotion, when the investor is in a good (bad) mood. This premium (discount) is directly related to the extent of emotion. Since the emotion parameter enters the asset price linearly, for $\eta=0$, I obtain the same price as in the Bayesian Benchmark.

\subsection{Comparative Statics}
I are interested in the price adjustment dynamics from period 0 to period 1: 
	\begin{equation} 
	 \Delta q^{Bayesian} -\Delta q^{EM} = - \eta m
	 \end{equation}

Thus, compared with the Bayesian benchmark, which also corresponds to neutral valuation, the price of asset A responds more to the signals when investors draw a negative emotion shock with respect to asset A. This is a testable implication of the model, which I investigate in Section \ref{sec:earnings_returns} empirically. In particular, I examine asset price movements surrounding quarterly earnings announcements (this stands for the idiosyncratic shock).

\section{Text Processing}\label{app:text_processing}

I first remove images, hyperlinks, and tags from the text. I discard tweeted at (e.g., @dvamossy), cashtags (e.g., \$FORD), and the retweet indicator (i.e., ``RT'') where applicable. I set text to lower case, translate emojis and emoticons (e.g., ``:)'' substituted with ``happyface''), fix contractions (e.g., ``i've'' changed to ``i have''), and correct common misspellings\footnote{I provide a description of my misspell correction in the Online Appendix.}. I replace numberspreceded by a \$ sign with ``isdollarvalue'', other numbers with ``isnumbervalue'', and the \% sign with ``ispercentage''. This feature is important for distinguishing between general chat versus stock trading related messages. I then remove any non-word tokens, such as punctuation marks. I include the 60,000 most frequent words in the model dictionary, changing all other tokens to ``NONE''. The messages are then tokenized (i.e., words are changed to numbers) and split into sentences using keras.

\section{Measuring Emotions with Deep Learning}\label{app:NLP}

My deep learning model operates by sequentially learning a latent representation. These reflect features such asword order, word usage, and local context. Minimization of prediction error\footnote{I use a categorical-cross entropy loss function.} drives feature extraction. I use a Bidirectional-GRU model, which can be defined as the composition of several functions (layers):

\begin{equation}
\text{f(X}_{j,T};w) = \text{S}\circ \text{D} \circ \text{O} \circ \text{BiGRU} \circ \text{Emb(X}_{j,T})
\end{equation}

where $\text{X}_{j,t}$ is the jth message of length T, Emb is the embedding, BiGRU is the Bidirectional Gated Recurrent Unit, O is a linear layer, D is a two-layered NN with ReLu activation, and S is the final softmax layer, which ensures that the output is between 0 and 1. I next define each component of the model.

\subsection{Message}

I define a message as a vector $\mathbf{X} = [x_1 \hdots x_T]$, where x$_k$ is the index of the kth word in the model dictionary, and T is the maximum document length (30 in my case). For documents shorter than the maximum document length, I fill the extra space with special padding words.

\subsection{Embedding}

Embedding (Emb) assigns vectors to individual words. I obtain the starting value for my word embeddings from \citeN{pennington2014glove}, leveraging 2 billion tweets, 27 billion tokens, and a vocabulary of 1.2 million words. These embedding vectors are then updated during estimation via backpropagation. I denote the embedding of the word x$_i$ as e(x$_i$) = e$_i \in$ $\mathbb{R}^{d(E)}$, where d(E) is the embedding size (200 in my case). Thus, the document can be represented as:

\begin{equation}
 \text{Emb(X}_{j,T}) = \begin{bmatrix} e_1 \\ e_2 \\ \vdots \\ e_T \end{bmatrix} = \begin{bmatrix} e_{1,1} & e_{1,2} & \hdots & e_{1,d(E)} \\ e_{2,1} & e_{2,2} & \hdots & e_{1,d(E)} \\ \vdots & \vdots & \vdots & \vdots \\ e_{T,1} & e_{T,2} & \hdots & e_{T,d(E)} \end{bmatrix}
\end{equation}

Words frequently used interchangeably are prone to cluster in the embedding space.\footnote{This property allows me to capture word similarities without imposing any additional structure.}

\subsection{Gated Recurrent Unit (GRU)}

\begin{figure}[htbp]
\centering
\includegraphics[clip, trim=0cm 11.5cm 1cm 1cm, scale=0.75]{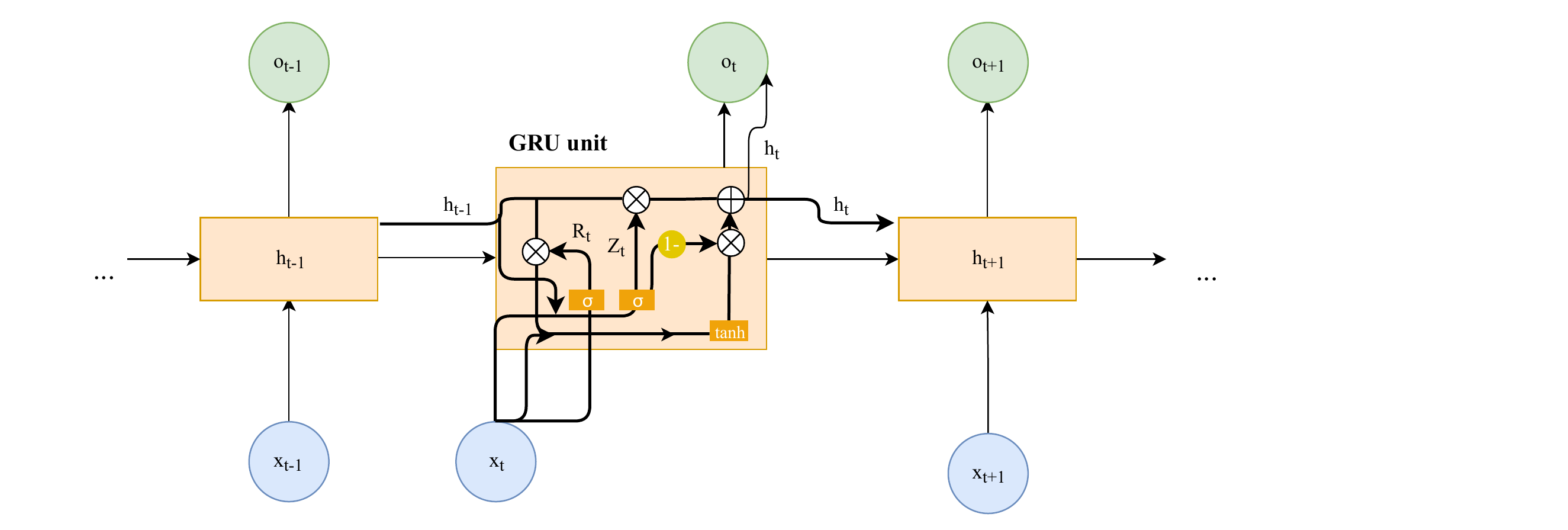}
\caption{The architecture of the GRU unit} \label{fig:gru}
\end{figure}

Introduced by \citeN{chung2014empirical}, the Gated Recurrent Unit (GRU) is a slight variation on the LSTM (see \citeN{lstm}). It addresses the high memory requirements imposed by the LSTM by combining the forget and input gates into a single ``update gate'', and by merging the cell state with the hidden state. The resulting model requires less computation, and has enjoyed growing popularity. I illustrate the GRU cell in Figure \ref{fig:gru}. 

The update gate decides which parts of the previous hidden state are updated (or discarded). By selecting valuable parts from the previous hidden state, the reset gate determines which parts are used to compute new content. This is then used along with current input to compute the hidden state update. Notice that the update gate controls both what is kept from the previous hidden state, and what is taken from the hidden state update. The sigmoid function ensures that the output is between zero and one. To illustrate this as a sequence of operations, consider time-step t (t-th word) and input $\mathbf{X_t} \in \mathbb{R}^{n \times d}$ (n is sample size, d denotes input). The computations (forward-propagation) for the GRU unit can be summarized as:

\begin{equation}\label{eq:gru_update}
\mathbf{Z_t} = \sigma_g(\mathbf{W_{xz}} \mathbf{X_t} + \mathbf{W_{hz}} \mathbf{H_{t-1}} + \mathbf{b_z})
\end{equation}
\begin{equation}
\mathbf{R_t} = \sigma_g (\mathbf{W_{xr}} \mathbf{X_t} + \mathbf{W_{hr}} \mathbf{H_{t-1}} + \mathbf{b_r})
\end{equation}
\begin{equation}\label{eq:gru_output}
\mathbf{H_t} = \mathbf{Z_t} \odot \mathbf{H_{t-1}} + (1-\mathbf{Z_t}) \odot \sigma_h(\mathbf{W_{xh}} \mathbf{X_t} + \mathbf{W_{hh}} (\mathbf{R_t} \odot \mathbf{H_{t-1}}) + \mathbf{b_h})
\end{equation}

where $\odot$ denotes elementwise multiplication, $\mathbf{X_t}$ denotes the input, $\mathbf{H_t} \in \mathbb{R}^{n \times h}$ the output, $\mathbf{Z_t} \in \mathbb{R}^{n \times h}$ the update gate, $\mathbf{R_t} \in \mathbb{R}^{n \times h}$ the reset gate, and h the number of hidden states.
Here, $\mathbf{W_{xr}}$, $\mathbf{W_{xz}} \in \mathbb{R}^{d \times h}$ and $\mathbf{W_{hr}}$, $\mathbf{W_{hz}} \in \mathbb{R}^{h \times h}$ are weight matrices, while $\mathbf{b_r}$, $\mathbf{b_z} \in \mathbb{R}^{1\times h}$ are bias parameters. Typically $\sigma_g$ is a sigmoid function to transform input values to the interval (0,1), while $\sigma_h$ is tanh.

\subsection{Bidirectional GRU}

Bidirectional RNNs were developed by \citeN{schuster1997bidirectional}. The key feature of the bidirectional architecture is that dependencies and training can go forwards and backwards in time.\footnote{For instance, if I were to ingest ``oil and gas'' with a forward architecture, ``oil'' would receive signal from ``gas'' during backpropagation but not the reverse. The bidirectional architecture allows for both relationships. For an in-depth discussion of different bidirectional architectures see \citeN{graves2005framewise}.} Before I move forward, let me denote my previous operations defined in Equations (\ref{eq:gru_update})-(\ref{eq:gru_output}) as $\mathbf{Z_t}=\overrightarrow{\mathbf{Z_t}}$, $\mathbf{R_t}=\overrightarrow{\mathbf{R_t}}$, $\mathbf{H_t}=\overrightarrow{\mathbf{H_t}}$. The key addition of the bidirectional architecture is that for a given timestep t, I also compute hidden state updates as follows:

\begin{equation}
\overleftarrow{\mathbf{Z_t}} = \sigma_g(\mathbf{W_{xz}^f} \mathbf{X_t} + \mathbf{W_{hz}^f} \overleftarrow{\mathbf{H_{t+1}}} + \mathbf{b_z^f})
\end{equation}
\begin{equation}
\overleftarrow{\mathbf{R_t}} = \sigma_g \mathbf{W_{xr}^f} \mathbf{X_t} + \mathbf{W_{hr}^f} \overleftarrow{\mathbf{H_{t+1}}} + \mathbf{b_r^f})
\end{equation}
\begin{equation}
\overleftarrow{\mathbf{H_t}} = \overleftarrow{\mathbf{Z_t}} \odot \overleftarrow{\mathbf{H_{t+1}}} + (1-\overleftarrow{\mathbf{Z_t}}) \odot \sigma_h(\mathbf{W_{xh}^f} \mathbf{X_t} + \mathbf{W_{hh}^f} (\overleftarrow{\mathbf{R_t}} \odot \overleftarrow{\mathbf{H_{t+1}})} + b_h^f)
\end{equation}

I then concatenate the forward and backward hidden states $\overleftarrow{\mathbf{H_t}}$ and $\overrightarrow{\mathbf{H_t}}$ to obtain the hidden state $\mathbf{H_t} \in \mathbb{R}^{n\times 2h}$.

\subsection{Linear Layer, Neural Network and Softmax Activation}

My next step is to apply another set of weights and bias terms and pass it to two-layered Neural Network:

\begin{equation}
\mathbf{O_t}=\mathbf{H_t} \mathbf{W_{h,q}}+\mathbf{b_q}
\end{equation}
\begin{equation}
 \mathbf{D} =\sigma_d(\mathbf{O_t} \mathbf{W_o} + \mathbf{b_o})
\end{equation}
\begin{equation}
 \mathbf{D'} =\sigma_d(\mathbf{D} W_d + b_d )
\end{equation}

the output of the GRU is $\mathbf{O_t} \in \mathbb{R}^{n\times q}$, the output of the first dense layer is $\mathbf{D} \in \mathbb{R}^{n\times q'}$, while the output of the second dense layer is $\mathbf{D'} \in \mathbb{R}^{n\times q''}$where q, q', q'' denote the number of hidden units for each of the layers, and $\sigma_d$ is the RELU activation function in my case, defined as:

\begin{equation}
\textrm{RELU}(x)= \left\{ \begin{array}{cl}
x & \textrm{if }x \ge 0 \\
0 & \textrm{otherwise}\\
\end{array}\right.
\end{equation}

This is then passed to another hidden layer, followed by a softmax layer to obtain the final output:

\begin{equation}
\mathbf{\hat{y}} = \text{softmax}(\mathbf{D'} W_y + b_y )
\end{equation}

where $\mathbf{\hat{y}}$ denotes the final output with $\mathbf{\hat{y}} \in \mathbb{R}^{n\times y}$, and y denotes the number of outputs, 7 in my case for the emotion classification, and 3 for the chat type classification.

\subsection{Training Data Sources}\label{online_app:data_sources}

Since performing textual analysis using any word classification scheme is inherently imprecise (see, e.g., \citeN{loughran2011liability}), I train two different models based on different data sources. My first, and preferred model is similar to \citeN{li2016can}. It relies on building the training data from dictionaries. In particular, I define dictionaries for each emotional states. My dictionaries include both emojis and emoticons, and consists of 2,250 words. To map emojis and emoticons to emotions I use \url{https://unicode.org/Public/emoji/13.0/emoji-test.txt}. I translate emoticons into categories such as ``happyface'', while I retain emojis in their original format, such as ``facewithopenmouth''. I do this to keep the diversity of my emotional labels, which has been shown to improve predictive power (e.g, \citeN{felbo2017using}). It is important to note that the Naive Bayes based sentiment methodology assigns a score of 0.5 (i.e., neutral) for each of the emoticons and emojis included in my dictionaries, both in its original format (i.e., :)) and its changed format (i.e., ``happyface''). I then prepare a training data with messages containing such words, and augment this with messages not containing any of these words while having zero sentiment as neutral ones.\footnote{I further require positive (negative) emotions to have positive (negative) sentiment, classified by \url{https://textblob.readthedocs.io/en/dev/}. I do not impose this for surprise. This reduced the coverage of my labeling model, but increased the accuracy.} I then use these dictionaries and \citeN{ratner2020snorkel} to generate my training data. The second classification scheme builds a model from pre-compiled emotion datasets based on Twitter messages. I construct this training data from \url{https://github.com/sarnthil/unify-emotion-datasets/tree/master/datasets}. I compare the performance of these two models in Appendix \ref{app:comparison}, and discuss further limitations of using the Twitter based model in the Online Appendix.

For my information-based classification, my ``fundamental'' data comes from StockTwits data with messages containing
earnings or fundamental information, while my ``chat'' data comes from my Twitter training data, excluding messages containing such information. This allows me to isolate general chat-like messages from those containing financial information. I rely on these models instead of using a dictionary-based method since it gives me a probabilistic assessment whether a message belongs to a certain class. Additionally, trained word-embeddings learn words often co-occurring with entries from my dictionaries, so that words not included in the dictionary but containing financial information could be picked up by my model.

\subsection{Implementation}

I include 30 words for each message and train roughly 47 million messages for my emotion classification\footnote{Training data messages by class: 19M neutral, 13.9M happy, 5M fear, 4.2M surprise, 2.6M sad, 1.5M disgust, 1.2M anger.}. I use a batch-size of 4,096, a learning rate of 0.01 (0.001 for Twitter), an early-stopping parameter of 1 (20 for Twitter), an embedding dropout of 0.25, 256 hidden units for my GRU, and 256-128 hidden units respectively for my dense layers. 

My deep learning models are made up of millions of free parameters. Since the estimation procedure relies on computing gradients via backpropagation, which tends to be time and memory intensive, using conventional computing resources (e.g., desktop) would be impractical (if not infeasible). Acknowledging the impact of GPUs in deep learning (see \citeN{schmidhuberdeep}), I train my models on a GPU cluster GPU cluster (1-2 NVIDIA GeForceGTX1080 GPUs proved to be sufficient). 
I conduct my analysis using Python 3.6.3 (Python Software Foundation), building on the packages numpy (\citeN{walt2011numpy}), pandas (\citeN{mckinney2010data}) and matplotlib (\citeN{hunter2007matplotlib}). I develop my bidirectional gru model with keras (\citeN{chollet2015keras}) running on top of Google TensorFlow, a powerful library for large-scale machine learning on heterogenous systems (\citeN{abadi2016tensorflow}).

\section{Model Comparison \& Output}\label{app:comparison}

I contrast my model trained on Twitter with the one trained on StockTwits data. Each methodology presents strengths and weaknesses. Given that the training data for the first model was built using dictionaries, it may miss words that my emotional in nature but were not included in the dictionary. To alleviate this concern, I report the accuracy and coverage of my dictionary based training data preparation on a sample of 5,000 hand-tagged messages from StockTwits in Table \ref{tab:labeling}. 

\input{Tables/labeling}

This shows that approximately 2/3 of my messages can be tagged with this approach with an accuracy of 89.5\%, suggesting that this issue might not be severe. The second model, however, was developed using Twitter messages, and it is unclear whether this model would be directly applicable to messages about stocks and companies (we provide mixed evidence in the Online Appendix). In addition, a large number of words are not accounted for using this technique: while in terms of word frequencies my Twitter words cover 96.2\% of my StockTwits words, they only cover 5.02\% of the vocabulary. Thus, trained on this data, my model discards potentially important words for classification.

\subsection{Classifier Performance}

To directly compare these two models, I test the accuracy of both classifiers on a sample of 5,000 hand-tagged messages from StockTwits. I report the results in Table \ref{tab:classification_test}. This shows that the StockTwits trained model performs significantly better in terms of accuracy (roughly 31\% better), but lower in terms of loss. The worse performance in terms of loss is due to the StockTwits based model's tendency to classify non-neutral messages as neutral with almost certainty.

\input{Tables/five_fold}

\begin{figure}[!h] 

\begin{center}
\subfloat[]{\includegraphics[scale=0.5]{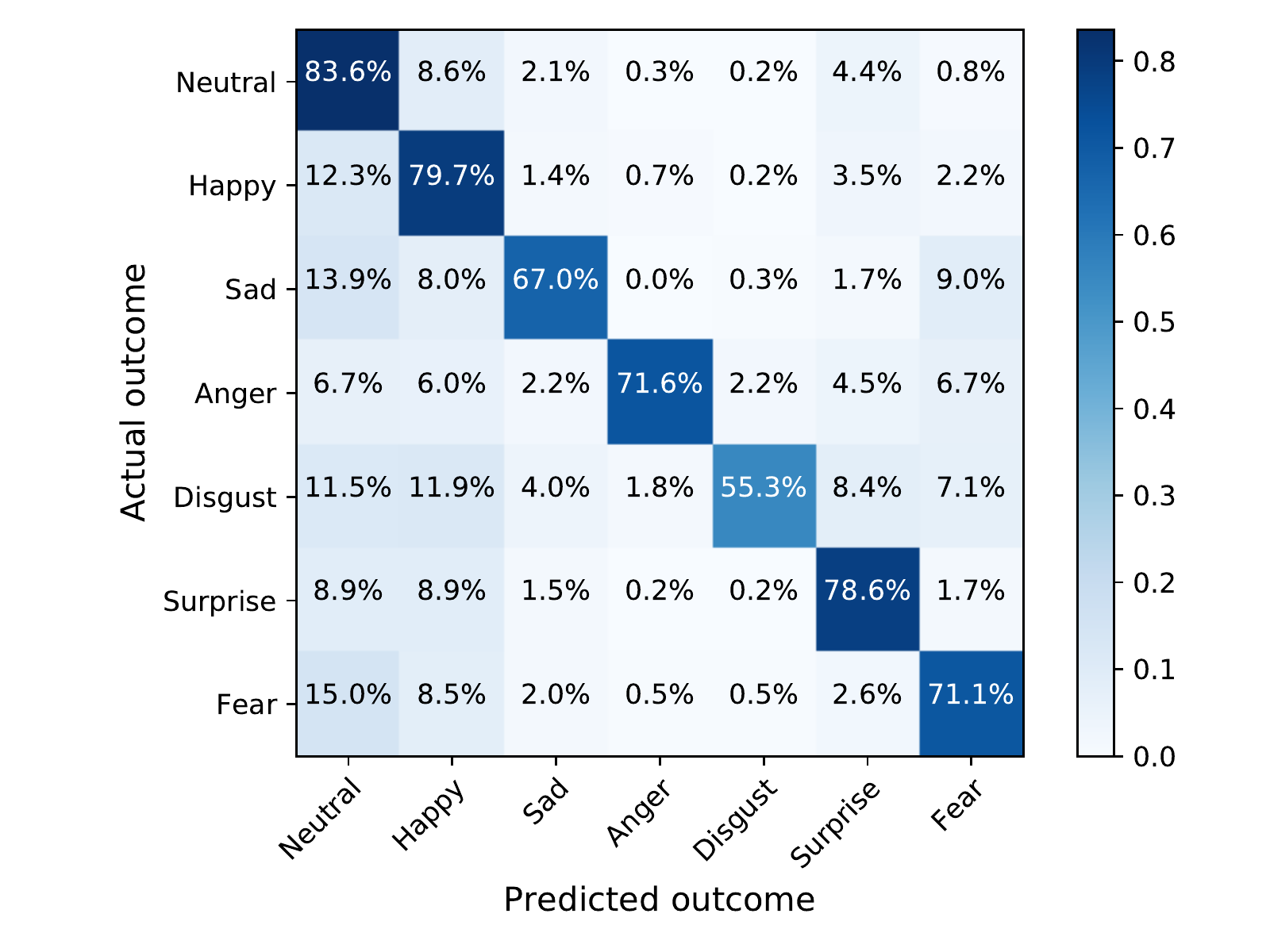}}
\subfloat[]{\includegraphics[scale=0.5]{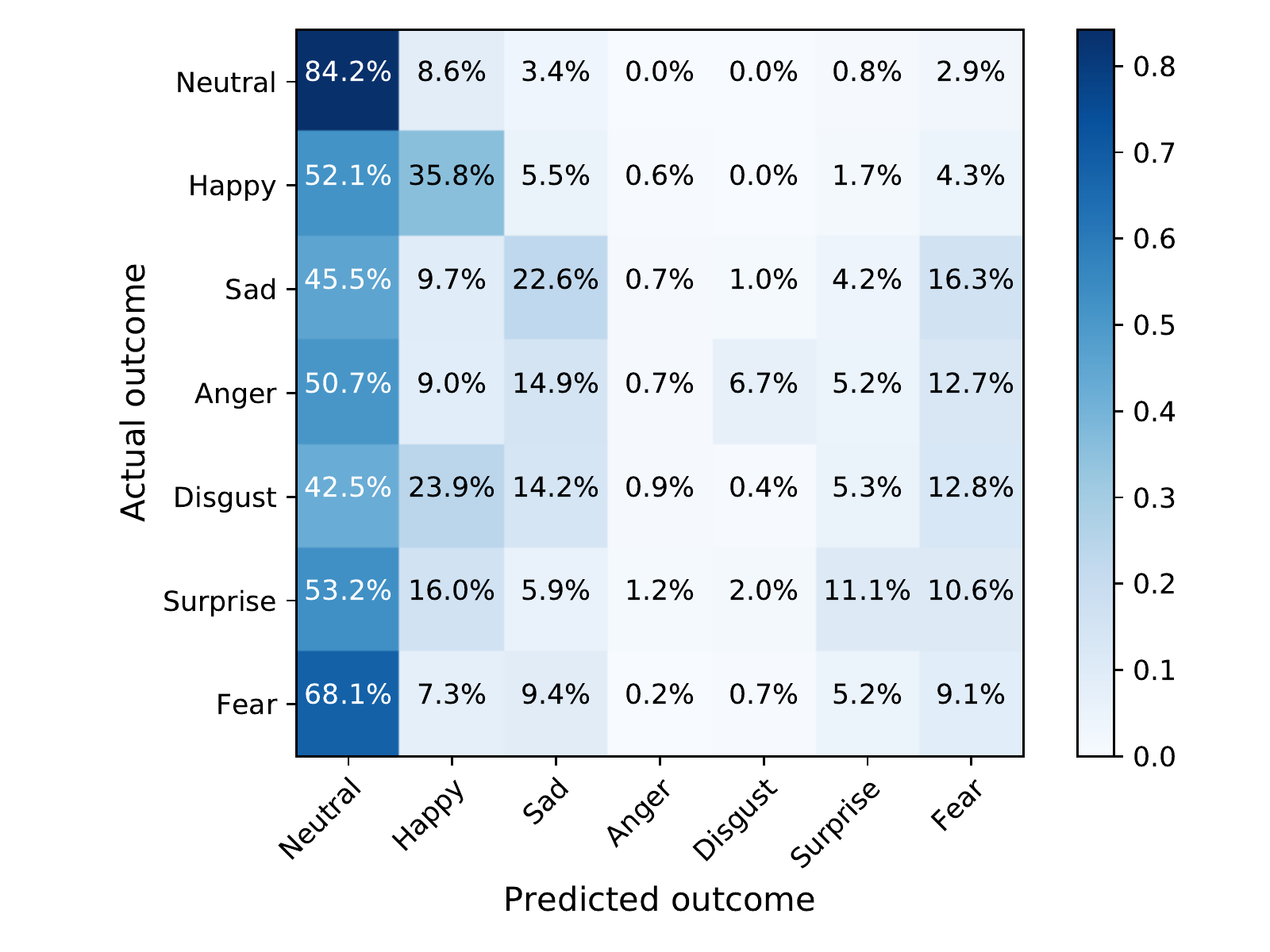}}

\caption{Confusion Matrices for Emotion Classification Models.}\label{fig:confusion_matrix}
\end{center}
\squeezeup \squeezeup  \begin{flushleft}
	\footnotesize{Notes: (a) StockTwits based model, (b) Twitter based model. Results reported are based on performance on the hand-tagged sample for the best performing model on the validation set during five-fold CV.}
\squeezeup \squeezeup  \end{flushleft}
\end{figure}

I confirm this with first plotting the classification errors for each emotions in the data for each of my models. Panel (a) of Figure \ref{fig:confusion_matrix} plots it for my StockTwits based model, while Panel (b) does so for the Twitter based model. The diagonal entries represent the precision of the classifier. For instance, the 83.6\% in the upper right corner of Panel (a) implies that my StockTwits classifier accurately classified 83.6\% of all neutral messages as neutral, while the 12.3\% in the second row first column represents that my classifier mistakenly tagged 12.3\% happy messages as neutral. As we can see, majority of my mistakes with the StockTwits based model is classifying non-neutral messages as neutral. Since I take neutral as my benchmark group, these types of mistakes likely bias my coefficients towards zero, but retain the true ordinal ranking. Particularly important for my results is the lack of misclassification between positive and negative valence emotions. This is not the case for my Twitter based model. Though the most mistakes are towards neutral, there is a large degree of misclassification from sad, angry, and disgust to both happy, and fear, I mainly use this model for a robustness check, and use my StockTwits model for most of my analyses. 

\begin{figure}[!h] 
\squeezeup \squeezeup
\begin{center}
	 \subfloat[]{\includegraphics[scale=0.4]{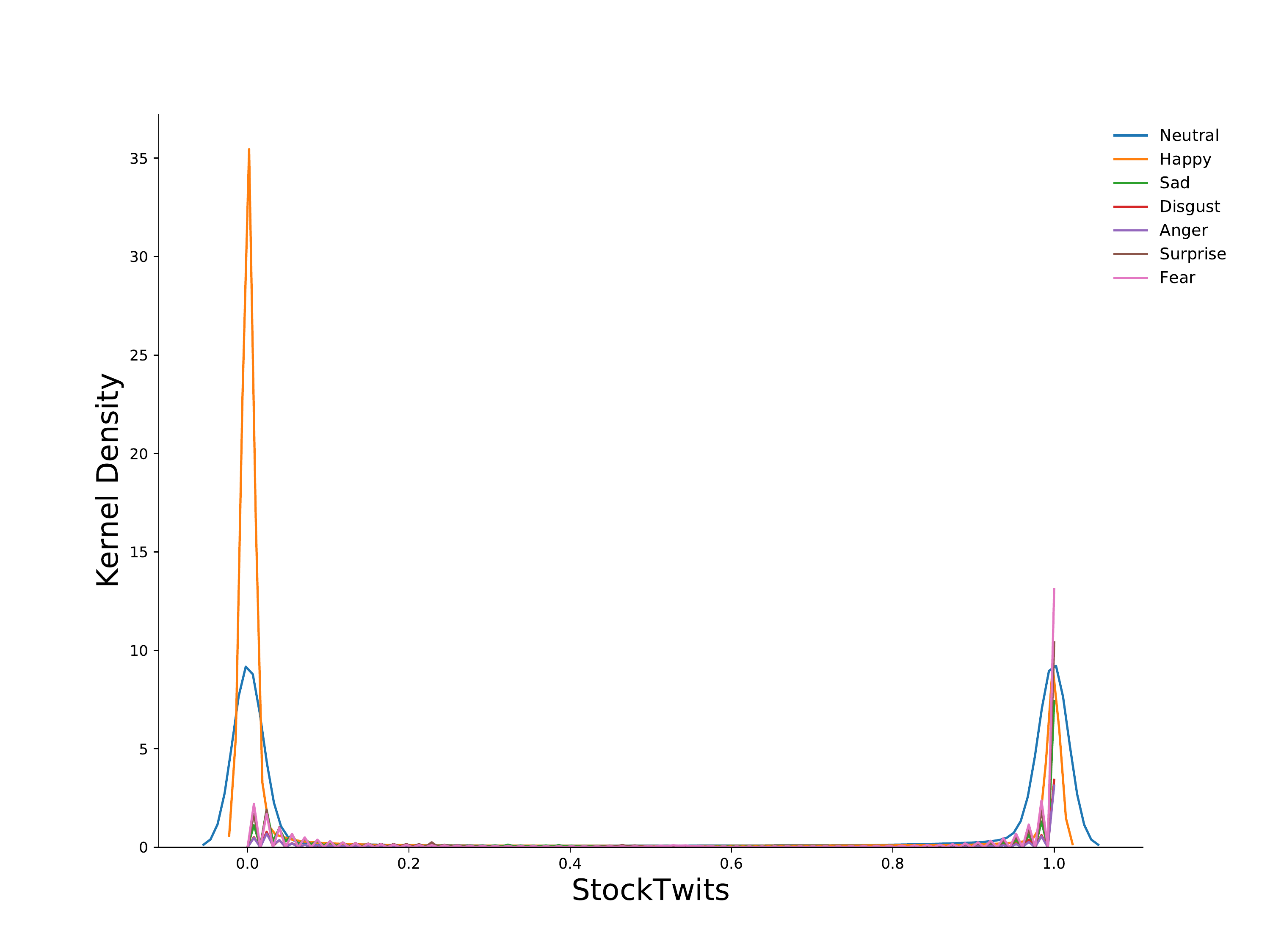}}
	 
\squeezeup \squeezeup 	 
	 \subfloat[]{\includegraphics[scale=0.4]{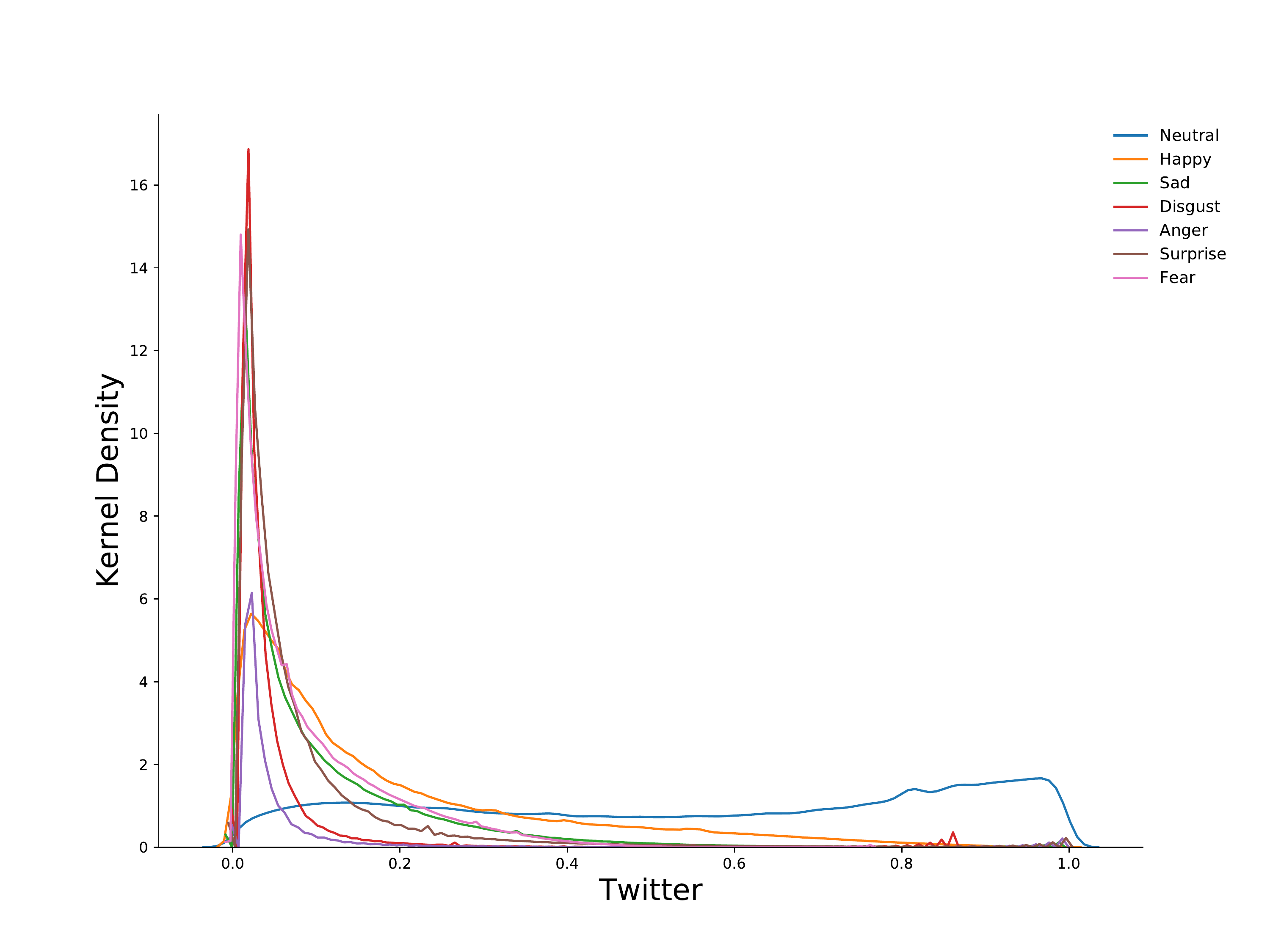}}
	\caption{Kernel Density Distributions of Emotional States.}\label{fig:kde_distributions}
\end{center}
\squeezeup \squeezeup  \begin{flushleft}
	\footnotesize{Notes: Panel (a) plots the Kernel Density Distribution for the StockTwits model, while Panel (b) displays it for the Twitter model.}
\squeezeup \squeezeup  \end{flushleft}
\end{figure}

I also plot the distributions of each emotions in the data in Figure \ref{fig:kde_distributions}. Combining Figure \ref{fig:kde_distributions} and Figure \ref{fig:confusion_matrix}: the mistakes the StockTwits based model makes is to classify non-neutral messages as neutral with almost certainty.\footnote{I find few errors in my chat type classification, but it was not evaluated on a hand-tagged sample, and hence the results are not surprising.}

\clearpage 

\subsection{Examples of Messages \& Outputs}

I provide examples of my model's predictions in Table \ref{tab:twit_examples}. 

\input{Tables/twit_examples}

\section{Model Explanations}

I next uncover associations between the explanatory variables (words) and my model's predictions. I implement SHapley Additive exPlanations (SHAP), a unified framework for interpreting predictions, to explain the output of my GRU model\footnote{For a detailed description of the approach see \citeN{SHAP}.}. SHAP leverages a game theoretical concept to give each feature (word) a local importance value for a given prediction. Shapley values are local by design, yet they can be combined into global explanations by averaging the absolute Shapley values word-wise. Then, I can compare words based on their absolute average Shapley values, with higher values implying higher word importance. 

\begin{figure}[htbp] 
\begin{center}
\subfloat[]{\includegraphics[scale=0.25]{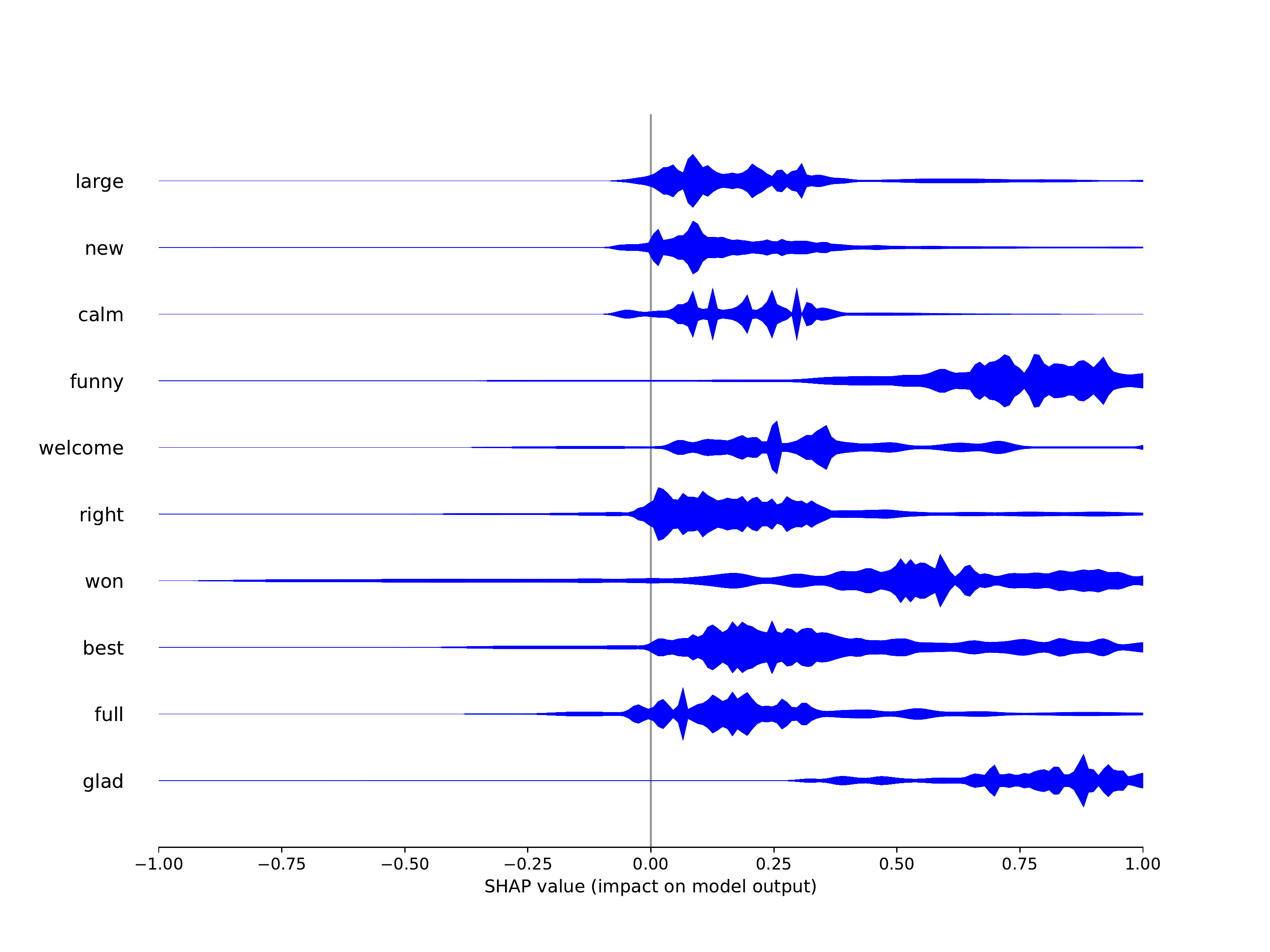}}
\subfloat[]{\includegraphics[scale=0.25]{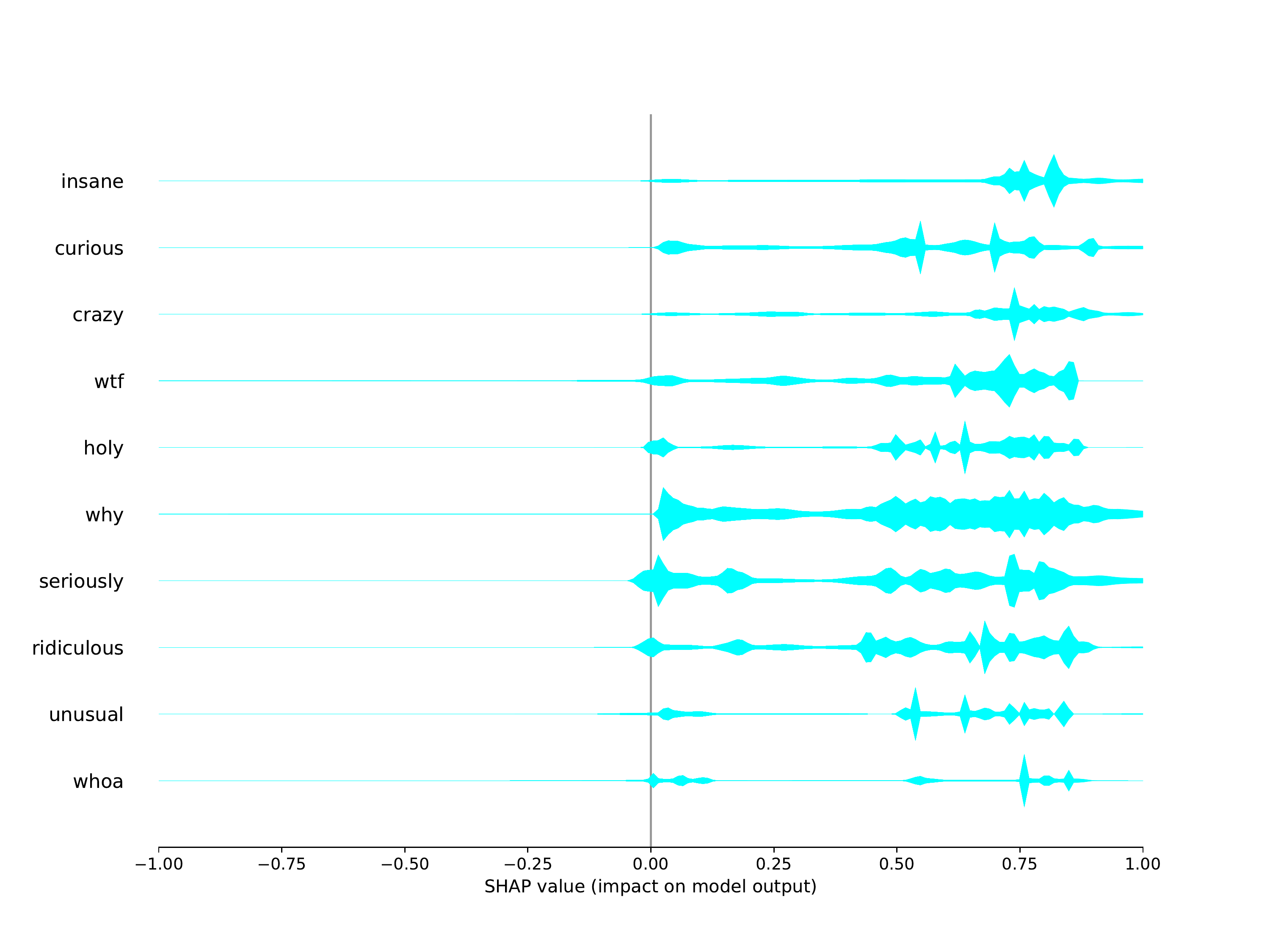}}

 \subfloat[]{\includegraphics[scale=0.25]{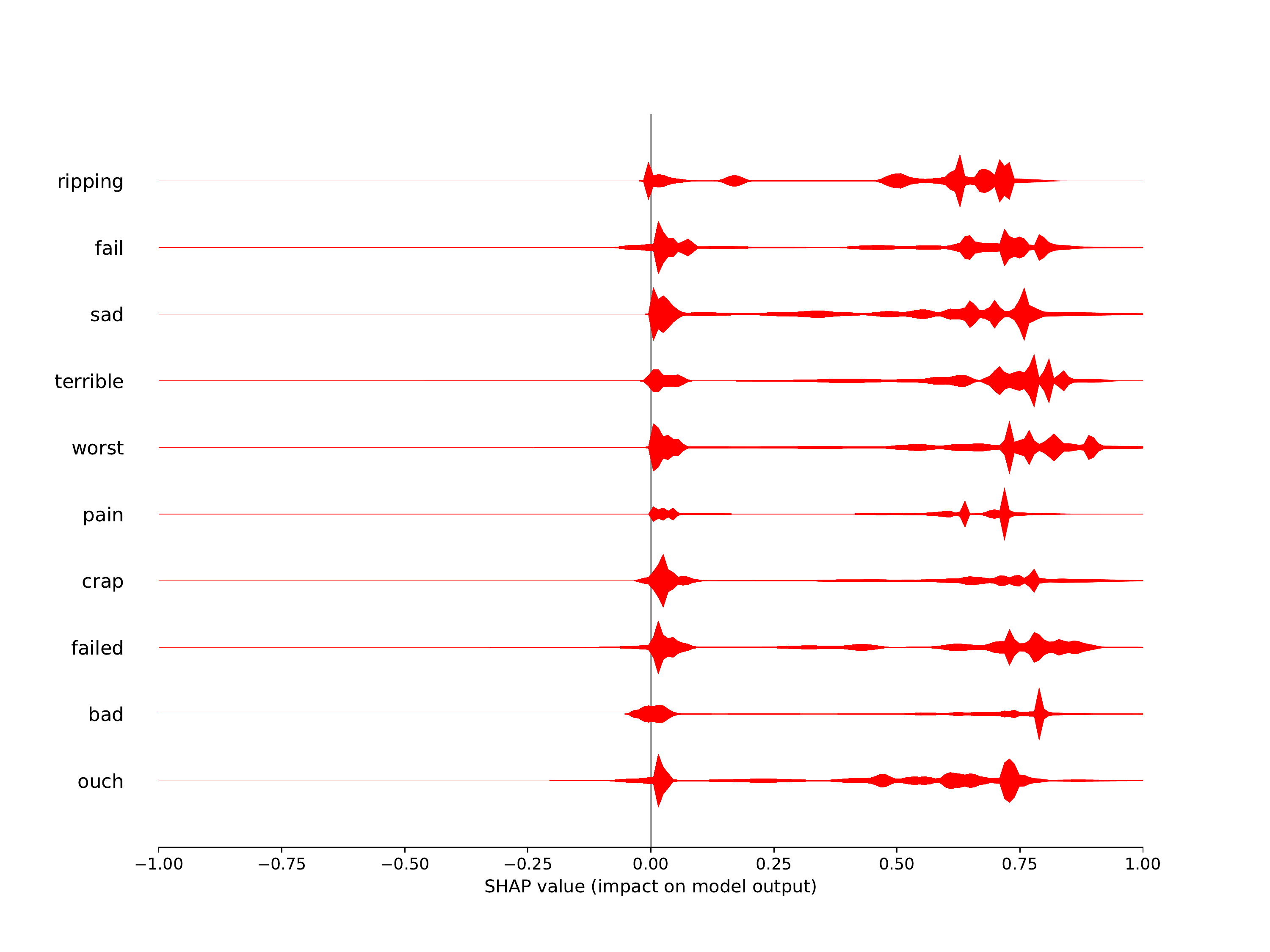}}
\subfloat[]{\includegraphics[scale=0.25]{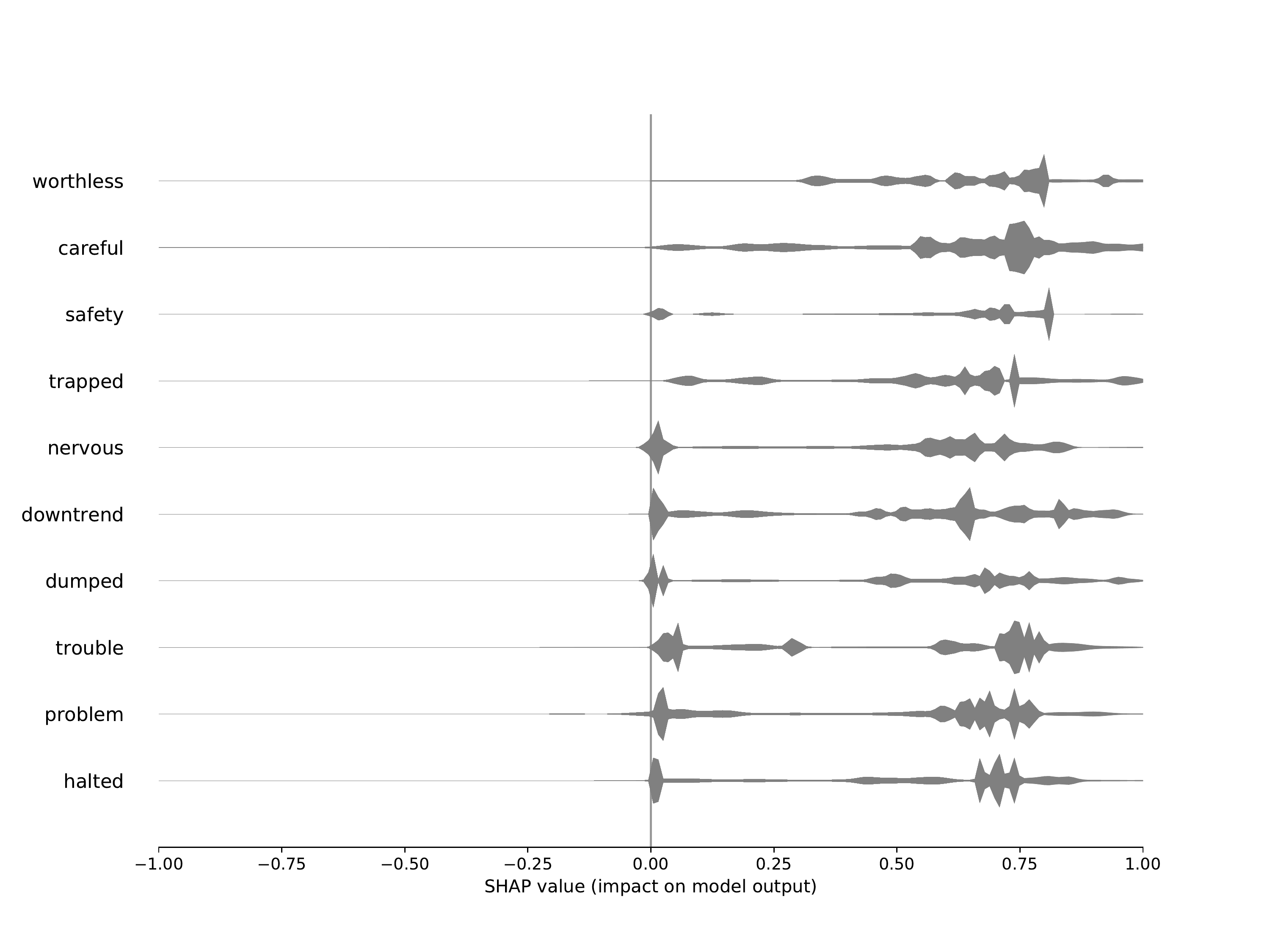}}

 \subfloat[]{\includegraphics[scale=0.25]{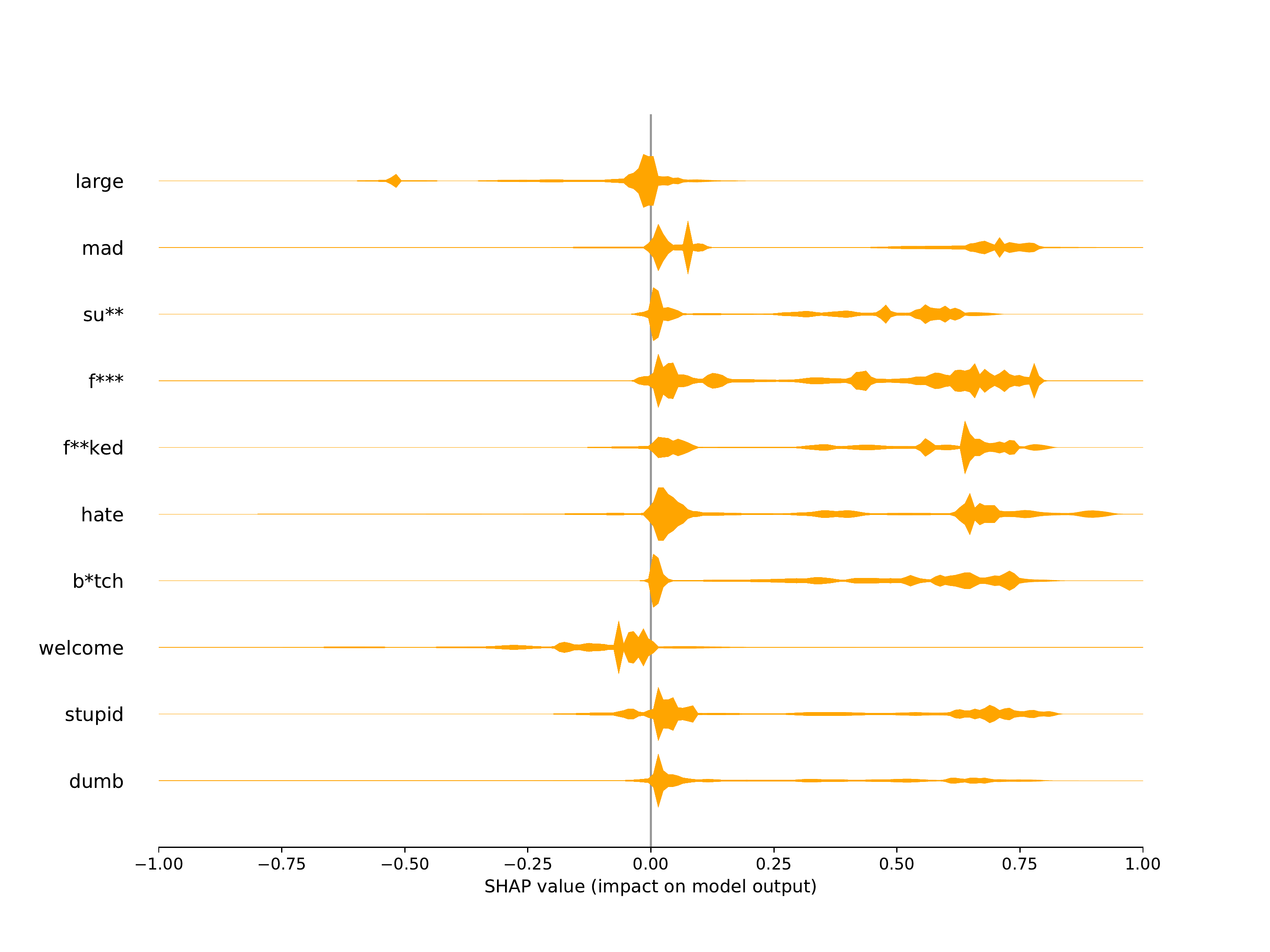}}
 \subfloat[]{\includegraphics[scale=0.25]{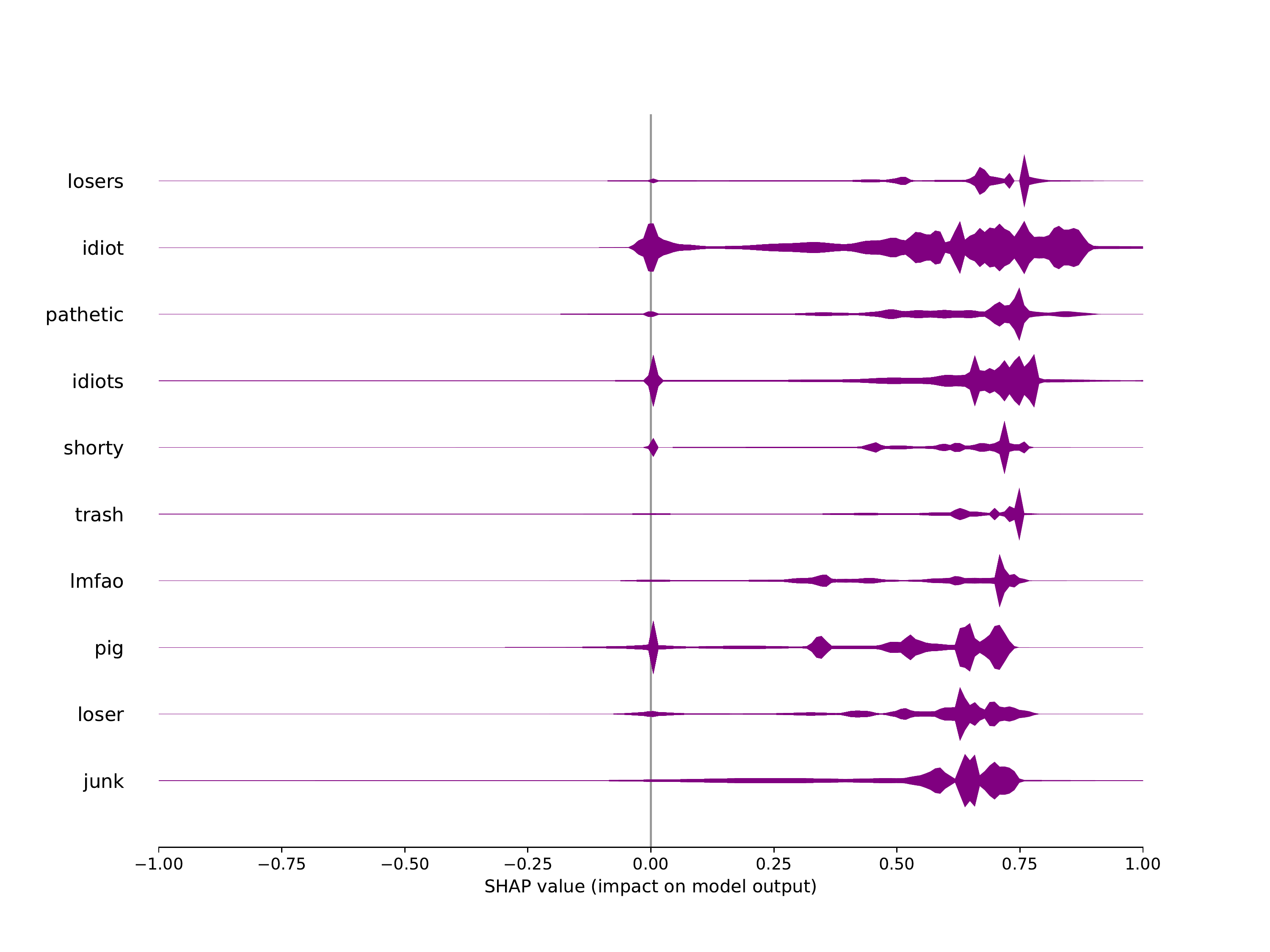}}

\caption{Selected Distribution of Word Importances for StockTwits Emotion Model.}\label{fig:shap_stocktwits}
\end{center}
\squeezeup \squeezeup  \begin{flushleft}
	\footnotesize{Notes: SHAP values evaluated on a random sample of 100,000 StockTwits messages. (a) happy, (b) surprise, (c) sad, (d) fear, (e) anger, (f) disgust. Not shown here: neutral.}
\squeezeup \squeezeup  \end{flushleft}
\end{figure}

To do the SHAP analysis, I a draw a random sample of 100,000 StockTwits messages. Table \ref{tab:shap_stocktwits} reports average absolute SHAP for my StockTwits model, while Figure \ref{fig:shap_stocktwits} plots the distribution of the ten most important words for each emotions. For instance, the first entry in Panel (b) of Figure \ref{fig:shap_stocktwits} is ``insane'', and this word is strongly associated with an increased model output for the surprise class. As another example, the ninth entry of Panel (d) is ``problem'', which has a dispersed distribution, illustrating that in certain cases the word ``problem'' nudges the model's prediction towards fear by relatively insignificantly\footnote{These are typically sentences where ``problem'' is surrounded by other words that are associated with fear.}. A quick inspection of Figure \ref{fig:shap_stocktwits} and Table \ref{tab:shap_stocktwits} confirm that my StockTwits model relates words to emotions correctly. When looking at the SHAP values of the Twitter model, however, we can see some of the roots of my misclassifications. For instance, ``marketing'' is a strong predictor for the happy class, while ``natural'' is a strong predictor for surprise. 

\begin{figure}[htbp] 
\begin{center}
\subfloat[]{\includegraphics[scale=0.6]{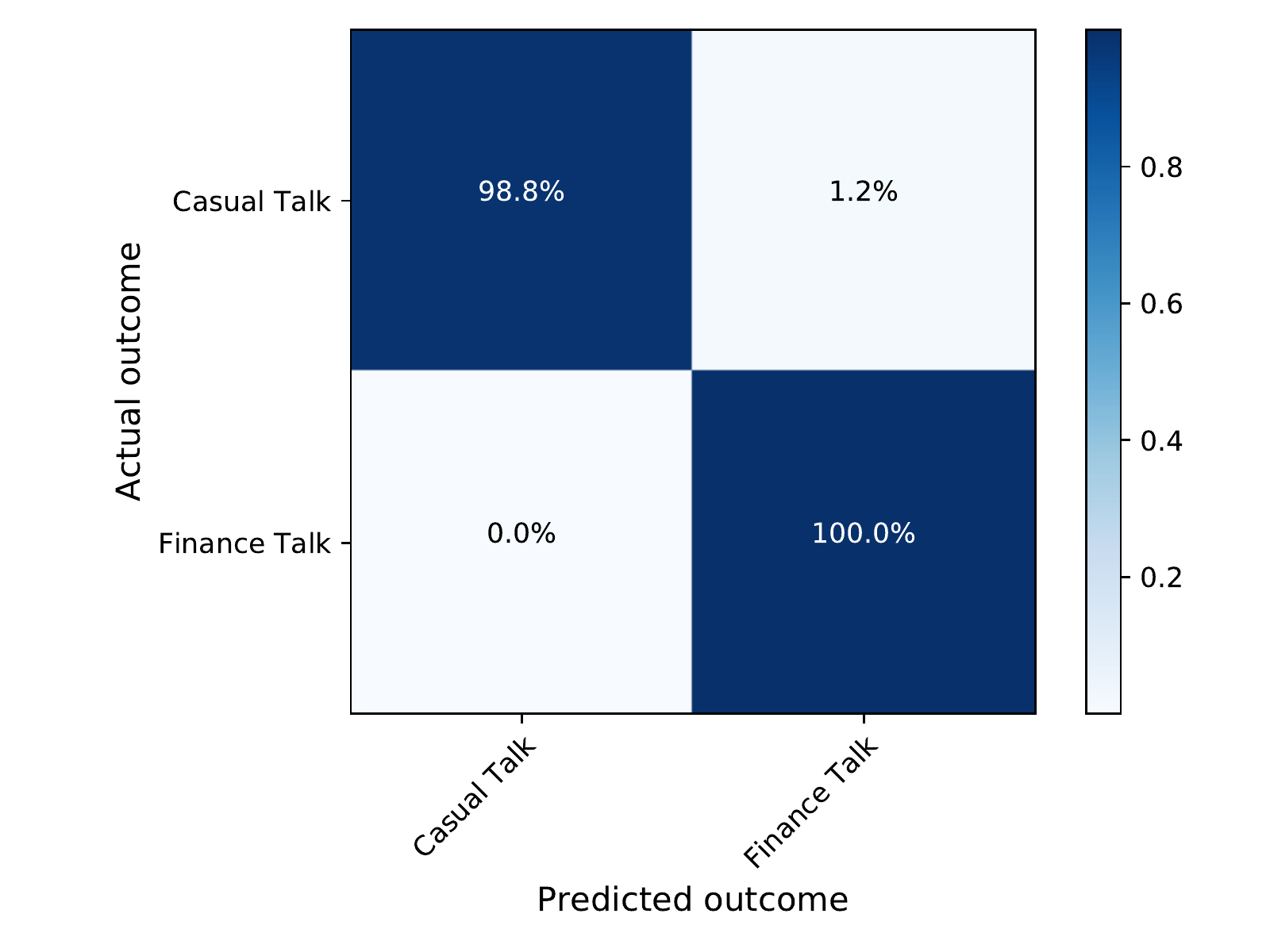}}
 
\subfloat[]{\includegraphics[scale=0.5]{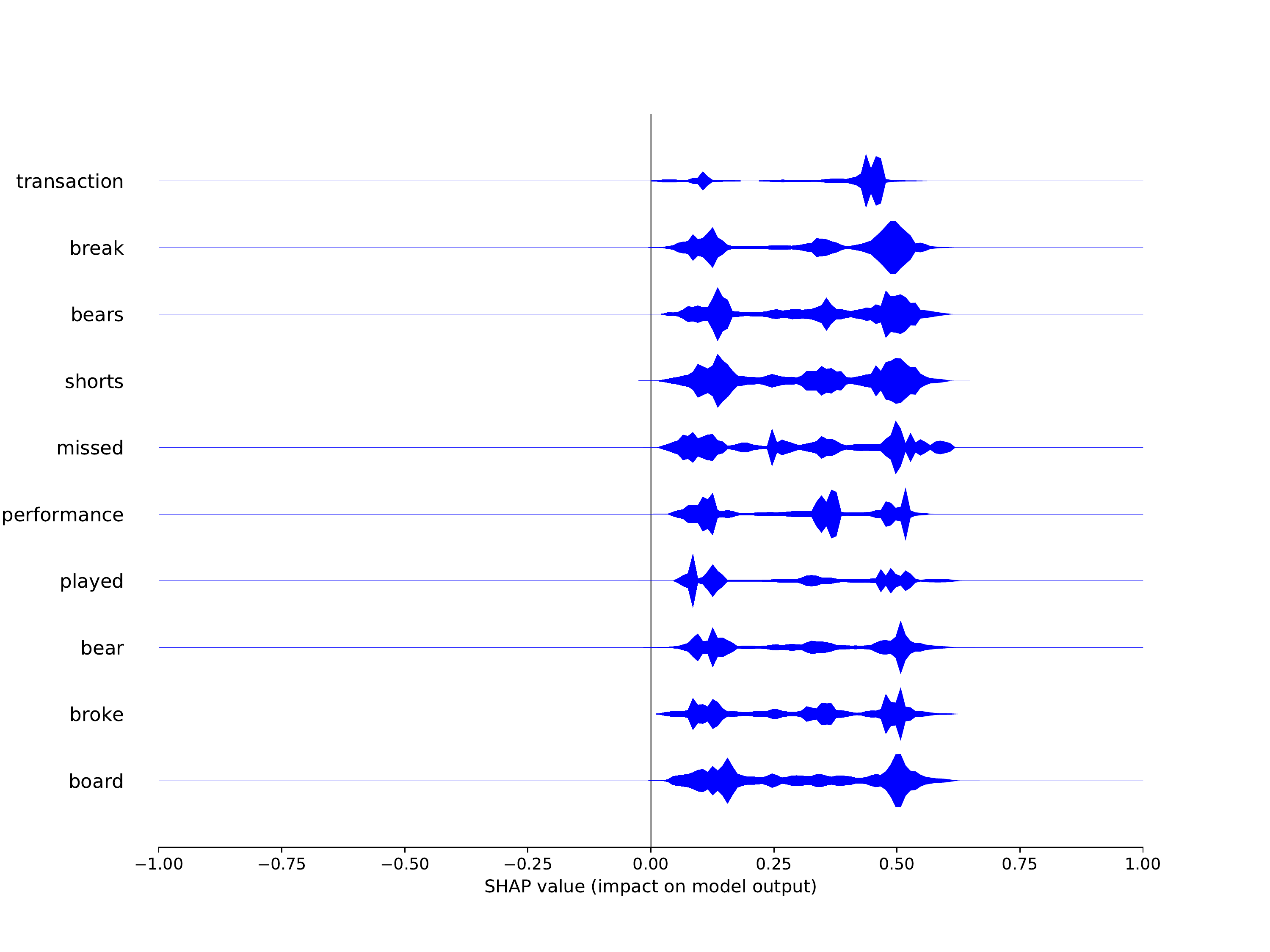}}

\caption{Distribution of Word Importances \& Confusion Matrix for Information Content Model.}\label{fig:shap_finance}
\end{center}
\squeezeup \squeezeup  \begin{flushleft}
	\footnotesize{Notes: Confusion Matrix results reported on the test set are based on the best performing model on the validation set. SHAP values plotted for the ``fundamental'' class. Given that the ranking is based on absolute average SHAP values, the ``chat'' class values are the negative of the ``fundamental'' class values.}
\squeezeup \squeezeup  \end{flushleft}
\end{figure}

The interpretability results for my ``chat type'' model are as expected (see Figure \ref{fig:shap_finance} and Table \ref{tab:shap_finance}). For instance, words such as transaction, bankruptcy are associated with a lower (higher) predicted probability for my ``chat'' (``fundamental'') class. 

\begin{figure}[htbp] 
\begin{center}
 \subfloat[]{\includegraphics[scale=0.25]{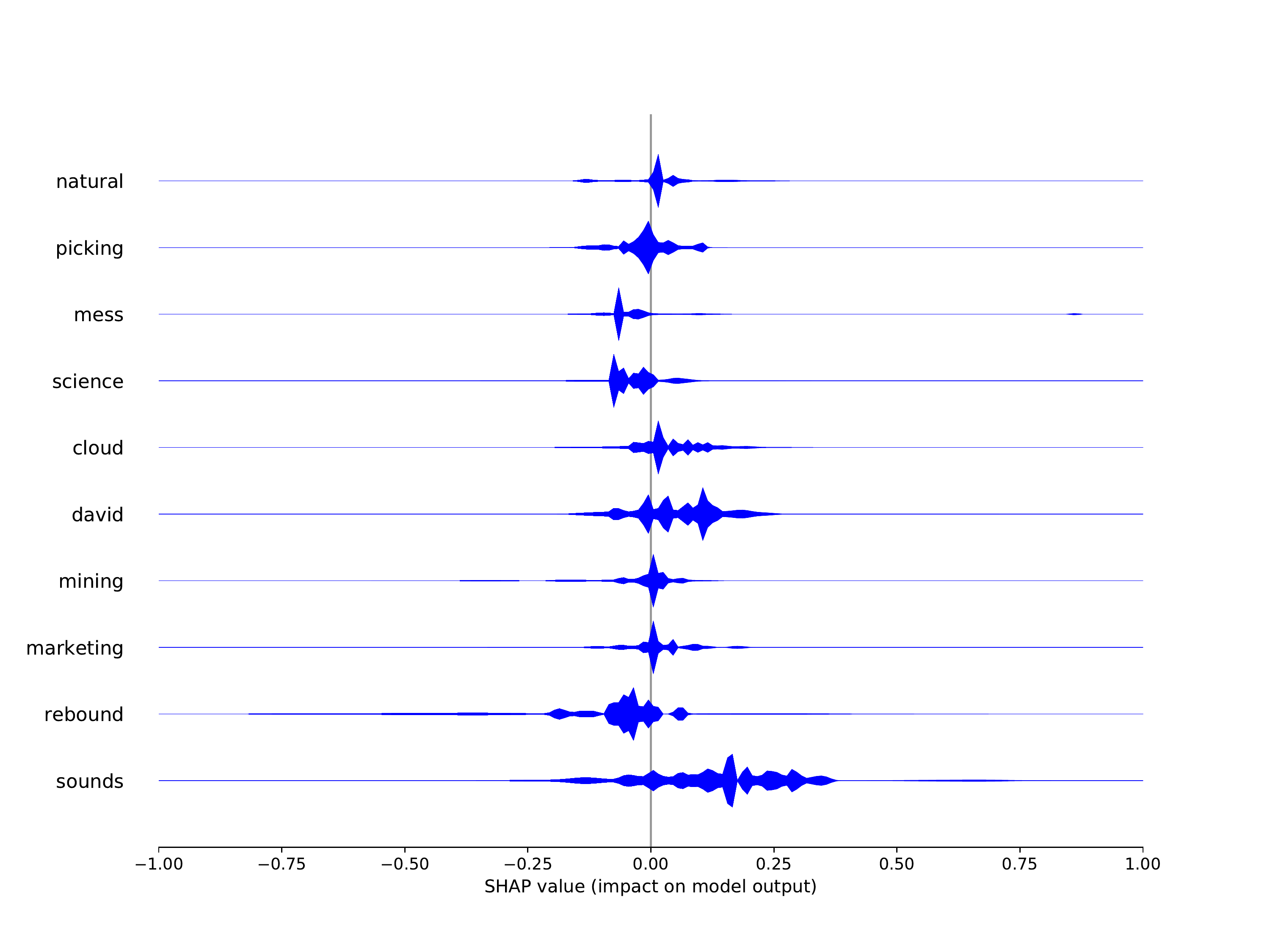}}
 \subfloat[]{\includegraphics[scale=0.25]{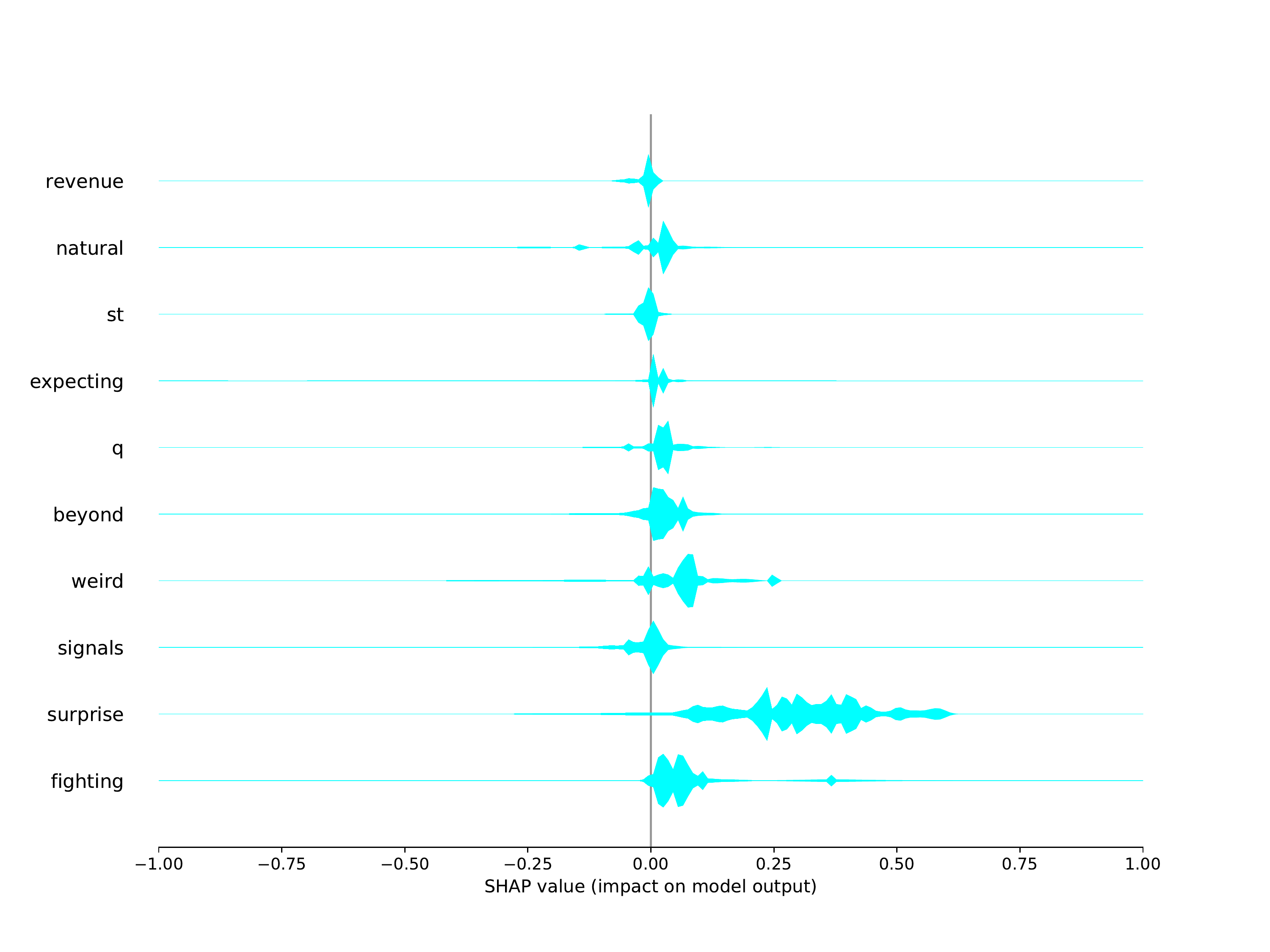}}
 
 \subfloat[]{\includegraphics[scale=0.25]{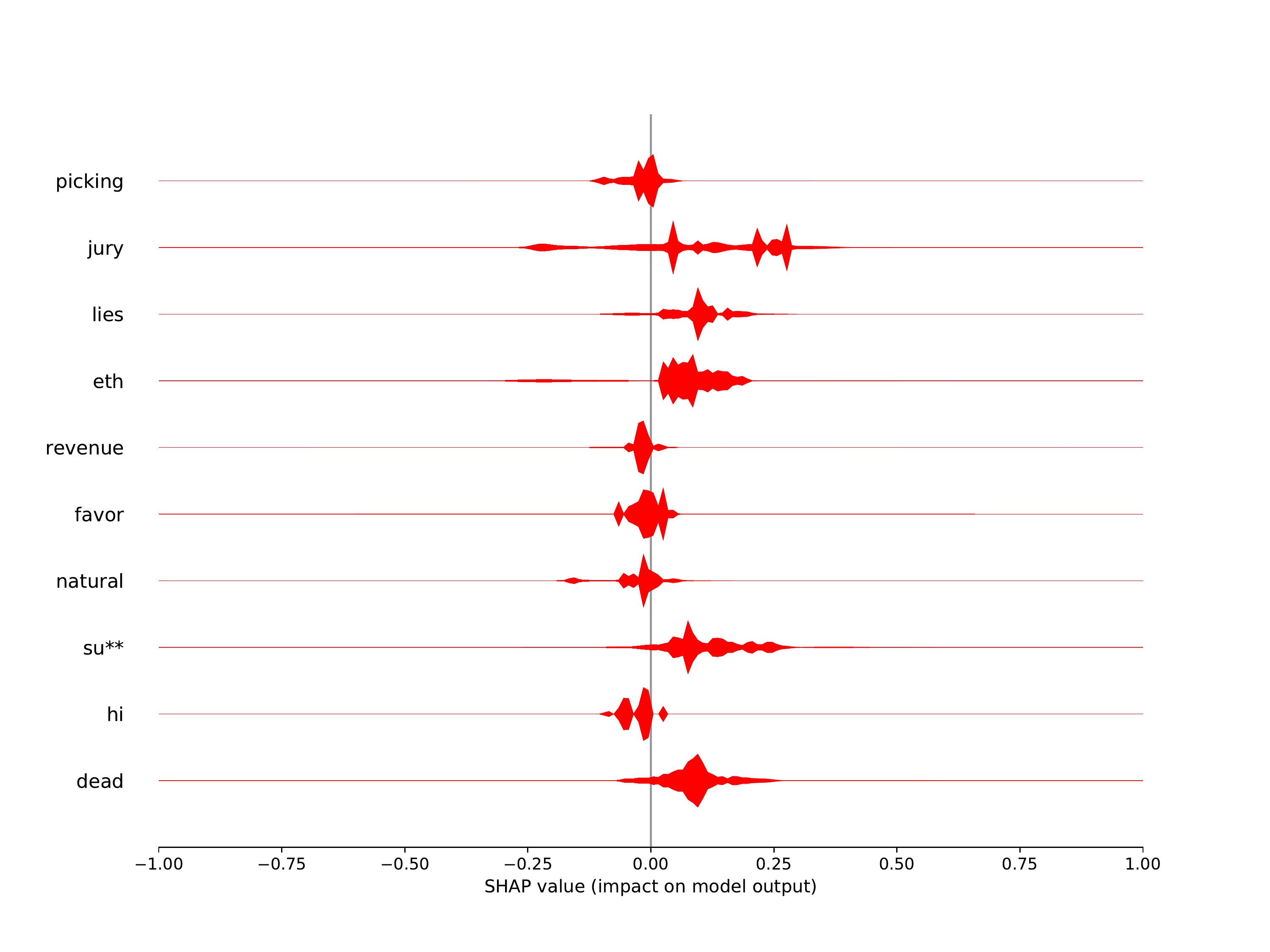}}
\subfloat[]{\includegraphics[scale=0.25]{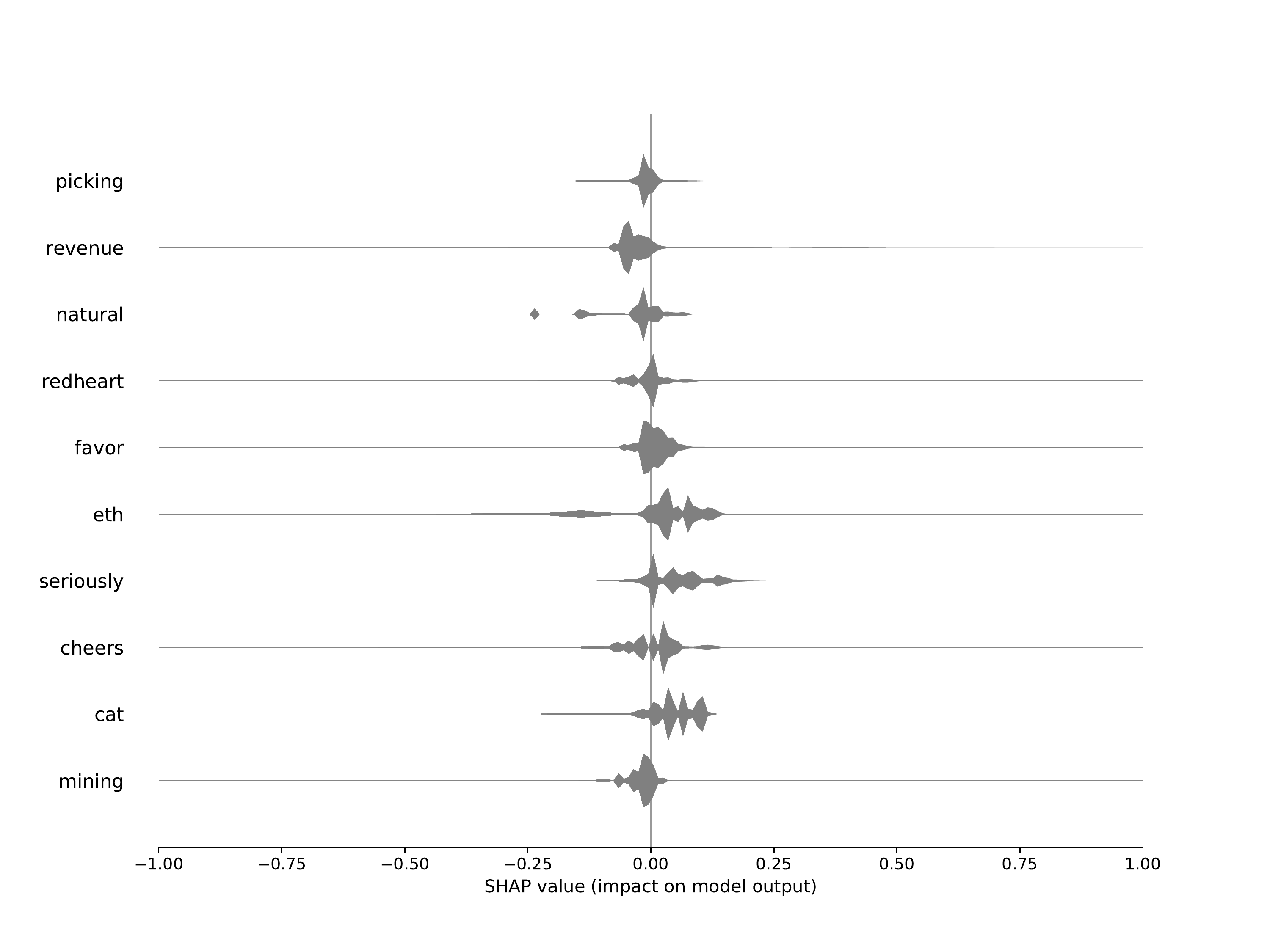}}
 
 \subfloat[]{\includegraphics[scale=0.25]{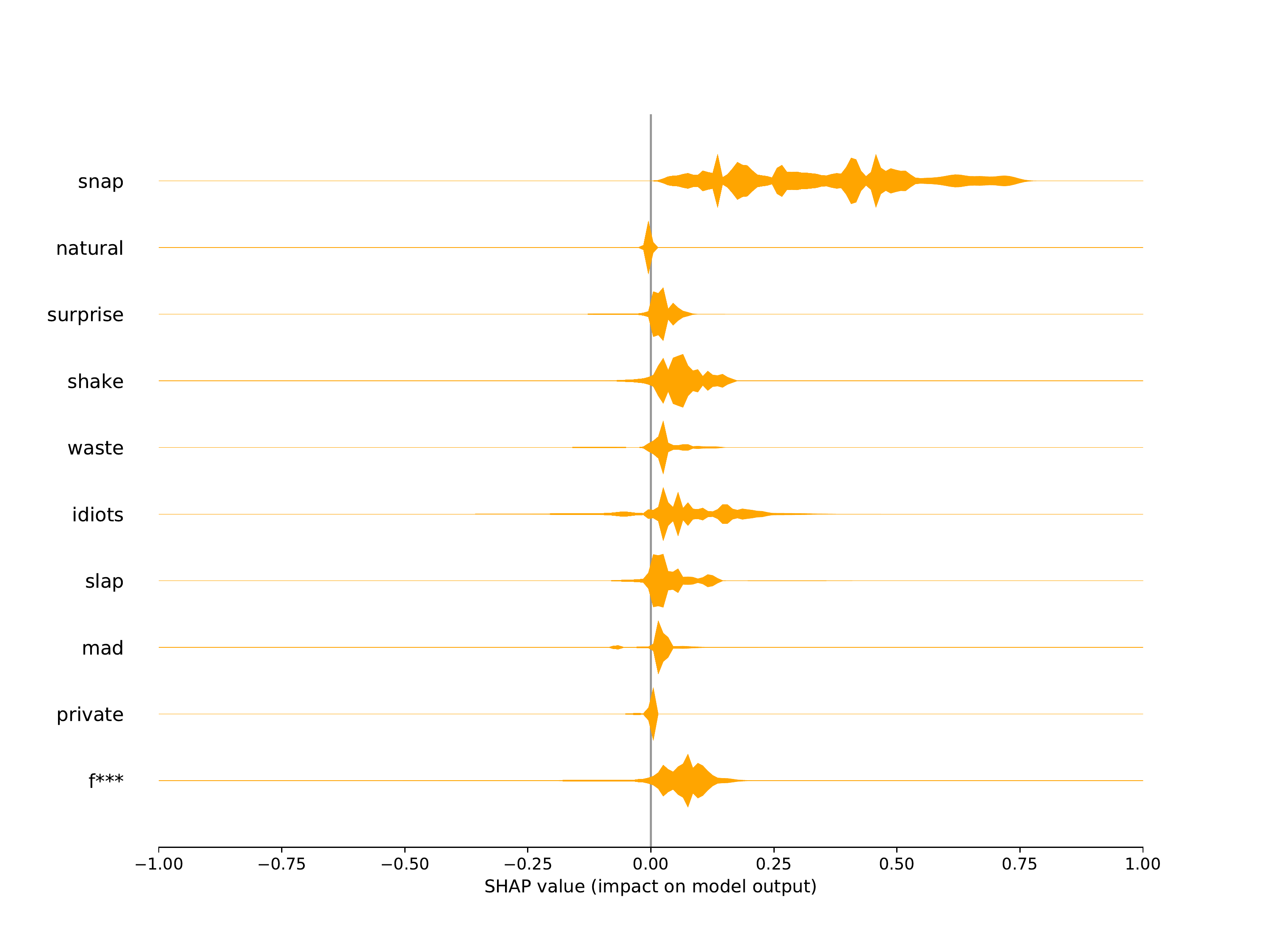}}
 \subfloat[]{\includegraphics[scale=0.25]{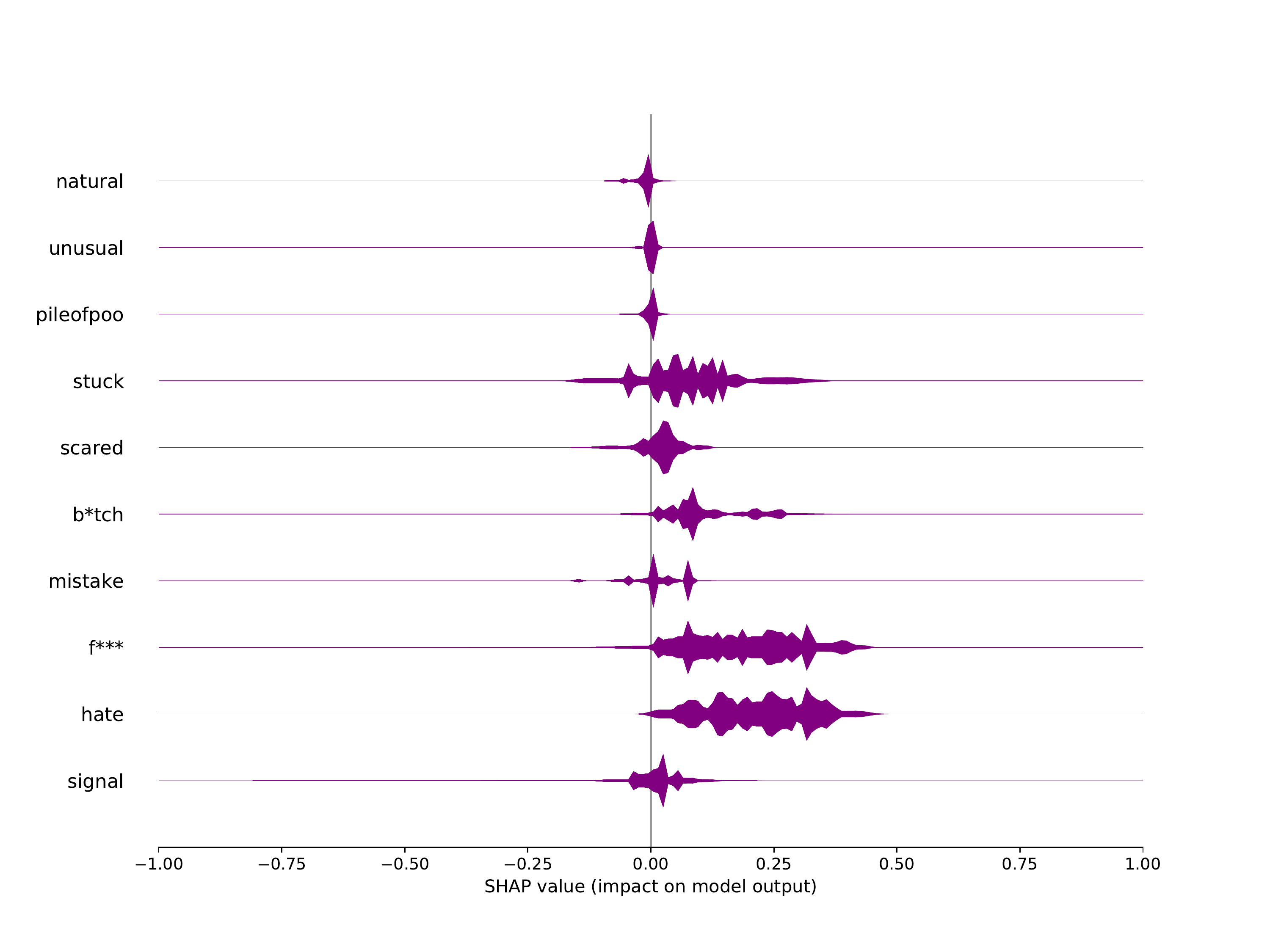}}

\caption{Distribution of Word Importances for Twitter Emotion Model.}\label{fig:shap_twitter}
\end{center}
\squeezeup \squeezeup  \begin{flushleft}
	\footnotesize{Notes: SHAP values evaluated on a random sample of 100,000 StockTwits messages. (a) happy, (b) surprise, (c) sad, (d) fear, (e) anger, (f) disgust. Not shown here: neutral.}
\squeezeup \squeezeup  \end{flushleft}
\end{figure}

\input{Tables/shap_stocktwits}
\input{Tables/shap_twitter}
\input{Tables/shap_finance}

\clearpage 

\section{Computing Excess Returns: Windows}
\begin{figure}[htbp] 
\begin{center}
\subfloat[]{\includegraphics[scale=0.4]{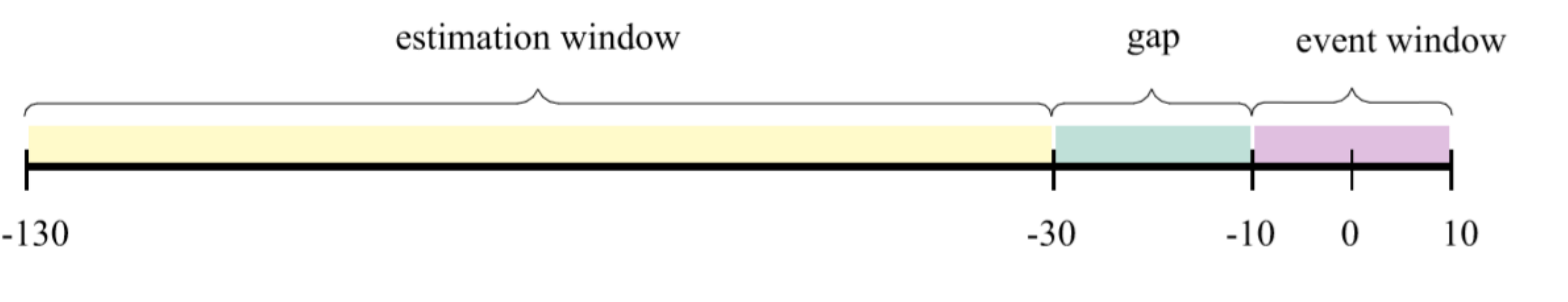}}

\subfloat[]{\includegraphics[scale=0.4]{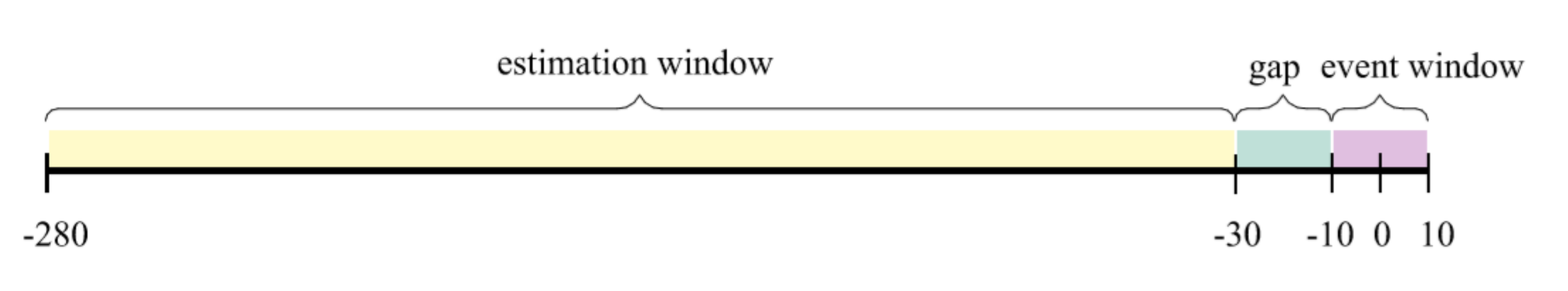}}
	\caption{Excess Return Calculation Windows}\label{fig:excess_calculation}
\end{center}
\squeezeup \squeezeup  \begin{flushleft}
	\footnotesize{Notes: Panel (a) reports estimation windows for my preferred specification, while Panel (b) presents the alternative estimation window for robustness checks.}
\squeezeup \squeezeup  \end{flushleft}
\end{figure}

Figure \ref{fig:excess_calculation} illustrates the event and estimation windows for the event study calculations.

\section{StockTwits Activity \& Sample Distributions}
Table \ref{tab:twits_by_quarter} presents the descriptive statistics on StockTwits activity for my sample. In particular, it reports the frequency distributions by calendar quarter. We can see a dramatic rise in StockTwits activity over my sample period: starting from 8,961 messages in 2010Q1 to 186,742 messages in 2019Q4. This pattern demonstrates the increased popularity of social media during the decade of 2010. Similarly, my coverage also expands to more firms, from 587 in 2010Q1 to 2,663 in the 2019Q4. 

I also compare the sample distributions of messages and firm-quarters by the Fama-French 48-industry groupings with the CRSP universe during my sample period. The results are reported in Table \ref{tab:fama_fench}. My sample spans all 48 industries, and my firm-quarter distribution is fairly similar CRSP's. Therefore, I find little evidence of industry clustering in my sample.

\input{Tables/event_quarterly} \clearpage 
\input{Tables/event_crsp_tab} \clearpage

\appendix

\end{document}

%% file: Tables/sample_restrictions.tex
\begin{table}[htbp]\centering
\small  
\begin{threeparttable}
\caption{Itemized Sample Restrictions}\label{tab:sample_restrictions}
\begin{tabular}{p{2.5in}p{1in}p{1.5in}}
 &  & Messages \\ \hline 
   &  &  \\
StockTwits Data 2010-2019 &  & 117,354,459
 \\ \hline
     &  &  \\
 Keep & &  \\ \hline 
   &  &  \\
NASDAQ/NYSE Ticker
 &  & 101,484,559
 \\
Single Ticker
&  & 74,648,778
 \\ 
Not Automated
 &  & 68,305,130

 \\
IBES/CRSP Ticker & &  60,963,143
 \\  \hline 
   &  &  \\
Final Announcement Sample & & 4,467,461 \\
   & & \\ \hline \hline 
\end{tabular}
\begin{tablenotes}
\footnotesize
\item I restrict my sample to firms with posts from at least 2 users for the period between 10 trading days before the earnings announcement until 2 trading days before the announcement. 
\end{tablenotes}
\end{threeparttable}
\end{table}

%% file: Tables/variable_definitions.tex
\begin{center}
\onehalfspacing
\small 
\begin{ThreePartTable}
\begin{TableNotes}
\footnotesize 
\item For the excess return calculation, see the event and estimation windows in Figure \ref{fig:excess_calculation}. CSHOQ left in millions.
\end{TableNotes}

\begin{longtable}[H]{p{0.75in}p{4in}p{1.25in}}
\caption{Variable Definitions}\label{tab:variable_definitions} \\

Variable & Definition & Source \\ \hline 
\endfirsthead

\multicolumn{3}{c}%
{{\bfseries \tablename\ \thetable{} -- continued from previous page}} \\
Variable & Definition & Source \\ \hline 
& & \\[\dimexpr-\normalbaselineskip+5pt]
\endhead

\insertTableNotes\\
\endlastfoot

Analysts & Natural logarithm of 1 plus the number of analysts in the latest I/B/E/S consensus analyst quarterly earnings per share forecast prior to the   quarter-end date. & I/B/E/S \\
Emotion & Each message is classified by a many-to-one deep learning model into one of the seven categories (i.e., neutral, happy, sad, anger, disgust, surprise, fear), so that the corresponding probabilities sum up to 1. For each emotions separately, we then take the weighted average of these probabilities during the nine trading-day window $[-10,-2]$, where day 0 is the quarterly earnings announcement date and the weights correspond to the number of followers of the user 1+$\log$(1 + \# of Followers).  & StockTwits \\
Exret (\%) & Buy-and-hold abnormal returns measured using \citeN{carhart1997persistence}'s four-factor model for the window specified multiplied by 100. Unless stated otherwise, we compute buy-and-hold abnormal returns for firm i for event window $[t,t+n]$ as follows: $$Exret_{i; t,t+n} = \prod_{k=t}^{t+n} (1+R_{ik}) - \prod_{k=t}^{t+n} (1+ER_{ik})$$ & WRDS (U.S. Daily Event Study) \\
Inst & Number of shares held by institutional investors scaled by total shares outstanding as of the quarter-end date & Thomson Reuters Institutional Holdings (13F) \\
Loss & Indicator variable equal to 1 if earnings before extraordinary items   (IBQ) is strictly negative in the prior quarter, and 0 otherwise & Compustat (Quarterly) \\
MB & Ratio of market value to book value of equity (CSHOQ$*$PRCCQ$/$CEQQ) & Compustat (Quarterly) \\
Sentiment & Twits classified as positive minus twits classified as negative during the nine trading-day window $[-10,-2]$, where day 0  is the quarterly earnings announcement date, using an enhanced naive Bayes classifier. Each positive or negative message is first weighted by the corresponding probability and by the number of followers of the user 1+$\log$(1 + \# of Followers). The measure is scaled by 1 plus the sum of the probability levels. & StockTwits \\
Size & Natural logarithm of market value of equity ($\log$(CSHOQ$*$PRCCQ)). & Compustat (Quarterly) \\
SUE & Standardized unexpected earnings (suescore from I/B/E/S). & I/B/E/S \\
Volatility & Standard deviation of daily returns during the half-year until 10 trading days before the announcement. & CRSP \\
& & \\ 
 \hline \hline 
\end{longtable}
\end{ThreePartTable}
\end{center}

%% file: Tables/summary_stats.tex
\begin{center}
\onehalfspacing
\small 
\begin{ThreePartTable}
\begin{TableNotes}
\footnotesize  
\item Note: $\sigma_{within}$ denotes the within-firm (demeaned) standard deviations. Continuous variables winsorized at the 1\% and 99\% level. 
Emotion classifications based on StockTwits model.
\end{TableNotes}

\begin{longtable}[H]{llllllll}
\caption{Descriptive Statistics}\label{tab:summary_stats} \\

 & Observations & Mean & $\sigma$ & $\sigma_{within}$ & Median & 10\% & 90\%  \\ \hline 
\endfirsthead

\multicolumn{8}{c}%
{{\bfseries \tablename\ \thetable{} -- continued from previous page}} \\
 & Observations & Mean & $\sigma$ & $\sigma_{within}$ & Median & 10\% & 90\%  \\ \hline 
  &  &  &  &  &  & & \\ [\dimexpr-\normalbaselineskip+2pt]
\endhead

\insertTableNotes\\
\endlastfoot

  &  &  &  &  &  & & \\ [\dimexpr-\normalbaselineskip+2pt]
\multicolumn{8}{l}{\uline{\textbf{Panel A: CRSP/IBES/Compustat/Thomson Reuters (13F)}}} \\ 
  &  &  &  &  & &  &  \\ [\dimexpr-\normalbaselineskip+2pt]
EXRET$_{-1,1}$ & 81886 & -0.088 & 8.447 & 8.128 & -0.048 & -9.783 & 9.493 \\
EXRET$_{2,4}$ & 81886 & -0.04 & 3.982 & 3.811 & -0.098 & -4.37 & 4.311 \\
EXRET$_{-10,-2}$ & 81886 & -0.152 & 6.252 & 5.996 & -0.182 & -7.042 & 6.66 \\
SUE & 81886 & 1.052 & 3.861 & 3.417 & 0.707 & -2.315 & 4.95 \\
Loss & 80808 & 0.293 & 0.455 & 0.306 & 0 & 0 & 1 \\
Analysts & 81886 & 2.052 & 0.644 & 0.234 & 2.033 & 1.186 & 2.931 \\
Institutional & 81886 & 0.585 & 0.361 & 0.231 & 0.714 & 0 & 0.982 \\
Size & 80765 & 7.668 & 1.801 & 0.475 & 7.637 & 5.358 & 10.035 \\
MB & 80499 & 4.044 & 6.084 & 4.271 & 2.315 & 0.8 & 7.979 \\
Volatility & 81861 & 0.025 & 0.013 & 0.008 & 0.021 & 0.012 & 0.042 \\ 
\multicolumn{7}{l}{\uline{\textbf{Panel B: StockTwits}}} \\ 
  &  &  &  &  &  &  \\ [\dimexpr-\normalbaselineskip+2pt]
\# of Messages$_{-10,-2}$  & 81886 & 54.557 & 413.773 & 300.613 & 9 & 3 & 62 \\
\# of Distinct Users$_{-10,-2}$ & 81886 & 16.812 & 62.141 & 41.544 & 6 & 2 & 29 \\
Sentiment$_{-10,-2}$ & 81886 & 1.867 & 2.525 & 2.119 & 1.773 & -1.26 & 5.209 \\
Sentiment$_{-1,1}$ & 81077 & 2.03 & 2.416 & 1.795 & 2.005 & -0.979 & 5.183 \\
Anger$_{-1,1}$ & 81077 & 0.005 & 0.013 & 0.012 & 0 & 0 & 0.016 \\
Anger$_{-10,-2}$ & 81886 & 0.005 & 0.014 & 0.013 & 0 & 0 & 0.014 \\
Disgust$_{-1,1}$ & 81077 & 0.006 & 0.016 & 0.014 & 0 & 0 & 0.019 \\
Disgust$_{-10,-2}$ & 81886 & 0.006 & 0.019 & 0.017 & 0 & 0 & 0.018 \\
Fear$_{-1,1}$ & 81077 & 0.04 & 0.059 & 0.052 & 0.008 & 0 & 0.114 \\
Fear$_{-10,-2}$ & 81886 & 0.044 & 0.077 & 0.071 & 0.002 & 0 & 0.131 \\
Happy$_{-1,1}$ & 81077 & 0.135 & 0.127 & 0.105 & 0.117 & 0 & 0.29 \\
Happy$_{-10,-2}$ & 81886 & 0.155 & 0.162 & 0.142 & 0.12 & 0 & 0.366 \\
Neutral$_{-1,1}$ & 81077 & 0.749 & 0.19 & 0.14 & 0.771 & 0.489 & 0.997 \\
Neutral$_{-10,-2}$ & 81886 & 0.727 & 0.223 & 0.184 & 0.749 & 0.431 & 1 \\
Sad$_{-1,1}$ & 81077 & 0.031 & 0.046 & 0.039 & 0.009 & 0 & 0.087 \\
Sad$_{-10,-2}$ & 81886 & 0.029 & 0.051 & 0.046 & 0.004 & 0 & 0.084 \\
Surprise$_{-1,1}$ & 81077 & 0.029 & 0.047 & 0.042 & 0.007 & 0 & 0.084 \\
Surprise$_{-10,-2}$ & 81886 & 0.026 & 0.047 & 0.043 & 0.002 & 0 & 0.081 \\  \hline \hline 

\end{longtable}
\end{ThreePartTable}
\end{center}

%% file: Tables/correlations.tex
\begin{sidewaystable}[htbp]\centering
\begin{singlespacing}
\caption{Correlation Matrix}\label{tab:corr_matrix}
\begin{threeparttable}
\tiny
\def\sym#1{\ifmmode^{#1}\else\(^{#1}\)\fi}
\begin{tabular}{l*{16}{c}}
\hline\hline                                                                                                                                   \\
                &\multicolumn{16}{c}{}                                                                                                                                                                                                                                                                                          \\
     &EXRET$_{-1,1}$         &     SUE$_t$           &SUE$_{t-1}$         &EXRET$_{-10,-2}$         &  Loss         & Anl         &     Inst         &     Size         &       MB         &Sent       &    Happy         &      Sad         &    Anger         &  Disgust         & Surprise         &     Fear         \\
\hline
&&&&&&&&&&&&&&& \\[\dimexpr-\normalbaselineskip+2pt]
EXRET$_{-1,1}$ &     1.00         &                  &                  &                  &                  &                  &                  &                  &                  &                  &                  &                  &                  &                  &                  &                  \\
                &                  &                  &                  &                  &                  &                  &                  &                  &                  &                  &                  &                  &                  &                  &                  &                  \\
&&&&&&&&&&&&&&& \\[\dimexpr-\normalbaselineskip+2pt]
SUE$_t$              &     0.26\sym{***}&     1.00         &                  &                  &                  &                  &                  &                  &                  &                  &                  &                  &                  &                  &                  &                  \\
                &                  &                  &                  &                  &                  &                  &                  &                  &                  &                  &                  &                  &                  &                  &                  &                  \\
&&&&&&&&&&&&&&& \\[\dimexpr-\normalbaselineskip+2pt]
SUE$_{t-1}$       &    -0.02\sym{***}&     0.25\sym{***}&     1.00         &                  &                  &                  &                  &                  &                  &                  &                  &                  &                  &                  &                  &                  \\
                &                  &                  &                  &                  &                  &                  &                  &                  &                  &                  &                  &                  &                  &                  &                  &                  \\
&&&&&&&&&&&&&&& \\[\dimexpr-\normalbaselineskip+2pt]
EXRET$_{-10,-2}$     &     0.01         &     0.03\sym{***}&    -0.01\sym{**} &     1.00         &                  &                  &                  &                  &                  &                  &                  &                  &                  &                  &                  &                  \\
                &                  &                  &                  &                  &                  &                  &                  &                  &                  &                  &                  &                  &                  &                  &                  &                  \\
&&&&&&&&&&&&&&& \\[\dimexpr-\normalbaselineskip+2pt]
Loss        &    -0.08\sym{***}&    -0.18\sym{***}&    -0.13\sym{***}&    -0.02\sym{***}&     1.00         &                  &                  &                  &                  &                  &                  &                  &                  &                  &                  &                  \\
                &                  &                  &                  &                  &                  &                  &                  &                  &                  &                  &                  &                  &                  &                  &                  &                  \\
&&&&&&&&&&&&&&& \\[\dimexpr-\normalbaselineskip+2pt]
Anl       &     0.01\sym{*}  &     0.09\sym{***}&     0.09\sym{***}&     0.00         &    -0.21\sym{***}&     1.00         &                  &                  &                  &                  &                  &                  &                  &                  &                  &                  \\
                &                  &                  &                  &                  &                  &                  &                  &                  &                  &                  &                  &                  &                  &                  &                  &                  \\
&&&&&&&&&&&&&&& \\[\dimexpr-\normalbaselineskip+2pt]
Inst            &     0.02\sym{***}&     0.07\sym{***}&     0.07\sym{***}&    -0.00         &    -0.17\sym{***}&     0.25\sym{***}&     1.00         &                  &                  &                  &                  &                  &                  &                  &                  &                  \\
                &                  &                  &                  &                  &                  &                  &                  &                  &                  &                  &                  &                  &                  &                  &                  &                  \\
&&&&&&&&&&&&&&& \\[\dimexpr-\normalbaselineskip+2pt]
Size            &     0.05\sym{***}&     0.13\sym{***}&     0.12\sym{***}&     0.01\sym{*}  &    -0.42\sym{***}&     0.61\sym{***}&     0.20\sym{***}&     1.00         &                  &                  &                  &                  &                  &                  &                  &                  \\
                &                  &                  &                  &                  &                  &                  &                  &                  &                  &                  &                  &                  &                  &                  &                  &                  \\
&&&&&&&&&&&&&&& \\[\dimexpr-\normalbaselineskip+2pt]
MB              &     0.04\sym{***}&     0.06\sym{***}&     0.06\sym{***}&    -0.01         &     0.07\sym{***}&     0.07\sym{***}&     0.03\sym{***}&     0.11\sym{***}&     1.00         &                  &                  &                  &                  &                  &                  &                  \\
                &                  &                  &                  &                  &                  &                  &                  &                  &                  &                  &                  &                  &                  &                  &                  &                  \\
&&&&&&&&&&&&&&& \\[\dimexpr-\normalbaselineskip+2pt]
Sent      &     0.01\sym{**} &     0.04\sym{***}&     0.04\sym{***}&     0.01         &    -0.13\sym{***}&     0.09\sym{***}&     0.03\sym{***}&     0.15\sym{***}&     0.00         &     1.00         &                  &                  &                  &                  &                  &                  \\
                &                  &                  &                  &                  &                  &                  &                  &                  &                  &                  &                  &                  &                  &                  &                  &                  \\
&&&&&&&&&&&&&&& \\[\dimexpr-\normalbaselineskip+2pt]
Happy           &    -0.01         &     0.01\sym{**} &     0.03\sym{***}&     0.08\sym{***}&     0.07\sym{***}&     0.07\sym{***}&    -0.04\sym{***}&     0.04\sym{***}&     0.05\sym{***}&    -0.07\sym{***}&     1.00         &                  &                  &                  &                  &                  \\
                &                  &                  &                  &                  &                  &                  &                  &                  &                  &                  &                  &                  &                  &                  &                  &                  \\
&&&&&&&&&&&&&&& \\[\dimexpr-\normalbaselineskip+2pt]
Sad             &    -0.01         &    -0.01         &    -0.01         &    -0.05\sym{***}&     0.07\sym{***}&     0.06\sym{***}&    -0.06\sym{***}&     0.01         &     0.03\sym{***}&    -0.09\sym{***}&     0.04\sym{***}&     1.00         &                  &                  &                  &                  \\
                &                  &                  &                  &                  &                  &                  &                  &                  &                  &                  &                  &                  &                  &                  &                  &                  \\
&&&&&&&&&&&&&&& \\[\dimexpr-\normalbaselineskip+2pt]
Anger           &    -0.01         &    -0.01\sym{**} &    -0.01         &    -0.02\sym{***}&     0.10\sym{***}&     0.03\sym{***}&    -0.06\sym{***}&    -0.01         &     0.03\sym{***}&    -0.08\sym{***}&     0.09\sym{***}&     0.14\sym{***}&     1.00         &                  &                  &                  \\
                &                  &                  &                  &                  &                  &                  &                  &                  &                  &                  &                  &                  &                  &                  &                  &                  \\
&&&&&&&&&&&&&&& \\[\dimexpr-\normalbaselineskip+2pt]
Disgust         &    -0.01         &    -0.02\sym{***}&    -0.02\sym{***}&    -0.04\sym{***}&     0.10\sym{***}&     0.02\sym{***}&    -0.08\sym{***}&    -0.02\sym{***}&     0.03\sym{***}&    -0.08\sym{***}&     0.06\sym{***}&     0.13\sym{***}&     0.14\sym{***}&     1.00         &                  &                  \\
                &                  &                  &                  &                  &                  &                  &                  &                  &                  &                  &                  &                  &                  &                  &                  &                  \\
&&&&&&&&&&&&&&& \\[\dimexpr-\normalbaselineskip+2pt]
Surprise        &    -0.01         &    -0.01\sym{*}  &    -0.01         &    -0.02\sym{***}&     0.11\sym{***}&     0.02\sym{***}&    -0.08\sym{***}&    -0.02\sym{***}&     0.04\sym{***}&    -0.09\sym{***}&     0.12\sym{***}&     0.10\sym{***}&     0.13\sym{***}&     0.11\sym{***}&     1.00         &                  \\
                &                  &                  &                  &                  &                  &                  &                  &                  &                  &                  &                  &                  &                  &                  &                  &                  \\
&&&&&&&&&&&&&&& \\[\dimexpr-\normalbaselineskip+2pt]
Fear            &    -0.01         &    -0.00         &    -0.00         &    -0.06\sym{***}&     0.05\sym{***}&     0.08\sym{***}&    -0.03\sym{***}&     0.05\sym{***}&     0.03\sym{***}&    -0.04\sym{***}&     0.00         &     0.16\sym{***}&     0.08\sym{***}&     0.09\sym{***}&     0.07\sym{***}&     1.00         \\
                &                  &                  &                  &                  &                  &                  &                  &                  &                  &                  &                  &                  &                  &                  &                  &                  \\
\hline
Observations    &    81886         &                  &                  &                  &                  &                  &                  &                  &                  &                  &                  &                  &                  &                  &                  &                  \\
\hline\hline
\end{tabular}
\begin{tablenotes}
\item Inst refers to institutional, Sent refers to sentiment, Anl refers to Analysts. Continuous variables winsorized at the 1\% and 99\% level. \sym{*} \(p<0.05\), \sym{**} \(p<0.01\), \sym{***} \(p<0.001\)
\end{tablenotes}
\end{threeparttable}
\end{singlespacing}
\end{sidewaystable}

%% file: Tables/cross_sectional_reg.tex
\begin{table}[h]\centering
\begin{threeparttable}
\small 
\def\sym#1{\ifmmode^{#1}\else\(^{#1}\)\fi}
\caption{Emotions, Earnings Surprises and Announcement Returns \label{tab:reg_cross_section}}
\begin{tabular}{l*{4}{c}}
\hline\hline
& & & &  \\[\dimexpr-\normalbaselineskip+2pt]
                    &\multicolumn{1}{c}{(1)}&\multicolumn{1}{c}{(2)}&\multicolumn{1}{c}{(3)}&\multicolumn{1}{c}{(4)}\\
& & & &  \\[\dimexpr-\normalbaselineskip+2pt]
                    &\multicolumn{1}{c}{SUE}&\multicolumn{1}{c}{SUE}&\multicolumn{1}{c}{EXRET$_{-1,1}$}&\multicolumn{1}{c}{EXRET$_{-1,1}$}\\
\hline
& & & &  \\[\dimexpr-\normalbaselineskip+2pt]
Happy$_{-10,-2}$                &      0.4359\sym{***}&      0.3101\sym{***}&     -0.3902\sym{**} &     -0.4423\sym{**} \\
                    &    (0.0993)         &    (0.0935)         &    (0.1833)         &    (0.1804)         \\
& & & &  \\[\dimexpr-\normalbaselineskip+2pt]
Sad$_{-10,-2}$                  &     -0.4343         &     -0.0468         &     -0.9214         &     -0.0858         \\
                    &    (0.3087)         &    (0.2725)         &    (0.6212)         &    (0.6067)         \\
& & & &  \\[\dimexpr-\normalbaselineskip+2pt]
Disgust$_{-10,-2}$              &     -2.6191\sym{***}&     -0.1508         &     -0.8860         &      1.9378         \\
                    &    (0.7429)         &    (0.6628)         &    (1.6837)         &    (1.6766)         \\
& & & &  \\[\dimexpr-\normalbaselineskip+2pt]
Anger$_{-10,-2}$                &     -3.1294\sym{***}&     -1.1747         &     -5.3028\sym{**} &     -1.4846         \\
                    &    (1.0204)         &    (0.9393)         &    (2.3774)         &    (2.3273)         \\
& & & &  \\[\dimexpr-\normalbaselineskip+2pt]
Fear$_{-10,-2}$                 &     -0.0272         &      0.0816         &     -0.3175         &     -0.0400         \\
                    &    (0.1935)         &    (0.1796)         &    (0.3821)         &    (0.3685)         \\
& & & &  \\[\dimexpr-\normalbaselineskip+2pt]
Surprise$_{-10,-2}$             &     -0.6937\sym{**} &      0.3012         &     -0.2477         &      0.8543         \\
                    &    (0.3078)         &    (0.2952)         &    (0.6945)         &    (0.6769)         \\
& & & &  \\[\dimexpr-\normalbaselineskip+2pt]
Constant            &      1.0474\sym{***}&      0.1720\sym{*}  &      0.0499         &      0.0472         \\
                    &    (0.0349)         &    (0.1004)         &    (0.0491)         &    (0.1211)         \\
\hline
& & & &  \\[\dimexpr-\normalbaselineskip+2pt]
Year, Month, Day of Week FE&           X         &           X         &           X         &           X         \\
Control Variables &  & X &  &X \\
Mean of DV          &      1.0518         &      1.0732         &     -0.0877         &     -0.0801         \\
Std. of DV          &      3.8609         &      3.8215         &      8.4472         &      8.4445         \\
Observations        &  81886         &  77563         &  81886         &  80808         \\
adj. $R^2$          &      0.0080         &      0.0896         &      0.0015         &      0.0730         \\
\hline\hline
\end{tabular}
\begin{tablenotes}
\footnotesize 
\item Notes:  Robust standard errors clustered at the firm (SUE) and industry-quarter (EXRET) level  are in parentheses. \sym{*} \(p<0.10\), \sym{**} \(p<0.05\), \sym{***} \(p<0.01\). Continuous variables winsorized at the 1\% and 99\% level to mitigate the impact of outliers.
\end{tablenotes}
\end{threeparttable}
\end{table}

%% file: Tables/surprise_preannouncement_reg.tex
\begin{sidewaystable}[htbp]\centering
\begin{threeparttable}
\small
\def\sym#1{\ifmmode^{#1}\else\(^{#1}\)\fi}
\caption{Pre-Announcement Emotions and Earnings Surprises \label{tab:reg_pre_announcement_mood_surprise}}
\begin{tabular}{l*{6}{c}}
\hline\hline
     &             &           &           &            &           &    \\[\dimexpr-\normalbaselineskip+2pt]
                    &\multicolumn{1}{c}{(1)}&\multicolumn{1}{c}{(2)}&\multicolumn{1}{c}{(3)}&\multicolumn{1}{c}{(4)}&\multicolumn{1}{c}{(5)}&\multicolumn{1}{c}{(6)}\\
     &             &           &           &            &           &    \\[\dimexpr-\normalbaselineskip+2pt]
                    & & &  \multicolumn{2}{c}{\uline{Chat Type}} & \multicolumn{2}{c}{\uline{Information Type}}  \\ 
                     & & &  \multicolumn{1}{c}{Chat}&\multicolumn{1}{c}{Fundamental} & \multicolumn{1}{c}{Original}&\multicolumn{1}{c}{Dissemination}\\
                                        \multicolumn{7}{l}{Dependent Variable: SUE$_t$} \\ 
\hline
     &             &           &           &            &           &    \\[\dimexpr-\normalbaselineskip+2pt]
Happy$_{-10,-2}$                &      0.1385         &      0.1141         &      0.0721         &      0.0475         &      0.0648         &     -0.0097         \\
                    &    (0.0893)         &    (0.1234)         &    (0.0611)         &    (0.0769)         &    (0.0593)         &    (0.0798)         \\
     &             &           &           &            &           &    \\[\dimexpr-\normalbaselineskip+2pt]
Sad$_{-10,-2}$                  &     -0.3764         &     -0.3107         &      0.2420         &     -0.6728\sym{***}&     -0.1211         &     -0.3434         \\
                    &    (0.2568)         &    (0.3600)         &    (0.1620)         &    (0.2469)         &    (0.1553)         &    (0.2718)         \\
     &             &           &           &            &           &    \\[\dimexpr-\normalbaselineskip+2pt]
Disgust$_{-10,-2}$              &     -0.6156         &     -0.6320         &     -0.2484         &     -0.9476         &     -0.0678         &     -1.0491         \\
                    &    (0.6551)         &    (0.9976)         &    (0.4848)         &    (0.6798)         &    (0.4409)         &    (0.8083)         \\
     &             &           &           &            &           &    \\[\dimexpr-\normalbaselineskip+2pt]
Anger$_{-10,-2}$                &     -1.7652\sym{*}  &     -0.4885         &     -0.6513         &     -1.7059         &     -1.5067\sym{***}&      0.9194         \\
                    &    (0.9717)         &    (1.4546)         &    (0.6497)         &    (1.0473)         &    (0.5283)         &    (1.9162)         \\
     &             &           &           &            &           &    \\[\dimexpr-\normalbaselineskip+2pt]
Fear$_{-10,-2}$                 &      0.0009         &      0.3366         &     -0.0380         &      0.0425         &     -0.0950         &      0.0516         \\
                    &    (0.1740)         &    (0.2483)         &    (0.1572)         &    (0.1535)         &    (0.1027)         &    (0.1621)         \\
     &             &           &           &            &           &    \\[\dimexpr-\normalbaselineskip+2pt]
Surprise$_{-10,-2}$             &     -0.0423         &     -0.4844         &     -0.1267         &      0.1717         &      0.0064         &      0.2992         \\
                    &    (0.2883)         &    (0.4028)         &    (0.1470)         &    (0.2915)         &    (0.1313)         &    (0.3220)         \\
     &             &           &           &            &           &    \\[\dimexpr-\normalbaselineskip+2pt]
Constant            &      0.2984         &     -1.6463\sym{***}&      0.6890\sym{**} &      0.2824         &      0.7121\sym{**} &      0.2620         \\
                    &    (0.2823)         &    (0.4570)         &    (0.3037)         &    (0.2853)         &    (0.3063)         &    (0.2871)         \\
\hline
     &             &           &           &            &           &    \\[\dimexpr-\normalbaselineskip+2pt]
Firm FE             &           X         &           X         &           X         &           X         &           X         &           X         \\
Year, Month, Day of Week FE&           X         &           X         &           X         &           X         &           X         &           X         \\
Controls & X & X & X &X &X&X \\
S\&P 1500 Firms     &                     &           X         &                     &                     &                     &                     \\
$\ge$ median users  &                     &                     &                     &                     &                     &                     \\
$\sigma_{y, within}$&      3.4196         &      3.4346         &      3.3624         &      3.4191         &      3.3307         &      3.4208         \\
Observations        &  77242         &  36663         &  58379         &  76750         &  55936         &  75778         \\
adj. $R^2$          &      0.1902         &      0.1609         &      0.2048         &      0.1900         &      0.2040         &      0.1902         \\
\hline\hline
\end{tabular}
\begin{tablenotes}
\footnotesize 
\item Notes:  Robust standard errors clustered at the firm level are in parentheses. \sym{*} \(p<0.10\), \sym{**} \(p<0.05\), \sym{***} \(p<0.01\). Continuous variables winsorized at the 1\% and 99\% level to mitigate the impact of outliers. I report the within-firm (demeaned) standard deviation of the dependent variable. 
\end{tablenotes}
\end{threeparttable}
\end{sidewaystable}

%% file: Tables/announcement_preannouncement_reg.tex
\begin{sidewaystable}[htbp]\centering
\small 
\begin{threeparttable}
\def\sym#1{\ifmmode^{#1}\else\(^{#1}\)\fi}
\caption{Pre-Announcement Emotions and Announcement Returns \label{tab:reg_pre_announcement_mood_returns}}
\begin{tabular}{l*{7}{c}}
\hline\hline
  &     &     &             &           &           &            &             \\[\dimexpr-\normalbaselineskip+2pt]
                    &\multicolumn{1}{c}{(1)}&\multicolumn{1}{c}{(2)}&\multicolumn{1}{c}{(3)}&\multicolumn{1}{c}{(4)}&\multicolumn{1}{c}{(5)}&\multicolumn{1}{c}{(6)}&\multicolumn{1}{c}{(7)}\\
  &     &     &             &           &           &            &               \\[\dimexpr-\normalbaselineskip+2pt]
                    & & & & \multicolumn{2}{c}{\uline{Chat Type}} & \multicolumn{2}{c}{\uline{Information Type}}  \\ 
                     & & & & \multicolumn{1}{c}{Chat}&\multicolumn{1}{c}{Fundamental} & \multicolumn{1}{c}{Original}&\multicolumn{1}{c}{Dissemination}\\
                    \multicolumn{8}{l}{Dependent Variable: EXRET$_{-1,1}$} \\ 
\hline
     &     &             &           &           &            &           &    \\[\dimexpr-\normalbaselineskip+2pt]
Happy$_{-10,-2}$               &     -0.6243\sym{***}&     -0.7001\sym{***}&     -1.0927\sym{***}&     -0.2457\sym{*}  &     -0.4849\sym{***}&     -0.3814\sym{***}&     -0.4656\sym{**} \\
                    &    (0.1975)         &    (0.2633)         &    (0.4066)         &    (0.1379)         &    (0.1754)         &    (0.1304)         &    (0.1819)         \\
     &     &             &           &           &            &           &    \\[\dimexpr-\normalbaselineskip+2pt]
Sad$_{-10,-2}$                 &      0.3823         &     -0.5314         &      0.7295         &      0.7799\sym{*}  &     -0.7828         &      0.2393         &      0.2127         \\
                    &    (0.6652)         &    (0.8586)         &    (1.1672)         &    (0.4013)         &    (0.6459)         &    (0.3983)         &    (0.6995)         \\
     &     &             &           &           &            &           &    \\[\dimexpr-\normalbaselineskip+2pt]
Disgust$_{-10,-2}$             &      2.5680         &      3.5904         &      3.8202         &     -0.9432         &      4.8074\sym{***}&      1.7455         &      0.9779         \\
                    &    (1.7652)         &    (2.5799)         &    (2.3828)         &    (1.5008)         &    (1.7696)         &    (1.2119)         &    (2.2292)         \\
     &     &             &           &           &            &           &    \\[\dimexpr-\normalbaselineskip+2pt]
Anger$_{-10,-2}$               &      0.0334         &      2.0671         &      0.9995         &      0.1685         &     -1.7922         &      0.0446         &     -7.6382         \\
                    &    (2.4488)         &    (3.6239)         &    (3.2609)         &    (1.7560)         &    (2.7051)         &    (1.4306)         &    (4.7303)         \\
     &     &             &           &           &            &           &    \\[\dimexpr-\normalbaselineskip+2pt]
Fear$_{-10,-2}$                &      0.4902         &      0.3964         &      1.3151\sym{*}  &      0.4197         &      0.4748         &      0.1099         &      0.2370         \\
                    &    (0.3891)         &    (0.5210)         &    (0.7202)         &    (0.3667)         &    (0.3390)         &    (0.2222)         &    (0.3683)         \\
     &     &             &           &           &            &           &    \\[\dimexpr-\normalbaselineskip+2pt]
Surprise$_{-10,-2}$            &      1.0479         &      0.9908         &      1.3625         &     -0.4526         &      1.8338\sym{***}&      0.3430         &      0.0424         \\
                    &    (0.7149)         &    (0.9850)         &    (1.1271)         &    (0.3756)         &    (0.6894)         &    (0.3389)         &    (0.7412)         \\
 \\[\dimexpr-\normalbaselineskip+2pt]
Constant            &      0.5219         &      0.3583         &     -0.0695         &      0.4829         &      0.5087         &      0.2963         &      0.5549\sym{*}  \\
                    &    (0.3322)         &    (0.5414)         &    (0.5386)         &    (0.3912)         &    (0.3345)         &    (0.4026)         &    (0.3317)         \\
\hline
&             &           &           &            &           &    \\[\dimexpr-\normalbaselineskip+2pt]
Firm FE             &           X         &           X         &           X         &           X         &           X         &           X         &           X         \\
Year, Month, Day of Week FE&           X         &           X         &           X         &           X         &           X         &           X         &           X         \\
Control Variables &  X & X & X & X & X & X & X \\
S\&P 1500 Firms     &                     &           X         &                     &                     &                     &                     &                     \\
$\ge$ median users  &                     &                     &           X         &                     &                     &                     &                     \\
$\sigma_{y, within}$&      8.1392         &      7.7552         &      8.6499         &      8.3875         &      8.1401         &      8.5494         &      8.1195         \\
Observations        &  80492         &  37509         &  41535         &  60773         &  79963         &  58260         &  78817         \\
adj. $R^2$          &      0.0854         &      0.1031         &      0.0765         &      0.0796         &      0.0857         &      0.0773         &      0.0861         \\ \hline \hline 
\end{tabular}
\begin{tablenotes}
\footnotesize 
\item Notes:  Robust standard errors clustered at the industry and quarter level are in parentheses. \sym{*} \(p<0.10\), \sym{**} \(p<0.05\), \sym{***} \(p<0.01\). Continuous variables winsorized at the 1\% and 99\% level to mitigate the impact of outliers. I report the within-firm (demeaned) standard deviation of the dependent variable. 
\end{tablenotes}
\end{threeparttable}
\end{sidewaystable}

%% file: Tables/pre_announcement_preannouncement_reg.tex
\begin{sidewaystable}[htbp]\centering
\small
\begin{threeparttable}
\def\sym#1{\ifmmode^{#1}\else\(^{#1}\)\fi}
\caption{Pre-Announcement Emotions and Pre-Announcement Returns \label{tab:reg_pre_announcement_mood_prereturns}}
\begin{tabular}{l*{7}{c}}
\hline\hline
  &     &     &             &           &           &            &             \\[\dimexpr-\normalbaselineskip+2pt]
                    &\multicolumn{1}{c}{(1)}&\multicolumn{1}{c}{(2)}&\multicolumn{1}{c}{(3)}&\multicolumn{1}{c}{(4)}&\multicolumn{1}{c}{(5)}&\multicolumn{1}{c}{(6)}&\multicolumn{1}{c}{(7)}\\
  &     &     &             &           &           &            &               \\[\dimexpr-\normalbaselineskip+2pt]
                    & & & & \multicolumn{2}{c}{\uline{Chat Type}} & \multicolumn{2}{c}{\uline{Information Type}}  \\ 
                     & & & & \multicolumn{1}{c}{Chat}&\multicolumn{1}{c}{Fundamental} & \multicolumn{1}{c}{Original}&\multicolumn{1}{c}{Dissemination}\\
                    \multicolumn{7}{l}{Dependent Variable: EXRET$_{-10,-2}$} \\ \hline 
  &     &     &             &           &           &            &             \\[\dimexpr-\normalbaselineskip+2pt]
\hline
Happy$_{-10,-2}$              &      3.5511\sym{***}&      2.4076\sym{***}&      6.7529\sym{***}&      1.5568\sym{***}&      2.4344\sym{***}&      1.0133\sym{***}&      1.8890\sym{***}\\
                    &    (0.1768)         &    (0.1861)         &    (0.3895)         &    (0.1121)         &    (0.1504)         &    (0.0972)         &    (0.1519)         \\
  &     &     &             &           &           &            &             \\[\dimexpr-\normalbaselineskip+2pt]
Sad$_{-10,-2}$                 &     -4.8477\sym{***}&     -3.2560\sym{***}&    -10.9442\sym{***}&     -2.7631\sym{***}&     -2.9521\sym{***}&     -2.8623\sym{***}&     -3.0090\sym{***}\\
                    &    (0.5168)         &    (0.6058)         &    (0.9473)         &    (0.3429)         &    (0.5091)         &    (0.2962)         &    (0.5242)         \\
  &     &     &             &           &           &            &             \\[\dimexpr-\normalbaselineskip+2pt]
Disgust$_{-10,-2}$               &     -9.5653\sym{***}&     -9.5326\sym{***}&    -12.9609\sym{***}&     -4.3280\sym{***}&    -10.7611\sym{***}&     -5.8141\sym{***}&    -10.4051\sym{***}\\
                    &    (1.4950)         &    (1.8327)         &    (1.9634)         &    (1.0027)         &    (1.6383)         &    (0.9697)         &    (1.8998)         \\
  &     &     &             &           &           &            &             \\[\dimexpr-\normalbaselineskip+2pt]
Anger$_{-10,-2}$                 &     -4.7888\sym{***}&     -3.0401         &     -8.7994\sym{***}&     -4.9647\sym{***}&     -3.8246\sym{*}  &     -2.5236\sym{**} &      0.0341         \\
                    &    (1.7639)         &    (2.4596)         &    (2.2383)         &    (1.2621)         &    (2.1306)         &    (1.0315)         &    (3.3809)         \\
  &     &     &             &           &           &            &             \\[\dimexpr-\normalbaselineskip+2pt]
Fear$_{-10,-2}$                  &     -4.2870\sym{***}&     -3.7990\sym{***}&     -8.3775\sym{***}&     -2.6132\sym{***}&     -3.3360\sym{***}&     -1.9855\sym{***}&     -3.7747\sym{***}\\
                    &    (0.3256)         &    (0.3857)         &    (0.6514)         &    (0.3129)         &    (0.2890)         &    (0.1779)         &    (0.3240)         \\
  &     &     &             &           &           &            &             \\[\dimexpr-\normalbaselineskip+2pt]
Surprise$_{-10,-2}$              &     -2.7351\sym{***}&     -1.2561\sym{*}  &     -5.6229\sym{***}&     -1.7570\sym{***}&     -1.5261\sym{***}&     -1.9512\sym{***}&      0.6704         \\
                    &    (0.5804)         &    (0.6704)         &    (0.9729)         &    (0.2815)         &    (0.5546)         &    (0.2557)         &    (0.6197)         \\
  &     &     &             &           &           &            &             \\[\dimexpr-\normalbaselineskip+2pt]
Constant            &     -0.4264         &      0.3664         &     -0.3876         &     -0.0911         &     -0.3600         &      0.1014         &     -0.2248         \\
                    &    (0.2911)         &    (0.3820)         &    (0.4945)         &    (0.3665)         &    (0.2955)         &    (0.3846)         &    (0.2915)         \\
\hline
&             &           &           &            &           &    \\[\dimexpr-\normalbaselineskip+2pt]
Firm FE             &           X         &           X         &           X         &           X         &           X         &           X         &           X         \\
Year, Month, Day of Week FE&           X         &           X         &           X         &           X         &           X         &           X         &           X         \\
Control Variables &  X & X & X & X & X & X & X \\
S\&P 1500 Firms     &                     &           X         &                     &                     &                     &                     &                     \\
$\ge$ median users  &                     &                     &           X         &                     &                     &                     &                     \\
$\sigma_{y, within}$&      5.9999         &      5.0492         &      6.7689         &      6.3793         &      6.0034         &      6.5221         &      5.9796         \\
Observations        &  80492         &  37509         &  41535         &  60773         &  79963         &  58260         &  78817         \\
adj. $R^2$          &      0.0293         &      0.0180         &      0.0447         &      0.0210         &      0.0245         &      0.0212         &      0.0231         \\
\hline\hline
\end{tabular}
\begin{tablenotes}
\footnotesize 
\item Notes:  Robust standard errors clustered at the industry-quarter level are in parentheses. \sym{*} \(p<0.10\), \sym{**} \(p<0.05\), \sym{***} \(p<0.01\). Continuous variables winsorized at the 1\% and 99\% level to mitigate the impact of outliers. I report the within-firm (demeaned) standard deviation of the dependent variable. 
\end{tablenotes}
\end{threeparttable}
\end{sidewaystable}

%% file: Tables/announcement_announcement_reg.tex
\begin{sidewaystable}[htbp]\centering
\begin{threeparttable}
\small 
\def\sym#1{\ifmmode^{#1}\else\(^{#1}\)\fi}
\caption{Announcement Emotions and Announcement Returns\label{tab:reg_announcement_day}}
\begin{tabular}{l*{7}{c}}
\hline\hline
  &     &     &             &           &           &            &             \\[\dimexpr-\normalbaselineskip+2pt]
                    &\multicolumn{1}{c}{(1)}&\multicolumn{1}{c}{(2)}&\multicolumn{1}{c}{(3)}&\multicolumn{1}{c}{(4)}&\multicolumn{1}{c}{(5)}&\multicolumn{1}{c}{(6)}&\multicolumn{1}{c}{(7)}\\
  &     &     &             &           &           &            &               \\[\dimexpr-\normalbaselineskip+2pt]
                    & & & & \multicolumn{2}{c}{\uline{Chat Type}} & \multicolumn{2}{c}{\uline{Information Type}}  \\ 
                     & & & & \multicolumn{1}{c}{Chat}&\multicolumn{1}{c}{Fundamental} & \multicolumn{1}{c}{Original}&\multicolumn{1}{c}{Dissemination}\\
                    \multicolumn{7}{l}{Dependent Variable: EXRET$_{-1,1}$} \\ \hline 
\hline
  &     &     &             &           &           &            &             \\[\dimexpr-\normalbaselineskip+2pt]
Happy$_{-1,1}$               &     12.3042\sym{***}&     11.3934\sym{***}&     22.0977\sym{***}&      4.6221\sym{***}&      9.6529\sym{***}&      3.2759\sym{***}&      7.0904\sym{***}\\
                    &    (0.4095)         &    (0.4530)         &    (0.6941)         &    (0.2018)         &    (0.3524)         &    (0.1620)         &    (0.3299)         \\
  &     &     &             &           &           &            &             \\[\dimexpr-\normalbaselineskip+2pt]
Sad$_{-1,1}$                 &    -19.5434\sym{***}&    -17.3859\sym{***}&    -32.3191\sym{***}&    -10.1680\sym{***}&    -13.5993\sym{***}&     -6.6619\sym{***}&    -12.1201\sym{***}\\
                    &    (0.9214)         &    (1.1943)         &    (1.4882)         &    (0.4868)         &    (0.8466)         &    (0.3493)         &    (0.9798)         \\
  &     &     &             &           &           &            &             \\[\dimexpr-\normalbaselineskip+2pt]
Disgust$_{-1,1}$             &    -44.0120\sym{***}&    -43.9898\sym{***}&    -52.5682\sym{***}&    -16.7932\sym{***}&    -41.5975\sym{***}&    -15.2944\sym{***}&    -48.1765\sym{***}\\
                    &    (2.9657)         &    (4.2516)         &    (3.6714)         &    (1.5248)         &    (3.1578)         &    (1.3594)         &    (3.9800)         \\
  &     &     &             &           &           &            &             \\[\dimexpr-\normalbaselineskip+2pt]
Anger$_{-1,1}$               &    -14.4643\sym{***}&     -3.9460         &    -18.5428\sym{***}&     -9.9765\sym{***}&    -13.5246\sym{***}&    -10.2545\sym{***}&     10.5939\sym{*}  \\
                    &    (3.8101)         &    (5.0062)         &    (4.7128)         &    (1.9380)         &    (3.8342)         &    (1.6866)         &    (6.3188)         \\
  &     &     &             &           &           &            &             \\[\dimexpr-\normalbaselineskip+2pt]
Fear$_{-1,1}$                &    -17.2914\sym{***}&    -16.5278\sym{***}&    -30.5120\sym{***}&     -7.3773\sym{***}&    -14.8993\sym{***}&     -4.5893\sym{***}&    -21.2039\sym{***}\\
                    &    (0.7554)         &    (0.8928)         &    (1.2096)         &    (0.4906)         &    (0.7057)         &    (0.3500)         &    (0.8863)         \\
  &     &     &             &           &           &            &             \\[\dimexpr-\normalbaselineskip+2pt]
Surprise$_{-1,1}$            &     -5.1685\sym{***}&     -4.8144\sym{***}&    -13.5744\sym{***}&     -2.4853\sym{***}&     -4.7825\sym{***}&     -3.5753\sym{***}&      1.4096         \\
                    &    (0.8949)         &    (1.1746)         &    (1.5627)         &    (0.4883)         &    (0.8350)         &    (0.3600)         &    (0.8692)         \\
  &     &     &             &           &           &            &             \\[\dimexpr-\normalbaselineskip+2pt]
Constant            &      0.2622         &      0.0547         &     -0.2526         &      0.6980\sym{*}  &      0.3739         &      0.7434\sym{*}  &      0.4038         \\
                    &    (0.3334)         &    (0.5436)         &    (0.5525)         &    (0.3936)         &    (0.3364)         &    (0.4247)         &    (0.3369)         \\
\hline
&             &           &           &            &           &    \\[\dimexpr-\normalbaselineskip+2pt]
Firm FE             &           X         &           X         &           X         &           X         &           X         &           X         &           X         \\
Year, Month, Day of Week FE&           X         &           X         &           X         &           X         &           X         &           X         &           X         \\
Control Variables &  X & X & X & X & X & X & X \\
S\&P 1500 Firms     &                     &           X         &                     &                     &                     &                     &                     \\
$\ge$ median users  &                     &                     &           X         &                     &                     &                     &                     \\
$\sigma_{y, within}$&      8.5198         &      8.0162         &      9.7884         &      8.9351         &      8.5211         &      9.1901         &      8.5201         \\
Observations        &  78089         &  36598         &  42598         &  64890         &  78051         &  60174         &  77614         \\
adj. $R^2$          &      0.1389         &      0.1542         &      0.1974         &      0.1175         &      0.1252         &      0.1176         &      0.1239         \\
\hline\hline
\end{tabular}
\begin{tablenotes}
\footnotesize 
\item Notes:  Robust standard errors clustered at the industry and quarter level are in parentheses. \sym{*} \(p<0.10\), \sym{**} \(p<0.05\), \sym{***} \(p<0.01\). Continuous variables winsorized at the 1\% and 99\% level to mitigate the impact of outliers. I report the within-firm (demeaned) standard deviation of the dependent variable. 
\end{tablenotes}
\end{threeparttable}
\end{sidewaystable}

%% file: Tables/announcement_post_announcement_reg.tex
\begin{sidewaystable}[htbp]\centering
\small
\begin{threeparttable}
\def\sym#1{\ifmmode^{#1}\else\(^{#1}\)\fi}
\caption{Announcement Emotions and Post-Announcement Returns \label{tab:reg_post_announcement}}
\begin{tabular}{l*{7}{c}}
\hline\hline
  &     &     &             &           &           &            &             \\[\dimexpr-\normalbaselineskip+2pt]
                    &\multicolumn{1}{c}{(1)}&\multicolumn{1}{c}{(2)}&\multicolumn{1}{c}{(3)}&\multicolumn{1}{c}{(4)}&\multicolumn{1}{c}{(5)}&\multicolumn{1}{c}{(6)}&\multicolumn{1}{c}{(7)}\\
  &     &     &             &           &           &            &               \\[\dimexpr-\normalbaselineskip+2pt]
                    & & & & \multicolumn{2}{c}{\uline{Chat Type}} & \multicolumn{2}{c}{\uline{Information Type}}  \\ 
                     & & & & \multicolumn{1}{c}{Chat}&\multicolumn{1}{c}{Fundamental} & \multicolumn{1}{c}{Original}&\multicolumn{1}{c}{Dissemination}\\
                    \multicolumn{7}{l}{Dependent Variable: EXRET$_{2,4}$} \\ \hline 
&             &           &           &            &           &    \\[\dimexpr-\normalbaselineskip+2pt]
Happy$_{-1,1}$               &     -0.5222\sym{***}&     -0.4323\sym{**} &     -0.6795\sym{**} &     -0.1133         &     -0.4303\sym{***}&     -0.1867\sym{**} &     -0.3595\sym{**} \\
                    &    (0.1614)         &    (0.1932)         &    (0.2772)         &    (0.0863)         &    (0.1505)         &    (0.0725)         &    (0.1512)         \\
&             &           &           &            &           &    \\[\dimexpr-\normalbaselineskip+2pt]
Sad$_{-1,1}$                 &      0.5187         &      1.0992\sym{**} &      0.7152         &     -0.1234         &      0.6562         &      0.1506         &      0.3632         \\
                    &    (0.4271)         &    (0.5182)         &    (0.6573)         &    (0.2186)         &    (0.4023)         &    (0.1574)         &    (0.4620)         \\
&             &           &           &            &           &    \\[\dimexpr-\normalbaselineskip+2pt]
Disgust$_{-1,1}$             &      0.3158         &      1.6946         &      0.5729         &     -0.7885         &      0.7143         &     -0.5280         &      2.9407\sym{*}  \\
                    &    (1.3971)         &    (1.7116)         &    (1.6394)         &    (0.7627)         &    (1.4270)         &    (0.6765)         &    (1.6958)         \\
&             &           &           &            &           &    \\[\dimexpr-\normalbaselineskip+2pt]
Anger$_{-1,1}$               &     -0.5171         &      1.6903         &      0.1925         &     -0.7382         &      0.8352         &      0.1741         &      0.6276         \\
                    &    (1.6884)         &    (2.2690)         &    (2.2101)         &    (0.9413)         &    (1.8149)         &    (0.7439)         &    (3.0866)         \\
&             &           &           &            &           &    \\[\dimexpr-\normalbaselineskip+2pt]
Fear$_{-1,1}$                &      0.8190\sym{***}&      0.4908         &      1.4762\sym{***}&      0.6231\sym{***}&      0.5382\sym{*}  &      0.3450\sym{**} &      0.5006\sym{*}  \\
                    &    (0.3128)         &    (0.3733)         &    (0.5282)         &    (0.2040)         &    (0.2885)         &    (0.1571)         &    (0.2748)         \\
&             &           &           &            &           &    \\[\dimexpr-\normalbaselineskip+2pt]
Surprise$_{-1,1}$            &     -0.0184         &     -0.2576         &      0.3396         &      0.0036         &      0.0596         &      0.0312         &     -0.2071         \\
                    &    (0.4236)         &    (0.5326)         &    (0.7212)         &    (0.2338)         &    (0.3937)         &    (0.1718)         &    (0.4088)         \\
&             &           &           &            &           &    \\[\dimexpr-\normalbaselineskip+2pt]
Constant            &     -1.0104\sym{***}&     -0.8762\sym{***}&     -1.3892\sym{***}&     -1.0139\sym{***}&     -1.0321\sym{***}&     -1.1223\sym{***}&     -1.0162\sym{***}\\
                    &    (0.1746)         &    (0.2330)         &    (0.2803)         &    (0.2013)         &    (0.1751)         &    (0.2110)         &    (0.1744)         \\
\hline
&             &           &           &            &           &    \\[\dimexpr-\normalbaselineskip+2pt]
Firm FE             &           X         &           X         &           X         &           X         &           X         &           X         &           X         \\
Year, Month, Day of Week FE&           X         &           X         &           X         &           X         &           X         &           X         &           X         \\
Control Variables &  X & X & X & X & X & X & X \\
S\&P 1500 Firms     &                     &           X         &                     &                     &                     &                     &                     \\
$\ge$ median users  &                     &                     &           X         &                     &                     &                     &                     \\
$\sigma_{y, within}$&      3.9859         &      3.2876         &      4.1997         &      4.0925         &      3.9852         &      4.1632         &      3.9830         \\
Observations        &  78089         &  36598         &  42598         &  64890         &  78051         &  60174         &  77614         \\
adj. $R^2$          &      0.0228         &      0.0091         &      0.0298         &      0.0217         &      0.0227         &      0.0236         &      0.0226         \\
\hline\hline
\end{tabular}
\begin{tablenotes}
\footnotesize 
\item Notes:  Robust standard errors clustered at the industry and quarter level are in parentheses. \sym{*} \(p<0.10\), \sym{**} \(p<0.05\), \sym{***} \(p<0.01\). Continuous variables winsorized at the 1\% and 99\% level to mitigate the impact of outliers. I report the within-firm (demeaned) standard deviation of the dependent variable. 
\end{tablenotes}
\end{threeparttable}
\end{sidewaystable}

%% file: Tables/announcement_preannouncement_heterogeneity_reg.tex
\begin{table}[!h]\centering
\small 
\begin{threeparttable}
\def\sym#1{\ifmmode^{#1}\else\(^{#1}\)\fi}
\squeezeup \squeezeup 
\caption{Pre-Announcement Emotions, Announcement Returns and Earnings \label{tab:reg_pre_announcement_mood_returns_heterogeneity}}
\begin{tabular}{p{2in}p{1in}p{1in}p{1in}p{1in}}
\hline\hline
      &                        &            &           &    \\[\dimexpr-\normalbaselineskip+2pt]
                    &\multicolumn{1}{c}{(1)}&\multicolumn{1}{c}{(2)}&\multicolumn{1}{c}{(3)}&\multicolumn{1}{c}{(4)}\\
       &             &           &           &         \\[\dimexpr-\normalbaselineskip+2pt]
                     & & & \multicolumn{2}{c}{\uline{Earnings Surprise}} \\ 
                    & & &  \multicolumn{1}{c}{Negative}&\multicolumn{1}{c}{Positive}\\
                    \multicolumn{5}{l}{Dependent Variable: EXRET$_{-1,1}$} \\ \hline 
      &                        &            &           &    \\[\dimexpr-\normalbaselineskip+2pt]                  
Happy$_{-10,-2}$               &     -0.6243\sym{***}&     -0.5346\sym{***}&     -1.1025\sym{***}&     -0.6622\sym{***}\\
                    &    (0.1975)         &    (0.1917)         &    (0.3604)         &    (0.2434)         \\
      &                        &            &           &    \\[\dimexpr-\normalbaselineskip+2pt] 
Sad$_{-10,-2}$                 &      0.3823         &      0.2940         &      0.4018         &      0.5493         \\
                    &    (0.6652)         &    (0.6641)         &    (1.2452)         &    (0.8062)         \\
      &                        &            &           &    \\[\dimexpr-\normalbaselineskip+2pt] 
Disgust$_{-10,-2}$             &      2.5680         &      2.2372         &      2.6393         &      2.1848         \\
                    &    (1.7652)         &    (1.7677)         &    (3.0227)         &    (2.1069)         \\
      &                        &            &           &    \\[\dimexpr-\normalbaselineskip+2pt] 
Anger$_{-10,-2}$               &      0.0334         &     -0.3450         &      2.5157         &     -2.1957         \\
                    &    (2.4488)         &    (2.4459)         &    (4.4799)         &    (3.0352)         \\
      &                        &            &           &    \\[\dimexpr-\normalbaselineskip+2pt] 
Fear$_{-10,-2}$                &      0.4902         &      0.4003         &      0.5965         &      0.5223         \\
                    &    (0.3891)         &    (0.3887)         &    (0.7157)         &    (0.4832)         \\
      &                        &            &           &    \\[\dimexpr-\normalbaselineskip+2pt] 
Surprise$_{-10,-2}$            &      1.0479         &      0.9065         &      2.3052\sym{*}  &      0.6077         \\
                    &    (0.7149)         &    (0.7127)         &    (1.2193)         &    (0.8413)         \\
 &                        &            &           &    \\[\dimexpr-\normalbaselineskip+2pt]
Volatility  &                     &      1.4387\sym{***}&                     &                     \\
                    &                     &    (0.2785)         &                     &                     \\
      &                        &            &           &    \\[\dimexpr-\normalbaselineskip+2pt] 
Volatility $\times$ Happy$_{-10,-2}$&                     &     -1.8081\sym{*}  &                     &                     \\
                    &                     &    (1.0352)         &                     &                     \\
      &                        &            &           &    \\[\dimexpr-\normalbaselineskip+2pt] 
Constant            &      0.5219         &      0.3269         &     -1.2743\sym{**} &      1.6897\sym{***}\\
                    &    (0.3322)         &    (0.3308)         &    (0.5399)         &    (0.4328)         \\
\hline
 &                        &            &           &    \\[\dimexpr-\normalbaselineskip+2pt]
Firm FE             &           X         &           X         &           X         &           X         \\
Year, Month, Day of Week FE&           X         &           X         &           X         &           X         \\
Control Variables & X & X & X &X \\
$\sigma_{y, within}$&      8.1392         &      8.1387         &      8.1245         &      7.8016         \\
Observations        &  80492         &  80469         &  27448         &  51504         \\
adj. $R^2$          &      0.0854         &      0.0863         &      0.0898         &      0.0519         \\
\hline\hline
\end{tabular}
\begin{tablenotes}
\footnotesize 
\item Notes:  Robust standard errors clustered at the industry and quarter level are in parentheses. \sym{*} \(p<0.10\), \sym{**} \(p<0.05\), \sym{***} \(p<0.01\). Continuous variables winsorized at the 1\% and 99\% level to mitigate the impact of outliers. I report the within-firm (demeaned) standard deviation of the dependent variable. Indicator variable for experiencing volatility in the top 10\% leading up to the announcement (Volatility). 
\end{tablenotes}
\end{threeparttable}
\end{table}

%% file: Tables/announcement_preannouncement_heterogeneity_users.tex
\begin{sidewaystable}[htbp]\centering
\scriptsize 
\begin{threeparttable}
\def\sym#1{\ifmmode^{#1}\else\(^{#1}\)\fi}
\caption{Pre-Announcement Emotions and Announcement Returns across Users \label{tab:reg_pre_announcement_mood_returns_heterogeneity_users}}
\begin{tabular}{l*{11}{c}}
\hline\hline
& & & & & & & & & & & \\[\dimexpr-\normalbaselineskip+2pt]
                    &\multicolumn{1}{c}{(1)}&\multicolumn{1}{c}{(2)}&\multicolumn{1}{c}{(3)}&\multicolumn{1}{c}{(4)}&\multicolumn{1}{c}{(5)}&\multicolumn{1}{c}{(6)}&\multicolumn{1}{c}{(7)}&\multicolumn{1}{c}{(8)}&\multicolumn{1}{c}{(9)}&\multicolumn{1}{c}{(10)}&\multicolumn{1}{c}{(11)}\\
                    & \multicolumn{2}{c}{Trading Approach} & \multicolumn{2}{c}{Investment Horizon} & \multicolumn{3}{c}{Trading Experience} & \multicolumn{2}{c}{Popularity} & \multicolumn{2}{c}{Account Type} \\
                    &\multicolumn{1}{c}{Technical}&\multicolumn{1}{c}{Fundamental}&\multicolumn{1}{c}{Long-Term}&\multicolumn{1}{c}{Short-Term}&\multicolumn{1}{c}{Professional}&\multicolumn{1}{c}{Intermediate}&\multicolumn{1}{c}{Novice}&\multicolumn{1}{c}{Top 5\%}&\multicolumn{1}{c}{Rest}&\multicolumn{1}{c}{Institution}&\multicolumn{1}{c}{Trader}\\ \hline
& & & & & & & & & & & \\[\dimexpr-\normalbaselineskip+2pt]
Happy$_{-10,-2}$               &     -0.2518\sym{**} &     -0.3770\sym{**} &     -0.1995         &     -0.0147         &     -0.3922\sym{***}&     -0.0738         &     -0.4237\sym{**} &     -0.4732\sym{***}&     -0.5287\sym{***}&      0.0231         &     -0.5363\sym{***}\\
                    &    (0.1264)         &    (0.1549)         &    (0.1722)         &    (0.1826)         &    (0.1452)         &    (0.1574)         &    (0.1869)         &    (0.1709)         &    (0.1890)         &    (0.1760)         &    (0.1858)         \\
& & & & & & & & & & & \\[\dimexpr-\normalbaselineskip+2pt]
Sad$_{-10,-2}$                 &      0.3607         &      0.7488         &      0.2779         &      0.6395         &      0.4344         &      0.2652         &      0.0905         &      0.1569         &      0.5196         &     -1.0240         &      0.4570         \\
                    &    (0.3903)         &    (0.5447)         &    (0.5172)         &    (0.4630)         &    (0.5096)         &    (0.4691)         &    (0.4401)         &    (0.5868)         &    (0.6326)         &    (0.6360)         &    (0.6470)         \\
& & & & & & & & & & & \\[\dimexpr-\normalbaselineskip+2pt]
Disgust$_{-10,-2}$             &     -1.1741         &      0.7368         &      1.0984         &     -0.0098         &      0.4807         &     -1.7150         &      2.4639         &     -0.3395         &      2.6239         &      1.6832         &      1.8661         \\
                    &    (1.6891)         &    (1.8056)         &    (2.0663)         &    (2.3011)         &    (1.9836)         &    (1.9473)         &    (1.5957)         &    (2.7879)         &    (1.8238)         &    (1.8907)         &    (1.7735)         \\
& & & & & & & & & & & \\[\dimexpr-\normalbaselineskip+2pt]
Anger$_{-10,-2}$               &     -2.4514         &     -0.9718         &      2.7262         &     -0.5905         &     -1.5046         &      0.1033         &     -1.3494         &     -3.8852         &     -0.1071         &      4.0924         &     -0.6228         \\
                    &    (2.1222)         &    (2.7324)         &    (3.0251)         &    (2.8356)         &    (2.8813)         &    (2.1588)         &    (1.2150)         &    (3.9396)         &    (2.3984)         &    (2.6472)         &    (2.3500)         \\
& & & & & & & & & & & \\[\dimexpr-\normalbaselineskip+2pt]
Fear$_{-10,-2}$                &     -0.0468         &      0.0447         &     -0.2501         &      0.2915         &      0.3356         &      0.3687         &      0.0216         &     -0.1295         &      0.3173         &      0.4696         &      0.3853         \\
                    &    (0.1967)         &    (0.3584)         &    (0.3569)         &    (0.2970)         &    (0.3032)         &    (0.2844)         &    (0.3107)         &    (0.2878)         &    (0.3649)         &    (0.3036)         &    (0.3752)         \\
& & & & & & & & & & & \\[\dimexpr-\normalbaselineskip+2pt]
Surprise$_{-10,-2}$            &     -0.0907         &      0.3807         &     -0.0301         &      0.1403         &     -0.1350         &      0.5638         &     -0.0573         &     -0.1769         &      0.7662         &      0.2107         &      1.0712         \\
                    &    (0.5229)         &    (0.5388)         &    (0.4919)         &    (0.5070)         &    (0.6094)         &    (0.6188)         &    (0.3504)         &    (0.6512)         &    (0.6574)         &    (0.5916)         &    (0.6892)         \\
& & & & & & & & & & & \\[\dimexpr-\normalbaselineskip+2pt]
Constant            &      0.2669         &      0.0783         &      0.1693         &     -0.3485         &      0.3813         &      0.2132         &      0.4092         &      0.2085         &      0.5105         &     -0.5456         &      0.5247         \\
                    &    (0.3913)         &    (0.4000)         &    (0.4395)         &    (0.6031)         &    (0.3541)         &    (0.4113)         &    (0.8218)         &    (0.5414)         &    (0.3304)         &    (0.5812)         &    (0.3311)         \\
\hline
& & & & & & & & & & & \\[\dimexpr-\normalbaselineskip+2pt]
Firm FE             &           X         &           X         &           X         &           X         &           X         &           X         &           X         &           X         &           X         &           X         &           X         \\
Year, Month, Day of Week FE&           X         &           X         &           X         &           X         &           X         &           X         &           X         &           X         &           X         &           X         &           X         \\
Control Variables & X & X &X&X&X&X&X&X&X&X&X \\ 
$\sigma_{y, within}$&      8.3455         &      8.2059         &      8.2603         &      8.5650         &      8.1225         &      8.3073         &      9.0067         &      8.3757         &      8.1403         &      8.4991         &      8.1411         \\
Observations        &  59331         &  57302         &  49863         &  32927         &  71115         &  59831         &  22628         &  41081         &  80179         &  35014         &  80398         \\
adj. $R^2$          &      0.0818         &      0.0828         &      0.0814         &      0.0733         &      0.0839         &      0.0827         &      0.0785         &      0.0737         &      0.0851         &      0.0783         &      0.0854         \\
\hline\hline
\end{tabular}
\begin{tablenotes}
\footnotesize 
\item Notes:  Robust standard errors clustered at the industry and quarter level are in parentheses. \sym{*} \(p<0.10\), \sym{**} \(p<0.05\), \sym{***} \(p<0.01\). Continuous variables winsorized at the 1\% and 99\% level to mitigate the impact of outliers. I report the within-firm (demeaned) standard deviation of the dependent variable. 
\end{tablenotes}
\end{threeparttable}
\end{sidewaystable}

%% file: Tables/announcement_preannouncement_robustness.tex
\begin{sidewaystable}[htbp]\centering
\begin{threeparttable}
\small 
\def\sym#1{\ifmmode^{#1}\else\(^{#1}\)\fi}
\caption{Sensitivity Analysis \label{tab:reg_robustness}}
\begin{tabular}{l*{7}{c}}
\hline\hline
&      &             &           &           &            &           &    \\[\dimexpr-\normalbaselineskip+2pt]
                    &\multicolumn{1}{c}{(1)}&\multicolumn{1}{c}{(2)}&\multicolumn{1}{c}{(3)}&\multicolumn{1}{c}{(4)}&\multicolumn{1}{c}{(5)}&\multicolumn{1}{c}{(6)}&\multicolumn{1}{c}{(7)}\\
                    & 2010-2013 &\multicolumn{2}{c}{Alternative Dep. Var.}&\multicolumn{1}{c}{Longer Window}&\multicolumn{1}{c}{Twitter Model}&\multicolumn{1}{c}{Unweighted}&\multicolumn{1}{c}{Alternative Weighting}\\
                    \multicolumn{7}{l}{Dependent Variable: EXRET$_{-1,1}$} \\ 
\hline
&      &             &           &           &            &           &    \\[\dimexpr-\normalbaselineskip+2pt]
Happy               &                     &     -0.6022\sym{***}&     -0.6024\sym{***}&     -0.9121\sym{***}&     -0.9187\sym{**} &     -0.8015\sym{***}&     -0.5605\sym{***}\\
                    &                     &    (0.1944)         &    (0.1943)         &    (0.2654)         &    (0.3861)         &    (0.2022)         &    (0.1713)         \\
&      &             &           &           &            &           &    \\[\dimexpr-\normalbaselineskip+2pt]
Sad                 &                     &      0.2700         &      0.1995         &      0.0516         &      1.3213         &      0.8529         &      0.4125         \\
                    &                     &    (0.6616)         &    (0.6599)         &    (0.8061)         &    (1.2923)         &    (0.6524)         &    (0.4647)         \\
&      &             &           &           &            &           &    \\[\dimexpr-\normalbaselineskip+2pt]
Disgust             &                     &      1.8904         &      1.7418         &      3.4765\sym{*}  &      6.8876         &      3.1532\sym{*}  &      1.6039\sym{*}  \\
                    &                     &    (1.7589)         &    (1.7539)         &    (1.9949)         &    (8.5461)         &    (1.6381)         &    (0.8593)         \\
&      &             &           &           &            &           &    \\[\dimexpr-\normalbaselineskip+2pt]
Anger               &                     &     -0.8264         &     -0.7744         &      1.2356         &     -9.1694         &      0.8667         &      0.2786         \\
                    &                     &    (2.4123)         &    (2.4100)         &    (2.9537)         &    (8.6857)         &    (2.2006)         &    (1.0009)         \\
&      &             &           &           &            &           &    \\[\dimexpr-\normalbaselineskip+2pt]
Fear                &                     &      0.4674         &      0.4435         &      0.4154         &     -1.5889         &      0.4329         &      0.3064         \\
                    &                     &    (0.3881)         &    (0.3861)         &    (0.5047)         &    (1.6614)         &    (0.3767)         &    (0.2903)         \\
&      &             &           &           &            &           &    \\[\dimexpr-\normalbaselineskip+2pt]
Surprise            &                     &      0.9738         &      0.9188         &      1.4252         &      0.4809         &      0.6559         &      0.7545\sym{*}  \\
                    &                     &    (0.7074)         &    (0.7101)         &    (0.9251)         &    (1.3585)         &    (0.6586)         &    (0.4547)         \\
&      &             &           &           &            &           &    \\[\dimexpr-\normalbaselineskip+2pt]
Sentiment           &      0.0638\sym{**} &                     &                     &                     &                     &                     &                     \\
                    &    (0.0262)         &                     &                     &                     &                     &                     &                     \\
&      &             &           &           &            &           &    \\[\dimexpr-\normalbaselineskip+2pt]
Constant            &     -0.2561         &      0.6121\sym{*}  &      0.6052\sym{*}  &      0.5662\sym{*}  &      0.6313\sym{*}  &      0.5464         &      0.5370         \\
                    &    (0.2398)         &    (0.3291)         &    (0.3270)         &    (0.3330)         &    (0.3423)         &    (0.3336)         &    (0.3314)         \\
\hline
&      &             &           &           &            &           &    \\[\dimexpr-\normalbaselineskip+2pt]
Firm FE             &                     &           X         &           X         &           X         &           X         &                     &           X         \\
Year, Month, Day of Week FE&                     &           X         &           X         &           X         &           X         &                     &           X         \\
Control Variables & X & X & X & X & X & X &X \\
$\sigma_{y}$        &      8.4107         &      8.0730         &      8.0696         &      8.1072         &      8.1072         &      8.1072         &      8.1072         \\
Observations        &  21470         &  83385         &  83385         &  83385         &  83385         &  83385         &  83385         \\
adj. $R^2$          &      0.0820         &      0.0892         &      0.0894         &      0.0868         &      0.0867         &      0.0869         &      0.0869         \\
\hline\hline
\end{tabular}
\begin{tablenotes}
\footnotesize 
\item Notes:  Robust standard errors clustered at the industry and quarter level are in parentheses. \sym{*} \(p<0.10\), \sym{**} \(p<0.05\), \sym{***} \(p<0.01\). Continuous variables winsorized at the 1\% and 99\% level to mitigate the impact of outliers. Aside from Column (1), I report the within-firm (demeaned) standard deviation of the dependent variable. In Column (2), I use cumulative abnormal returns over the three-day trading window as my dependent variable, while in Column (3) I use a longer estimation window for excess returns (see Figure \ref{fig:excess_calculation}). The "Long Window" in Column (4) refers to $[-20,2]$, while my alternative weighting in Column (7) refers to my like-based weighting scheme: 1 + $\log(1+$\# of likes).
\end{tablenotes}
\end{threeparttable}
\end{sidewaystable}

%% file: Tables/labeling.tex
\begin{table}[htbp]\centering
\small 
\caption{Generating Training Data: Dictonary Based Labeling Accuracy}\label{tab:labeling}
\begin{threeparttable}
\begin{tabular}{p{3in}p{0.75in}p{0.75in}p{0.75in}}
Class & Correct & Incorrect & Empirical Accuracy \\ \hline 
& & &  \\[\dimexpr-\normalbaselineskip+2pt]
Neutral & 1251 & 36 & 97.2\% \\
Happy & 837 & 110 & 88.4\% \\
Sad & 163 & 51 & 76.2\% \\
Anger & 84 & 12 & 87.5\% \\
Disgust & 99 & 36 & 73.3\% \\
Surprise & 310 & 44 & 87.6\% \\
Fear & 330 & 73 & 81.9\% \\ \hline \hline
\end{tabular}
\begin{tablenotes}
\footnotesize 
\item Labeling accuracy evaluated on the hand-tagged 5,000 messages. Note: my labeling model would cover 68.7\% of these messages, so it would not classify 1,564 messages (i.e., would not label the message with any of our classes). 
\end{tablenotes}
\end{threeparttable}
\end{table}

%% file: Tables/five_fold.tex
\begin{table}[htbp]\centering
\small
\begin{threeparttable}
\caption{Five Fold Cross Validation with Hand-Tagged Test Sample}\label{tab:classification_test}
\begin{tabular}{lllllllll}
&\multicolumn{4}{c}{\uline{StockTwits}} & \multicolumn{4}{c}{\uline{Twitter}} \\ 
& & & & & & & &  \\[\dimexpr-\normalbaselineskip+2pt]
&\multicolumn{2}{c}{In-Sample} & \multicolumn{2}{c}{Test Sample} & \multicolumn{2}{c}{In-Sample} & \multicolumn{2}{c}{Test Sample} \\
Fold & Loss & Accuracy & Loss & Accuracy & Loss & Accuracy & Loss & Accuracy \\ \hline 
& & & & & & & &  \\[\dimexpr-\normalbaselineskip+2pt]
\#1 & 0.0218 & 99.13\% & 1.9602 & 80.38\% & 0.8669 & 69.60\% & 1.6034 & 47.96\% \\
\#2 & 0.0216 & 99.14\% & 1.9071 & 79.86\% & 0.8679 & 69.97\% & 1.6562 & 45.88\% \\
\#3 & 0.0216 & 99.11\% & 2.0113 & 79.82\% & 0.8569 & 70.49\% & 1.5904 & 47.68\% \\
\#4 & \textbf{0.0212} & \textbf{99.16\%} & \textbf{2.1130} & \textbf{78.14\%} & \textbf{0.8517} & \textbf{70.51\%} & \textbf{1.5862} & \textbf{47.74\%} \\
\#5 & 0.0216 & 99.14\% & 1.9954 & 80.94\% & 0.8518 & 70.31\% & 1.6575 & 46.74\% \\ \hline \hline 
\end{tabular}
\begin{tablenotes}
\footnotesize
\item Test sample refers to the hand-tagged 5,000 messages; these messages are never fed into the model during training. Model that gets selected based on in-sample loss in bold.
\end{tablenotes}
\end{threeparttable}
\end{table}

%% file: Tables/twit_examples.tex
\begin{table}[htbp]\centering
\footnotesize 
\caption{Examples of Model Outputs}\label{tab:twit_examples}
\begin{tabular}{p{4in}p{0.5in}p{0.5in}p{0.5in}}
Text & Emotion & Emotion (\%) & Type \\ \hline 
& & \\ \\[\dimexpr-\normalbaselineskip+2pt]
angryfacesymbolshead angryfacesymbolshead   angryfacesymbolshead angryfacesymbolshead & anger & 100.0\% & chat \\
i hate google chrome so buggy even with   windows isnumbervalue & anger & 100.0\% & chat \\
fu shorts f*** right the f*** off & anger & 100.0\% & finance \\
d*mn it i do not have funds in yob it how   long does it usually take & anger & 100.0\% & finance \\
this guy is f*g made accounts to shill   his own account lm*ao & disgust & 99.7\% & chat \\
those shorter are silent now say   something losers & disgust & 100.0\% & chat \\
new high the master mind price gouging   and manipulator martin ceo will push up & disgust & 68.7\% & finance \\
now we know which a**holes are shorting   crooks isnumbervalue & disgust & 97.4\% & finance \\
dump it & fear & 100.0\% & chat \\
big boys dumping ah & fear & 100.0\% & chat \\
these fools bashing instead of loading   all they can while still under dont want to make a lot of lol isnumbervalue & fear & 90.4\% & chat \\
er is nov what s the problem   isnumbervalue & fear & 100.0\% & finance \\
what would you do price may drop as it   breaks higher dillinger band view odds of downtrend & fear & 99.4\% & finance \\
a gift & happy & 100.0\% & chat \\
love the fear they try to spread it s   literally a discord channel for them they try unbelievably hard i ll give   them that isnumbervalue & happy & 98.0\% & chat \\
trend reversed in complete bull momentum   and will continue to rally hard going up thumbsup & happy & 100.0\% & finance \\
way way undervalued here rocket   moneywithwings rocket moneywithwings rocket moneywithwings rocket   moneywithwings & happy & 100.0\% & finance \\
same patterns & neutral & 100.0\% & chat \\
brussels is the center of european union & neutral & 100.0\% & chat \\
bank of marin bangor ceo russell colombo   sells in isdollarvalue isnumbervalue & neutral & 100.0\% & finance \\
just filed a earnings release and a   financial exhibit & neutral & 100.0\% & finance \\
ding dong the witch is dead for now & sad & 100.0\% & chat \\
tough to watch having doubts will even   hold might have to not watch and check back in a few months brutal   isnumbervalue & sad & 100.0\% & chat \\
this stock is brutal & sad & 100.0\% & finance \\
stop bleeding they are reducing staff and   working on getting a billion dollars tax credit isnumbervalue & sad & 100.0\% & finance \\
this makes no sense lmao & surprise & 100.0\% & chat \\
this thing might even push holy crap   isdollarvalue & surprise & 100.0\% & chat \\
i seriously doubt that if you are still   holding or god forbid buying at these levels right before an earnings miss & surprise & 100.0\% & finance \\
gifted some shares at the opening bell   lol isdollarvalue & surprise & 86.3\% & finance \\ \hline 
\end{tabular}
\end{table}

%% file: Tables/shap_stocktwits.tex
\begin{sidewaystable}[htbp]\centering
\caption{Shap Values: StockTwits Model}\label{tab:shap_stocktwits}
\begin{threeparttable}
\scriptsize 
\begin{tabular}{llllllllllllll}
\multicolumn{2}{c}{Neutral}  & \multicolumn{2}{c}{Happy}  & \multicolumn{2}{c}{Sad}  & \multicolumn{2}{c}{Anger} & \multicolumn{2}{c}{Disgust} & \multicolumn{2}{c}{Surprise}  & \multicolumn{2}{c}{Fear}  \\ \hline 
& & & & & & & & & & & & & \\ [\dimexpr-\normalbaselineskip+2pt]
large & 1.916 & large & 4.808 & ripping & 0.680 & large & 1.615 & losers & 0.500 & insane & 0.808 & worthless & 0.801 \\
new & 0.858 & new & 1.810 & fail & 0.602 & mad & 0.600 & idiot & 0.458 & curious & 0.710 & careful & 0.698 \\
events & 0.554 & calm & 1.072 & sad & 0.588 & su** & 0.492 & pathetic & 0.440 & crazy & 0.706 & safety & 0.661 \\
calm & 0.504 & funny & 1.048 & terrible & 0.587 & f*** & 0.468 & idiots & 0.408 & wtf & 0.624 & trapped & 0.646 \\
full & 0.489 & welcome & 1.037 & worst & 0.567 & f**ked & 0.452 & shorty & 0.395 & holy & 0.595 & nervous & 0.631 \\
impressive & 0.478 & right & 1.003 & pain & 0.565 & hate & 0.416 & trash & 0.382 & why & 0.591 & downtrend & 0.616 \\
excellent & 0.466 & won & 0.982 & crap & 0.563 & b*tch & 0.413 & lmfao & 0.366 & seriously & 0.586 & dumped & 0.611 \\
insane & 0.453 & best & 0.977 & failed & 0.557 & welcome & 0.412 & pig & 0.365 & ridiculous & 0.570 & trouble & 0.609 \\
welcome & 0.445 & full & 0.971 & bad & 0.548 & stupid & 0.384 & loser & 0.358 & unusual & 0.552 & problem & 0.589 \\
winning & 0.443 & glad & 0.915 & ouch & 0.546 & dumb & 0.370 & junk & 0.357 & whoa & 0.546 & halted & 0.588 \\
latest & 0.437 & events & 0.913 & ugh & 0.542 & full & 0.359 & bullshit & 0.354 & surprised & 0.546 & falls & 0.580 \\
right & 0.433 & own & 0.908 & dead & 0.526 & right & 0.355 & fake & 0.352 & wow & 0.536 & worry & 0.577 \\
glad & 0.433 & kind & 0.904 & crushed & 0.521 & su**s & 0.353 & garbage & 0.348 & surprise & 0.527 & bashing & 0.576 \\
careful & 0.432 & congrats & 0.897 & scalp & 0.520 & largest & 0.341 & clown & 0.346 & jesus & 0.525 & tanking & 0.574 \\
great & 0.431 & loving & 0.890 & maxpain & 0.492 & f***ing & 0.332 & lie & 0.337 & nuts & 0.508 & rs & 0.574 \\
amazing & 0.431 & love & 0.883 & desperate & 0.480 & d*mn & 0.329 & turd & 0.334 & omg & 0.500 & dumping & 0.573 \\
awesome & 0.430 & impressive & 0.873 & silly & 0.473 & sh** & 0.320 & fool & 0.333 & god & 0.460 & correction & 0.571 \\
worst & 0.430 & fair & 0.872 & alone & 0.471 & tv & 0.295 & bags & 0.328 & new & 0.448 & manipulated & 0.569 \\
live & 0.429 & locked & 0.871 & sadface & 0.463 & pileofpoo & 0.286 & screwed & 0.327 & geez & 0.440 & worries & 0.558 \\
beautiful & 0.429 & amazing & 0.861 & rip & 0.459 & trades & 0.260 & scam & 0.321 & thinkingface & 0.387 & dropping & 0.558 \\
terrible & 0.429 & excellent & 0.860 & death & 0.449 & mostly & 0.259 & bashers & 0.317 & personshrugging & 0.369 & falling & 0.558 \\
fair & 0.426 & excited & 0.838 & facewithrollingeyes & 0.441 & own & 0.257 & fools & 0.309 & flushedface & 0.304 & panic & 0.556 \\
nice & 0.425 & hopefully & 0.835 & regret & 0.422 & a** & 0.255 & bs & 0.297 & sick & 0.287 & pumping & 0.554 \\
funny & 0.425 & storm & 0.833 & missed & 0.417 & best & 0.237 & pumpers & 0.295 & welcome & 0.253 & recession & 0.550 \\
loving & 0.421 & huge & 0.821 & losing & 0.409 & calm & 0.234 & bagholders & 0.293 & best & 0.224 & covering & 0.546 \\
safety & 0.418 & bullish & 0.819 & bleeding & 0.405 & kind & 0.232 & bag & 0.285 & lol & 0.215 & trap & 0.545 \\
easy & 0.416 & winning & 0.816 & smh & 0.373 & fair & 0.231 & fraud & 0.282 & calm & 0.206 & manipulation & 0.544 \\
golden & 0.416 & great & 0.811 & hurt & 0.357 & hilarious & 0.227 & pumper & 0.271 & interesting & 0.185 & contracts & 0.544 \\
kind & 0.414 & hope & 0.802 & decent & 0.337 & TRUE & 0.226 & lies & 0.270 & zanyface & 0.183 & selloff & 0.535 \\
failed & 0.413 & cheers & 0.799 & personfacepalming & 0.320 & tired & 0.223 & lmao & 0.256 & guess & 0.179 & dropped & 0.528 \\
best & 0.412 & interested & 0.797 & tired & 0.309 & new & 0.220 & stupid & 0.064 & full & 0.167 & crash & 0.525 \\
love & 0.409 & nice & 0.795 & events & 0.300 & freaking & 0.218 & holder & 0.053 & won & 0.163 & attack & 0.522 \\
good & 0.408 & moon & 0.792 & new & 0.281 & latest & 0.215 & f***ing & 0.045 & sure & 0.156 & bearish & 0.514 \\
worthless & 0.408 & golden & 0.791 & crude & 0.226 & live & 0.214 & peers & 0.043 & hopefully & 0.155 & bubble & 0.511 \\
nervous & 0.407 & live & 0.789 & base & 0.210 & fine & 0.209 & welcome & 0.039 & live & 0.154 & downside & 0.500 \\
nicely & 0.404 & awesome & 0.788 & poor & 0.185 & positive & 0.203 & special & 0.038 & tired & 0.147 & fear & 0.500 \\
trapped & 0.403 & program & 0.786 & exp & 0.184 & high & 0.202 & hate & 0.038 & wonder & 0.146 & scared & 0.495 \\
pain & 0.402 & beautiful & 0.781 & black & 0.184 & across & 0.202 & boring & 0.037 & happened & 0.144 & greedy & 0.495 \\
excited & 0.402 & lol & 0.772 & weak & 0.175 & fly & 0.200 & firm & 0.034 & huge & 0.142 & overbought & 0.488 \\
bad & 0.402 & happy & 0.771 & late & 0.171 & quick & 0.199 & poor & 0.033 & kind & 0.138 & pullback & 0.483 \\
own & 0.400 & decent & 0.763 & tough & 0.166 & rich & 0.197 & mad & 0.033 & storm & 0.138 & pump & 0.481 \\
wins & 0.400 & rich & 0.761 & tight & 0.165 & lol & 0.195 & pure & 0.025 & right & 0.130 & squeeze & 0.476 \\
ouch & 0.398 & finally & 0.761 & sorry & 0.165 & dillinger & 0.190 & su**s & 0.025 & funny & 0.129 & risky & 0.475 \\
mad & 0.398 & bargain & 0.758 & loudlycryingface & 0.160 & ignore & 0.182 & a** & 0.024 & TRUE & 0.129 & scare & 0.473 \\
happy & 0.397 & good & 0.757 & common & 0.159 & good & 0.182 & sound & 0.024 & outstanding & 0.126 & divergence & 0.454 \\
fly & 0.393 & exciting & 0.754 & wrong & 0.158 & perfect & 0.178 & dumb & 0.024 & wants & 0.125 & warning & 0.449 \\
lol & 0.393 & rally & 0.753 & loving & 0.157 & pro & 0.176 & stuckout & 0.023 & fly & 0.124 & worried & 0.430 \\
crazy & 0.393 & special & 0.750 & worse & 0.156 & top & 0.171 & faith & 0.023 & weird & 0.122 & bearface & 0.421 \\
sweet & 0.391 & positive & 0.749 & active & 0.154 & funny & 0.167 & noise & 0.022 & begin & 0.122 & dump & 0.399 \\
perfect & 0.389 & easy & 0.748 & flushedface & 0.148 & outstanding & 0.165 & favorite & 0.022 & good & 0.121 & bomb & 0.382\\ \hline \hline 
\end{tabular}
\begin{tablenotes}
\scriptsize 
\item Average absolute SHAP values evaluated on a random sample of 100,000 StockTwits messages. Words reported, followed by their corresponding average absolute SHAP values, are the 50 most important words that appear at least 50 times in the SHAP sample. stuckout = facewithstuckouttonguewinkingeye.
\end{tablenotes}
\end{threeparttable}
\end{sidewaystable}

%% file: Tables/shap_twitter.tex
\begin{sidewaystable}[htbp]\centering
\caption{Shap Values: Twitter Model}\label{tab:shap_twitter}
\begin{threeparttable}
\scriptsize 
\begin{tabular}{llllllllllllll}
\multicolumn{2}{c}{Neutral}  & \multicolumn{2}{c}{Happy}  & \multicolumn{2}{c}{Sad}  & \multicolumn{2}{c}{Anger} & \multicolumn{2}{c}{Disgust} & \multicolumn{2}{c}{Surprise}  & \multicolumn{2}{c}{Fear}  \\ \hline 
& & & & & & & & & & & & & \\ [\dimexpr-\normalbaselineskip+2pt]
natural & 3.814 & natural & 4.292 & picking & 1.169 & snap & 0.440 & natural & 0.692 & revenue & 4.417 & picking & 1.758 \\
picking & 2.464 & picking & 4.242 & jury & 1.090 & natural & 0.262 & unusual & 0.553 & natural & 2.307 & revenue & 1.319 \\
pressure & 2.028 & mess & 2.000 & lies & 0.779 & surprise & 0.159 & pileofpoo & 0.493 & st & 1.738 & natural & 1.175 \\
favor & 1.800 & science & 1.869 & eth & 0.773 & shake & 0.159 & stuck & 0.416 & expecting & 1.703 & redheart & 1.012 \\
david & 1.714 & cloud & 1.835 & revenue & 0.758 & waste & 0.155 & scared & 0.374 & q & 1.510 & favor & 0.800 \\
cloud & 1.651 & david & 1.805 & favor & 0.752 & idiots & 0.151 & b*tch & 0.359 & beyond & 0.979 & eth & 0.705 \\
revenue & 1.631 & mining & 1.784 & natural & 0.726 & slap & 0.145 & mistake & 0.320 & weird & 0.949 & seriously & 0.677 \\
marketing & 1.621 & marketing & 1.755 & su** & 0.722 & mad & 0.140 & f*** & 0.320 & signals & 0.857 & cheers & 0.657 \\
mess & 1.590 & rebound & 1.700 & hi & 0.716 & private & 0.120 & hate & 0.298 & surprise & 0.848 & cat & 0.649 \\
accumulation & 1.500 & sounds & 1.659 & dead & 0.690 & f*** & 0.115 & signal & 0.295 & fighting & 0.786 & mining & 0.636 \\
talks & 1.479 & presentation & 1.580 & radar & 0.680 & completely & 0.112 & su** & 0.284 & talks & 0.770 & fighting & 0.623 \\
presentation & 1.424 & talks & 1.520 & fade & 0.655 & a** & 0.099 & storm & 0.280 & winners & 0.740 & scared & 0.622 \\
sounds & 1.410 & holiday & 1.463 & itself & 0.620 & dude & 0.099 & net & 0.273 & swings & 0.687 & opposite & 0.614 \\
scared & 1.400 & itself & 1.440 & sounds & 0.617 & weird & 0.099 & statement & 0.269 & scared & 0.673 & gamble & 0.613 \\
b*tch & 1.354 & boring & 1.435 & stuck & 0.615 & accumulate & 0.098 & limited & 0.264 & fear & 0.647 & dead & 0.598 \\
itself & 1.340 & material & 1.430 & reading & 0.614 & facewithrollingeyes & 0.098 & themselves & 0.260 & panic & 0.629 & david & 0.571 \\
impressive & 1.305 & betting & 1.425 & boring & 0.572 & grab & 0.096 & fighting & 0.253 & bids & 0.618 & sounds & 0.546 \\
multi & 1.183 & impressive & 1.424 & fighting & 0.571 & trash & 0.094 & sh** & 0.253 & idiots & 0.597 & prob & 0.545 \\
serious & 1.152 & along & 1.331 & deep & 0.569 & smilingheart & 0.093 & idiots & 0.249 & owner & 0.572 & itself & 0.536 \\
somebody & 1.132 & ideas & 1.264 & both & 0.563 & mess & 0.092 & competition & 0.248 & science & 0.565 & radar & 0.529 \\
awesome & 1.113 & pressure & 1.263 & lost & 0.533 & join & 0.091 & favor & 0.246 & omg & 0.560 & boring & 0.518 \\
tried & 1.111 & ish & 1.246 & cat & 0.532 & hype & 0.090 & omg & 0.240 & nervous & 0.554 & st & 0.516 \\
pivot & 1.110 & whale & 1.235 & cancer & 0.511 & sh** & 0.088 & f***ing & 0.236 & pivot & 0.553 & expecting & 0.506 \\
impact & 1.088 & impact & 1.211 & mess & 0.509 & rocket & 0.088 & picking & 0.235 & basically & 0.532 & signals & 0.500 \\
beyond & 1.087 & awesome & 1.209 & trump & 0.505 & ridiculous & 0.086 & mad & 0.231 & flag & 0.532 & missing & 0.492 \\
fighting & 1.084 & okhand & 1.163 & ga & 0.500 & mentioned & 0.086 & bo & 0.229 & awesome & 0.531 & beyond & 0.491 \\
along & 1.083 & serious & 1.149 & scared & 0.499 & stuck & 0.085 & worthless & 0.226 & beautiful & 0.527 & impressive & 0.481 \\
ish & 1.069 & storm & 1.132 & rolling* & 0.493 & mins & 0.084 & ridiculous & 0.226 & scare & 0.526 & multi & 0.481 \\
rebound & 1.061 & somebody & 1.122 & weak & 0.488 & dumb & 0.083 & su**s & 0.223 & wall & 0.503 & feeling & 0.480 \\
stuck & 1.056 & pivot & 1.092 & dillinger & 0.487 & hate & 0.083 & pressure & 0.221 & vs & 0.496 & material & 0.478 \\
truly & 1.028 & amazing & 1.072 & sad & 0.473 & moneywithwings & 0.083 & d*mn & 0.214 & yes & 0.495 & acting & 0.470 \\
shake & 1.021 & multi & 1.064 & multi & 0.473 & joke & 0.082 & accumulation & 0.213 & multi & 0.492 & omg & 0.466 \\
betting & 1.004 & lots & 1.056 & side & 0.468 & beyond & 0.081 & boring & 0.212 & stocktwits & 0.490 & tried & 0.465 \\
president & 1.004 & google & 1.048 & loss & 0.464 & blow & 0.078 & ranks & 0.211 & somebody & 0.489 & q & 0.464 \\
omg & 0.999 & finally & 1.038 & strike & 0.461 & stanley & 0.077 & social & 0.210 & acquired & 0.484 & plz & 0.463 \\
movement & 0.985 & smell & 1.033 & alot & 0.457 & af & 0.077 & fair & 0.210 & lots & 0.465 & broke & 0.461 \\
boring & 0.975 & beyond & 1.029 & feeling & 0.457 & panic & 0.076 & multi & 0.210 & monster & 0.463 & accumulation & 0.455 \\
stuff & 0.968 & classic & 1.028 & worried & 0.457 & multi & 0.076 & mess & 0.208 & congrats & 0.461 & mess & 0.450 \\
exactly & 0.946 & stuff & 1.028 & canada & 0.441 & crap & 0.076 & part & 0.207 & movement & 0.449 & su** & 0.445 \\
lots & 0.946 & powell & 1.006 & tried & 0.431 & favor & 0.075 & blocked & 0.206 & portfolio & 0.436 & deep & 0.438 \\
science & 0.943 & car & 0.994 & missing & 0.430 & beating & 0.075 & grab & 0.205 & positions & 0.429 & storm & 0.422 \\
gamble & 0.942 & completely & 0.977 & intel & 0.425 & clown & 0.074 & session & 0.197 & holiday & 0.426 & recovery & 0.418 \\
dead & 0.910 & scam & 0.975 & momentum & 0.421 & talks & 0.074 & beyond & 0.192 & favor & 0.424 & f*** & 0.418 \\
stocktwits & 0.902 & gas & 0.973 & worst & 0.415 & momo & 0.072 & perfect & 0.187 & performance & 0.418 & burn & 0.411 \\
happyface & 0.899 & cool & 0.947 & slow & 0.411 & bull & 0.072 & surprised & 0.186 & tank & 0.417 & spread & 0.408 \\
ideas & 0.891 & congrats & 0.935 & smilingsun* & 0.411 & okhand & 0.071 & panic & 0.183 & momentum & 0.411 & minimum & 0.407 \\
amazing & 0.888 & dog & 0.931 & losers & 0.411 & literally & 0.071 & vs & 0.179 & become & 0.405 & biotech & 0.404 \\
owner & 0.886 & exactly & 0.928 & hand & 0.410 & boring & 0.071 & somebody & 0.179 & pig & 0.395 & extremely & 0.404 \\
radar & 0.882 & funny & 0.928 & rally & 0.404 & d*mn & 0.070 & intraday & 0.176 & grinning* & 0.390 & funds & 0.399 \\
signals & 0.878 & beautiful & 0.918 & signals & 0.401 & vs & 0.070 & ha & 0.176 & picking & 0.388 & scam & 0.395 \\ \hline \hline 
\end{tabular}
\begin{tablenotes}
\scriptsize 
\item Average absolute SHAP values evaluated on a random sample of 100,000 StockTwits messages. Words reported, followed by their corresponding average absolute SHAP values, are the 50 most important words that appear at least 50 times in the SHAP sample. rolling* = rollingonthefloorlaughing, smilingsun*=smilingfacewithsunglasses, smilingheart=smilingfacewithhearteyes, grinning*=grinningfacewithsmilingeyes.
\end{tablenotes}
\end{threeparttable}
\end{sidewaystable}

%% file: Tables/shap_finance.tex
\begin{table}[htbp]\centering \squeezeup
\scriptsize 
\caption{Shap Values: Chat Type}\label{tab:shap_finance}
\begin{threeparttable}
\begin{tabular}{p{4in}p{1in}}
\multicolumn{2}{c}{Most Important Words}   \\ \hline 
&  \\ [\dimexpr-\normalbaselineskip+2pt]
transaction & 0.366 \\
break & 0.340 \\
bears & 0.328 \\
shorts & 0.325 \\
missed & 0.319 \\
performance & 0.318 \\
played & 0.317 \\
bear & 0.313 \\
broke & 0.312 \\
board & 0.311 \\
downgraded & 0.310 \\
climb & 0.308 \\
longs & 0.308 \\
bulls & 0.304 \\
news & 0.304 \\
bought & 0.300 \\
bull & 0.298 \\
reversal & 0.297 \\
buying & 0.295 \\
halt & 0.294 \\
oversold & 0.291 \\
move & 0.291 \\
pumpers & 0.291 \\
stop & 0.290 \\
close & 0.289 \\
economy & 0.287 \\
highs & 0.286 \\
charts & 0.286 \\
rally & 0.285 \\
miss & 0.285 \\
profits & 0.284 \\
sells & 0.284 \\
support & 0.284 \\
spike & 0.283 \\
breaks & 0.283 \\
long & 0.282 \\
ceo & 0.282 \\
stops & 0.282 \\
moving & 0.281 \\
bankruptcy & 0.281 \\
bullish & 0.281 \\
sold & 0.280 \\
upgraded & 0.279 \\
production & 0.278 \\
premarket & 0.277 \\
fed & 0.277 \\
watchlist & 0.277 \\
plays & 0.276 \\
vix & 0.276 \\
bearish & 0.275 \\ \hline \hline 
\end{tabular}
\begin{tablenotes}
\scriptsize 
\item Average absolute SHAP values evaluated on a random sample of 100,000 StockTwits messages. Words reported, followed by their corresponding average absolute SHAP values, are the 50 most important words that appear at least 50 times in the SHAP sample. 
\end{tablenotes}
\end{threeparttable}
\end{table}

%% file: Tables/event_quarterly.tex
\begin{table}[htbp]\centering 
\footnotesize
\caption{Distribution of Twits by Calendar Quarter \label{tab:twits_by_quarter}}
\def\sym#1{\ifmmode^{#1}\else\(^{#1}\)\fi}
\begin{tabular}{l*{2}{c}}
\hline\hline
& & \\[\dimexpr-\normalbaselineskip+2pt]
                    &\multicolumn{1}{c}{(1)}&\multicolumn{1}{c}{(2)}\\
                    &\multicolumn{1}{c}{Firm-Quarter Observations} &\multicolumn{1}{c}{Twits} \\
\hline
& & \\[\dimexpr-\normalbaselineskip+2pt]
\hline
2010Q1                 &         518         &        8780         \\

& & \\[\dimexpr-\normalbaselineskip+2pt]
2010Q2                 &         717         &       12717         \\

& & \\[\dimexpr-\normalbaselineskip+2pt]
2010Q3                &         630         &       12390         \\

& & \\[\dimexpr-\normalbaselineskip+2pt]
2010Q4                 &         850         &       14126         \\

& & \\[\dimexpr-\normalbaselineskip+2pt]
2011Q1                 &        1231         &       20488         \\

& & \\[\dimexpr-\normalbaselineskip+2pt]
2011Q2                 &        1297         &       22567         \\

& & \\[\dimexpr-\normalbaselineskip+2pt]
2011Q3                 &        1332         &       23730         \\

& & \\[\dimexpr-\normalbaselineskip+2pt]
2011Q4                 &        1354         &       27495         \\

& & \\[\dimexpr-\normalbaselineskip+2pt]
2012Q1                 &        1495         &       35889         \\

& & \\[\dimexpr-\normalbaselineskip+2pt]
2012Q2                &        1220         &       47628         \\

& & \\[\dimexpr-\normalbaselineskip+2pt]
2012Q3                 &        1257         &       50147         \\

& & \\[\dimexpr-\normalbaselineskip+2pt]
2012Q4                 &        1206         &       58129         \\

& & \\[\dimexpr-\normalbaselineskip+2pt]
2013Q1                 &        1739         &       85368         \\

& & \\[\dimexpr-\normalbaselineskip+2pt]
2013Q2                 &        1610         &       58828         \\

& & \\[\dimexpr-\normalbaselineskip+2pt]
2013Q3                &        1813         &       52671         \\

& & \\[\dimexpr-\normalbaselineskip+2pt]
2013Q4                &        1803         &       85817         \\

& & \\[\dimexpr-\normalbaselineskip+2pt]
2014Q1                  &        2229         &      105835         \\

& & \\[\dimexpr-\normalbaselineskip+2pt]
2014Q2                 &        2075         &       64998         \\

& & \\[\dimexpr-\normalbaselineskip+2pt]
2014Q3                 &        2307         &       88872         \\

& & \\[\dimexpr-\normalbaselineskip+2pt]
2014Q4                 &        2523         &       81936         \\

& & \\[\dimexpr-\normalbaselineskip+2pt]
2015Q1                  &        2592         &      104294         \\

& & \\[\dimexpr-\normalbaselineskip+2pt]
2015Q2                 &        2626         &      111147         \\

& & \\[\dimexpr-\normalbaselineskip+2pt]
2015Q3                 &        2437         &      116640         \\

& & \\[\dimexpr-\normalbaselineskip+2pt]
2015Q4                &        2380         &      129342         \\

& & \\[\dimexpr-\normalbaselineskip+2pt]
2016Q1                 &        2627         &      113920         \\

& & \\[\dimexpr-\normalbaselineskip+2pt]
2016Q2                 &        2653         &      125894         \\

& & \\[\dimexpr-\normalbaselineskip+2pt]
2016Q3                  &        2611         &      137382         \\

& & \\[\dimexpr-\normalbaselineskip+2pt]
2016Q4                  &        2493         &      167053         \\

& & \\[\dimexpr-\normalbaselineskip+2pt]
2017Q1                   &        2835         &      198029         \\

& & \\[\dimexpr-\normalbaselineskip+2pt]
2017Q2                 &        2736         &      197646         \\

& & \\[\dimexpr-\normalbaselineskip+2pt]
2017Q3                 &        2791         &      202406         \\

& & \\[\dimexpr-\normalbaselineskip+2pt]
2017Q4                &        2608         &      203824         \\

& & \\[\dimexpr-\normalbaselineskip+2pt]
2018Q1                &        2606         &      222239         \\

& & \\[\dimexpr-\normalbaselineskip+2pt]
2018Q2                 &        2879         &      270343         \\

& & \\[\dimexpr-\normalbaselineskip+2pt]
2018Q3                 &        2587         &      207093         \\

& & \\[\dimexpr-\normalbaselineskip+2pt]
2018Q4                 &        2394         &      216816         \\

& & \\[\dimexpr-\normalbaselineskip+2pt]
2019Q1                &        2705         &      208230         \\

& & \\[\dimexpr-\normalbaselineskip+2pt]
2019Q2                 &        2777         &      210109         \\

& & \\[\dimexpr-\normalbaselineskip+2pt]
2019Q3                  &        2711         &      179989         \\

& & \\[\dimexpr-\normalbaselineskip+2pt]
2019Q4                 &        2632         &      186654         \\

& & \\[\dimexpr-\normalbaselineskip+2pt]
Total               &       81886         &     4467461         \\

\hline\hline
\end{tabular}
\end{table}

%% file: Tables/event_crsp_tab.tex
\begin{table}[htbp]\centering  \squeezeup
\begin{threeparttable}
\caption{Distribution of Twits Based on Fama-French 48-Industry Classification\label{tab:fama_fench}}
\footnotesize 
\def\sym#1{\ifmmode^{#1}\else\(^{#1}\)\fi}
\begin{tabular}{l*{3}{c}}
\hline\hline
& & & \\[\dimexpr-\normalbaselineskip+2pt]
                    &\multicolumn{1}{c}{(1)}&\multicolumn{1}{c}{(2)}&\multicolumn{1}{c}{(3)}\\
Fama-French industry code (48 industries)                    &         CRSP (\%)&         Twits (\%)& Firm-Quarters (\%) \\
\hline
& & & \\[\dimexpr-\normalbaselineskip+2pt]
Agriculture         &        0.26&        0.03&        0.14\\
Food Products       &        1.12&        0.46&        1.21\\
Candy \& Soda        &        0.35&        0.17&        0.34\\
Beer \& Liquor       &        0.28&        0.09&        0.27\\
Tobacco Products    &        0.13&        0.10&        0.16\\
Recreation          &        0.48&        0.13&        0.41\\
Entertainment       &        0.89&        2.14&        0.97\\
Printing and Publishing&        0.62&        0.10&        0.44\\
Consumer Goods      &        0.97&        0.35&        0.98\\
Apparel             &        0.70&        0.60&        0.81\\
Healthcare          &        1.28&        0.63&        1.25\\
Medical Equipment   &        2.43&        1.08&        2.23\\
Pharmaceutical Products&        4.75&       10.87&        4.89\\
Chemicals           &        1.70&        0.67&        1.95\\
Rubber and Plastic Products&        0.34&        0.08&        0.36\\
Textiles            &        0.22&        0.01&        0.13\\
Construction Materials&        0.97&        0.15&        0.91\\
Construction        &        1.04&        0.25&        1.21\\
Steel Works Etc     &        0.97&        0.77&        1.02\\
Fabricated Products &        0.22&        0.07&        0.16\\
Machinery           &        2.23&        1.13&        2.35\\
Electrical Equipment&        1.23&        1.79&        1.04\\
Automobiles and Trucks&        1.23&        0.61&        1.40\\
Aircraft            &        0.40&        0.48&        0.54\\
Shipbuilding, Railroad Equipment&        0.16&        0.04&        0.17\\
Defense             &        0.18&        0.14&        0.19\\
Precious Metals     &        0.36&        0.25&        0.45\\
Non-Metallic and Industrial Metal Mining&        0.50&        0.87&        0.63\\
Coal                &        0.29&        0.20&        0.35\\
Petroleum and Natural Gas&        4.18&        3.62&        5.21\\
Utilities           &        3.00&        1.04&        3.12\\
Communication       &        2.73&        1.60&        2.23\\
Personal Services   &        1.07&        0.36&        0.97\\
Business Services   &       10.41&       14.22&       10.86\\
Computers           &        2.16&        8.68&        2.22\\
Electronic Equipment&        5.02&       10.49&        5.10\\
Measuring and Control Equipment&        1.42&        0.71&        1.41\\
Business Supplies   &        0.74&        0.09&        0.65\\
Shipping Containers &        0.18&        0.04&        0.24\\
Transportation      &        3.14&        1.37&        3.45\\
Wholesale           &        2.64&        0.95&        2.64\\
Retail              &        3.82&        4.22&        4.71\\
Restaraunts, Hotels, Motels&        1.57&        1.34&        1.81\\
Banking             &        8.61&        1.86&        5.85\\
Insurance           &        2.98&        0.70&        2.89\\
Real Estate         &        0.75&        0.19&        0.46\\
Trading             &        7.65&        1.98&        6.73\\
Almost Nothing      &       11.62&       22.32&       12.53\\
\hline
Observations        &    11471550&     4467461&       81886\\
\hline\hline
 
\end{tabular}
\begin{tablenotes}
\scriptsize 
\item CRSP sample corresponds to 2010-2019 NASDAQ/NYSE subsample.
\end{tablenotes}
\end{threeparttable}
\end{table}